\documentclass[fleqn,usenatbib]{mnras}

\usepackage{newtxtext,newtxmath}

\usepackage[T1]{fontenc}

\DeclareRobustCommand{\VAN}[3]{#2}
\let\VANthebibliography\thebibliography
\def\thebibliography{\DeclareRobustCommand{\VAN}[3]{##3}\VANthebibliography}

\usepackage{graphicx}	
\usepackage{amsmath}	
\usepackage{rotating}
\usepackage{wasysym}    

\usepackage{orcidlink}
\usepackage{xspace}
\usepackage{ulem}
\makeatletter 
  \patchcmd{\NAT@citex}
    {\@citea\NAT@hyper@{%
      \NAT@nmfmt{\NAT@nm}%
      \hyper@natlinkbreak{\NAT@aysep\NAT@spacechar}{\@citeb\@extra@b@citeb}%
      \NAT@date}}
    {\@citea\NAT@nmfmt{\NAT@nm}%
    \NAT@aysep\NAT@spacechar\NAT@hyper@{\NAT@date}}{}{}

  \patchcmd{\NAT@citex}
    {\@citea\NAT@hyper@{%
      \NAT@nmfmt{\NAT@nm}%
      \hyper@natlinkbreak{\NAT@spacechar\NAT@@open\if*#1*\else#1\NAT@spacechar\fi}%
        {\@citeb\@extra@b@citeb}%
      \NAT@date}}
    {\@citea\NAT@nmfmt{\NAT@nm}%
    \NAT@spacechar\NAT@@open\if*#1*\else#1\NAT@spacechar\fi\NAT@hyper@{\NAT@date}}
    {}{}
\makeatother

\newcommand{\HM}{\ion{H}{$_2$}\xspace}
\newcommand{\HI}{\ion{H}{I}\xspace}
\newcommand{\HII}{\ion{H}{II}\xspace}
\newcommand{\HeI}{\ion{He}{I}\xspace}
\newcommand{\HeII}{\ion{He}{II}\xspace}
\newcommand{\HeIII}{\ion{He}{III}\xspace}
\newcommand\OIII{\ion{O}{III}\xspace} 

\newcommand{\thesan}{\textsc{thesan}\xspace}
\newcommand{\thzoom}{\mbox{\textsc{thesan-zoom}}\xspace}

\title[Chemical evolution in the early Universe]
{The \thzoom project: Mystery N/O more --- uncovering the origin of peculiar chemical abundances and a not-so-fundamental metallicity relation at $3<z<12$}

\author[W. McClymont et al.]{%
William McClymont $\orcidlink{0009-0009-5565-3790}$,$^{1,2}$\thanks{E-mail: \href{mailto:wjm50@cam.ac.uk}{wjm50@cam.ac.uk} (WM)}
Sandro Tacchella $\orcidlink{0000-0002-8224-4505}$,$^{1,2}$
Aaron Smith $\orcidlink{0000-0002-2838-9033}$,$^{3}$
Rahul Kannan $\orcidlink{0000-0001-6092-2187}$,$^{4}$
\newauthor
Enrico Garaldi $\orcidlink{0000-0002-6021-7020}$,$^{5}$
Ewald Puchwein $\orcidlink{0000-0001-8778-7587}$,$^{6}$
Yuki Isobe $\orcidlink{0000-0001-7730-8634}$,$^{1,2}$
Xihan Ji $\orcidlink{0000-0002-1660-9502}$,$^{1,2}$
Xuejian Shen $\orcidlink{0000-0002-6196-823X}$,$^{7}$
\newauthor
Zihao Wang,$^{7,8}$
Vasily Belokurov  $\orcidlink{0000-0002-0038-9584}$,$^{9}$
Josh Borrow  $\orcidlink{0000-0002-1327-1921}$,$^{10}$
Francesco D'Eugenio $\orcidlink{0000-0003-2388-8172}$,$^{1,2}$
Laura Keating $\orcidlink{0000-0001-5211-1958}$,$^{11}$
\newauthor
Roberto Maiolino $\orcidlink{0000-0002-4985-3819}$,$^{1,2}$
Stephanie Monty $\orcidlink{0000-0002-9225-5822}$,$^{9}$
Mark Vogelsberger $\orcidlink{0000-0001-8593-7692}$,$^{7}$
and Oliver Zier $\orcidlink{0000-0003-1811-8915}$$^{12,7}$
\\
\\
$^{1}$Kavli Institute for Cosmology, University of Cambridge, Madingley Road, Cambridge CB3 0HA, UK\\
$^{2}$Cavendish Laboratory, University of Cambridge, 19 JJ Thomson Avenue, Cambridge CB3 0HE, UK\\
$^3$ Department of Physics, The University of Texas at Dallas, Richardson, TX 75080, USA \\
$^4$ Department of Physics and Astronomy, York University, 4700 Keele Street, Toronto, ON M3J 1P3, Canada \\
$^5$ Kavli IPMU (WPI), UTIAS, The University of Tokyo, Kashiwa, Chiba 277-8583, Japan \\
$^6$ Leibniz-Institut f\"ur Astrophysik Potsdam, An der Sternwarte 16, 14482 Potsdam, Germany \\
$^{7}$ Department of Physics, Kavli Institute for Astrophysics and Space Research, Massachusetts Institute of Technology, Cambridge, MA 02139, USA \\
$^{8}$ School of Astronomy and Space Science, Nanjing University, Nanjing, Jiangsu 210093, People’s Republic of China \\
$^{9}$Institute of Astronomy, University of Cambridge, Madingley Road, Cambridge CB3 0HA, UK \\
$^{10}$ Department of Physics and Astronomy, University of Pennsylvania, 209 South 33rd Street, Philadelphia, PA 19104, USA \\
$^{11}$ Institute for Astronomy, University of Edinburgh, Blackford Hill, Edinburgh, EH9 3HJ, UK \\
$^{12}$ Center for Astrophysics $|$ Harvard $\&$ Smithsonian, 60 Garden Street, Cambridge, MA 02138, USA
}

\date{Accepted XXX. Received YYY; in original form ZZZ}

\pubyear{\the\year{}}

\begin{document}
\label{firstpage}
\pagerange{\pageref{firstpage}--\pageref{lastpage}}
\maketitle

\begin{abstract}
We present an analysis of metallicities and chemical abundances at $3<z<12$ in the \thzoom simulations. We find that smoothly curved gas-phase and stellar mass-metallicity relations (MZR) are already in place at $z\approx12$ and evolve slowly ($\sim$0.2\,dex increase for gas, $\sim$0.4\,dex increase for stars at a fixed stellar mass) down to $z=3$, governed largely by the efficiency with which galaxies retain their metals, rather than gas fraction. The canonical fundamental metallicity relation (FMR) survives in stars but breaks down and inverts for gas in low-mass galaxies ($M_\ast\lesssim10^{9}\mathrm{M_\odot}$) due to regular dilution by low-metallicity gas inflow. We find broad agreement of gas-phase N/O, Fe/O, and C/O with high-redshift observations, including the presence of nitrogen-rich galaxies (NRGs; $\log(\mathrm{N/O})>-0.6$) without the need for exotic yields in our chemical network. Instead, bursty star formation naturally generates order-of-magnitude excursions in N/O on $\lesssim$100\,Myr timescales due to temporally differential galactic winds; after a starburst, stellar feedback expels gas, leaving a large population of asymptotic-giant-branch stars to dominate the enrichment of the relatively low-mass interstellar medium. NRGs lie below the main sequence and typically exhibit EW[H$\beta$] $\lesssim$40\,\AA, in apparent tension with observed high-EW NRGs. This tension is reconciled if observed NRGs are in the initial stages of a subsequent starburst, illuminating previously enriched gas, which is supported by the finding of high SFR surface density nitrogen-rich giant molecular clouds.
\end{abstract}




\section{Introduction}
\label{sec:Introduction}

The primordial Universe was chemically simple, composed only of hydrogen, helium, and a sprinkle of lithium \citep{Hoyle:1964aa,Peebles:1966aa,Wagoner:1967aa}. The collapse of this pristine gas in dark matter mini-haloes precipitated the ignition of the Universe's first stars, which acted as metal factories through the nuclear fusion of primordial elements \citep{Klessen:2023aa}. These metals, defined as elements heavier than helium, were dispersed into the surrounding gas, from which subsequent generations of stars would form, marking the beginning of a new, chemically complex chapter of the Universe's history. Since then, the powerful cooling effect of metals, as well as the clumping of metals into dust grains, has had a dramatic impact on fundamental processes in galaxy evolution, including star formation and the growth of black holes \citep{Maiolino:2019aa,Kobayashi:2020aa}.

In addition to their direct impact on galaxies, metals act as an excellent tracer of galaxy evolution \citep{Maiolino:2019aa}. Metals are synthesized in stars and supernovae and expelled into the interstellar medium (ISM) through stellar feedback mechanisms, including supernovae and stellar winds, which means that metallicities are sensitive to star-formation histories, outflows, and the accretion of pristine and recycled gas \citep{Kobayashi:2020aa}. On galactic scales, the key relation in this context is the mass-metallicity relation (MZR), which is an empirical relationship linking the stellar mass of galaxies to their gas-phase metallicity \citep{Lequeux:1979aa,Tremonti:2004aa,Lee:2006aa}. Higher mass galaxies are found to have higher metallicities, before flattening toward a saturation metallicity \citep{Andrews:2013aa,Curti:2020aa}. The existence of the MZR has been attributed to a variety of effects, including that more massive galaxies are more chemically evolved \citep{Somerville:2015aa,Baker:2023ac} and that less massive galaxies are less able to hold onto their metal-enriched gas due to their shallow gravitational potentials \citep{Christensen:2016aa}. At a fixed mass, metallicity has been shown to increase with decreasing star-formation rate \citep[SFR;][]{Ellison:2008aa}. This led to the development of the Fundamental Metallicity Relation \citep[FMR;][]{Mannucci:2010aa,Yates:2012aa,Telford:2016aa}, which is a scaling relation between metallicity and both stellar mass and SFR. The scatter along the FMR is smaller by a factor of $\sim$30\% compared to the MZR scatter, indicating that the secondary SFR dependence is important \citep{Curti:2020aa}, and the FMR has been shown not to evolve significantly out to $z=3$ \citep{Salim:2015aa,Sanders:2018aa,Sanders:2021aa,Topping:2021aa}.

The advent of the \textit{James Webb Space Telescope} (\textit{JWST}) has revolutionized our understanding of galaxy evolution, especially at high redshift, by providing unprecedented sensitivity and resolution in the near-infrared regime \citep{Stark:2025aa}. \textit{JWST} has enabled the routine measurement of metallicities in the first two Gyr \citep{Curti:2023aa,Matthee:2023aa,Rhoads:2023aa,DEugenio:2024ab,Witten:2025aa}, with tentative measurements made even out to $z\approx14$, implying rapid metal enrichment \citep{Carniani:2025aa,Naidu:2025ab}. Recent \textit{JWST} observations have significantly extended our knowledge of the MZR into the early Universe, revealing only modest evolution in metallicities from $z\sim10$ to $z\sim3$ \citep{Langeroodi:2023aa,Nakajima:2023aa,Shapley:2023aa,Curti:2024ab,Pollock:2025aa}. Furthermore, deviations from the FMR have emerged at these early epochs, hinting at complex feedback and enrichment mechanisms unique to early galaxies \citep{Heintz:2023aa,Curti:2024ab,Pollock:2025aa,Scholte:2025aa}.

A variety of approaches have been used to advance our understanding of galactic chemical evolution. The most basic is the analytic closed-box model, in which metal enrichment is modeled in an isolated system as gas is converted to stars, enriching the ISM in the process \citep{Talbot:1971aa,Searle:1972aa,Tinsley:1980aa}. Improvements were later made in the form of accreting and leaky box models, which account for the inflow and outflow of gas and their effects on galaxies' metallicities \citep{Larson:1972aa,Larson:1974aa,Edmunds:1990aa}. These early models set the stage for more advanced descriptions of galaxy evolution, such as equilibrium and gas regulator models, which explicitly model the inflow and outflow of gas and enrichment of gas through star formation. In equilibrium models, galaxies achieve a quasi-steady state wherein metal production by star formation is balanced by metal loss through galactic winds \citep{Finlator:2008aa,Dave:2012aa}, and these models assume that the metallicity is largely set by the current rates of metal production and loss, rather than placing emphasis on the long-term history of galaxies. On the other hand, gas regulator models explicitly allow for the gas reservoir of galaxies to vary over time \citep{Lilly:2013aa,Peng:2014aa}. These models have been able to produce key features of the MZR, such as its slope and high-mass turnover. 

Modern cosmological simulations have successfully captured many features of observed galaxy populations \citep{Vogelsberger:2020aa}, including reproducing the broad trends of the MZR and demonstrating the critical roles of feedback-driven outflows and gas recycling in governing metallicity evolution \citep{Ma:2016aa,Torrey:2018aa,Marszewski:2024aa}. However, results from cosmological simulations have not reached a clear consensus on the evolution of the MZR, with predictions of both strong \citep{Torrey:2018aa} and weak \citep{Ma:2016aa,Langan:2020aa,Katz:2023ab,Wilkins:2023aa} evolution of the MZR with redshift. The evolution of the MZR with redshift is usually attributed to the evolution of gas fractions \citep{Ma:2016aa,Torrey:2018aa,Langan:2020aa}, however, other explanations, such as the evolution of inflow/outflow metallicity and mass-loading factors \citep{Bassini:2024aa,Marszewski:2025aa}, have been proposed.

While a great deal has been revealed through the study of bulk metallicities, the abundances of individual elements can provide more detailed information. Different elements are created through different nucleosynthetic channels, each with a different characteristic time delay from the onset of star formation. Massive stars ($M>8\,\mathrm{M_\odot}$) significantly contribute oxygen and other alpha-elements via core-collapse supernovae (CC SNe) on $\sim$Myr timescales after the onset of star formation. Intermediate-mass stars ($1-8\,\mathrm{M_\odot}$) contribute carbon and nitrogen through asymptotic giant branch (AGB) winds \citep{Chiappini:2003aa,Karakas:2014aa,Kobayashi:2020aa} on a timescales of $\sim$40\,Myr to $\sim$10\,Gyr, depending on the star's mass. Finally, Fe-peak elements are enriched through CC SNe, but primarily through the explosion of intermediate-mass stars through Type Ia Supernovae (SNe Ia) on similar timescales as AGB enrichment \citep{Maoz:2010aa}. Consequently, measuring and modeling chemical abundances provides critical insights into star formation histories and the relative importance of various stellar populations within galaxies \citep{Maiolino:2019aa}. 

\textit{JWST} has revealed unusual abundance patterns, notably enhanced nitrogen-to-oxygen (N/O) ratios in high-redshift galaxies identified with UV emission lines \citep{Bunker:2023aa,Cameron:2023ab,Isobe:2023aa,Ji:2024aa,Schaerer:2024aa}. Nitrogen-rich galaxies (NRGs; $\log(\mathrm{N/O})>0.6$) show a number of interesting properties, including high electron densities \citep{Pascale:2023aa,Ji:2024aa,Topping:2024aa,Topping:2025aa,Yanagisawa:2024ab,Isobe:2025aa}, stellar mass surface densities \citep{Schaerer:2024aa,Topping:2025aa}, and small sizes \citep{Harikane:2025aa}, although selection effects may bias these apparent associations \citep{Morel:2025aa}. Various scenarios have been invoked to understand the origin of NRGs: pristine gas inflows dilute the metallicity of the gas while leaving abundance ratios unaffected, causing low metallicity galaxies to show the higher C/O and N/O more typical of higher-metallicity galaxies \citep{DAntona:2023aa,Stiavelli:2025aa}; Wolf-Rayet stars can eject nitrogen-enriched gas via stellar winds which, along with intermittent star formation, could lead to the observed abundances of NRGs \citep{Kobayashi:2024aa}; differential galactic winds, whereby CCSNe-enriched gas is preferentially ejected can lead to nitrogen-rich abundances, even without intermittent star formation \citep{Rizzuti:2025aa}; very massive and supermassive stars formed via runaway collisions in dense stellar environments or extremely massive stars also lead to nitrogen-rich gas, and this explanation has also been used to link the nitrogen-enrichment of high-redshift galaxies to the nitrogen-rich generation of stars seen in local globular clusters \citep{Charbonnel:2023aa, Belokurov:2023aa,Senchyna:2024aa,Gieles:2025aa,Ji:2026aa}. Interestingly, \citet{Isobe:2025aa} find that stacked spectra of broad-line (BL) active galactic nuclei (AGN) show nitrogen enhancement, demonstrating a potential link between nitrogen-rich abundances and AGN activity. 

The \thzoom project \citep{Kannan:2025aa} is a suite of high-resolution zoom-in radiation-hydrodynamic simulations of high-redshift galaxies selected from the larger \thesan cosmological volume \citep{Kannan:2022aa,Smith:2022ab,Garaldi:2022aa,Garaldi:2024aa}. The \thzoom simulations incorporate a detailed multi-phase ISM model that includes explicit treatments of stellar feedback processes, radiative transfer, dust physics, and non-equilibrium thermochemistry, offering an unprecedented capability to explore the detailed interplay of feedback, metal enrichment, and galaxy growth processes. Notably, the \thzoom simulations explicitly track the abundances of nine chemical elements (H, He, C, N, O, Ne, Mg, Si, and Fe), tracing their injection from stellar particles and diffusion throughout gas \citep{Vogelsberger:2013aa,Marinacci:2019aa}. In this study, we use the \thzoom simulations to provide a detailed exploration of chemical abundances at high redshift. We aim to understand deviations from local scaling relations, the origin of enhanced nitrogen abundances, and their implications for the underlying star formation and feedback processes.

In Section~\ref{sec:Simulations}, we describe the simulation methodology in detail. In Section~\ref{sec:Metallicity}, we present our results on the evolution and scatter of the MZR and the FMR, including the breakdown of the FMR in low mass galaxies. In Section~\ref{sec:Chemical abundances}, we discuss chemical abundances and the underlying physical processes driving them, focusing in particular on the emergence of nitrogen-rich galaxies. In Section~\ref{sec:Discussion} we discuss the implications of our results in the context of the high-redshift baryon cycle and chemical enrichment networks. In Section~\ref{sec:Conclusions}, we summarize our key findings.

\section{Simulations}
\label{sec:Simulations}

This study examines the \thzoom suite of cosmological zoom-in simulations \citep{Kannan:2025aa}. The suite represents a high-resolution follow-up to the original, large-volume ($\sim$100\,cMpc) \thesan runs \citep{Kannan:2022aa, Smith:2022ab, Garaldi:2022aa}, which were built on the IllustrisTNG galaxy-formation model \citep{Pillepich:2018aa} and incorporated on-the-fly radiative transfer. Although the parent \thesan simulations have been widely exploited to probe galaxy evolution through the Epoch of Reionisation \citep[EoR; e.g.][]{Yeh:2023aa,Neyer:2024aa,Garaldi:2024aa,Shen:2024aa,Jamieson:2025aa}, their effective-equation-of-state treatment of the ISM limits their ability to address parsec-scale star-formation physics. By contrast, \thzoom retains the cosmological environment of \thesan while resolving the ISM and stellar feedback in detail, and has already served as the basis for studies of the high-redshift main sequence and burstiness \citep{McClymont:2025aa}, galaxy-scale star-formation efficiencies \citep[SFEs;][]{Shen:2025aa}, the imprint of reionisation on galaxies \citep{Zier:2025aa}, Population III star formation \citep{Zier:2025ab}, the evolution of galaxy sizes \citep{McClymont:2025ab}, and cloud-scale SFEs \citep{Wang:2025aa}.

\subsection{Simulation details}
\label{sec:Simulation details}

Each zoom region is selected from the parent \thesan box and inherits the time-dependent ionising background generated by the larger simulation, capturing the impact of external radiative feedback on \thzoom galaxies, which can have a strong impact on high-redshift galaxy properties \citep{Rosdahl:2018aa, Ocvirk:2020aa,Katz:2020aa,Borrow:2023aa}. A comprehensive description of the numerical set-up is given in \citet{Kannan:2025aa}; here we summarise only the aspects most relevant to this work. The calculations are performed with {\sc arepo-rt} \citep{Kannan:2019aa}, the radiation-hydrodynamics extension of the moving-mesh code {\sc arepo} \citep{Springel:2010aa}, now equipped with an efficient node-to-node communication scheme \citep{Zier:2024aa}. Radiation transport is solved on-the-fly using the M1 moment method with a reduced speed of light approximation, where we use $0.01c$. The non-equilibrium thermochemistry follows six species ($\HM$, $\HI$, $\HII$, $\HeI$, $\HeII$, $\HeIII$). Metal-line cooling assumes ionisation equilibrium in the \citet{Faucher-Giguere:2009aa} UV background and employs pre-computed \textsc{cloudy} \citep{Ferland:2017aa} look-up tables \citep{Vogelsberger:2013aa}.

Three resolution tiers are available, which are labeled ``4x'', ``8x'', and ``16x'', corresponding to spatial resolutions 4, 8 and 16 times finer than in \thesan and to baryonic particle masses of $9.09\times10^3\,\mathrm{M}_\odot$, $1.14\times10^3\,\mathrm{M}_\odot$, and $1.42\times10^2\,\mathrm{M}_\odot$ respectively. Throughout this study, we adopt the highest resolution available for each zoom-in region and restrict ourselves to the fiducial physics setup.

The \thzoom framework employs a multi-channel stellar-feedback scheme that incorporates photoionization, radiation pressure, stellar winds, and supernova feedback \citep{Kannan:2020aa, Kannan:2021aa}. Radiation from newly born stars is propagated on-the-fly, so that their ionising photons directly modify the local gas temperature and ionisation state. Supernovae deposit both thermal energy and radial momentum into neighbouring cells, while stellar winds follow the SMUGGLE prescription of \citet{Marinacci:2019aa} which deposits mass, momentum, and metals into gas cells surrounding young (OB) and AGB stars.  Acting together, these processes suppress runaway collapse, launch galactic outflows, and regulate the assembly of early galaxies. An additional “early feedback’’ channel disrupts dense star-forming clouds within a few Myr of the onset of star formation, further moderating the local star-formation efficiency \citep[][]{Kannan:2025aa,Wang:2025aa}.

\subsection{Sample selection}
\label{sec:Sample selection}

\begin{figure*} 
\centering
	\includegraphics[width=\textwidth]{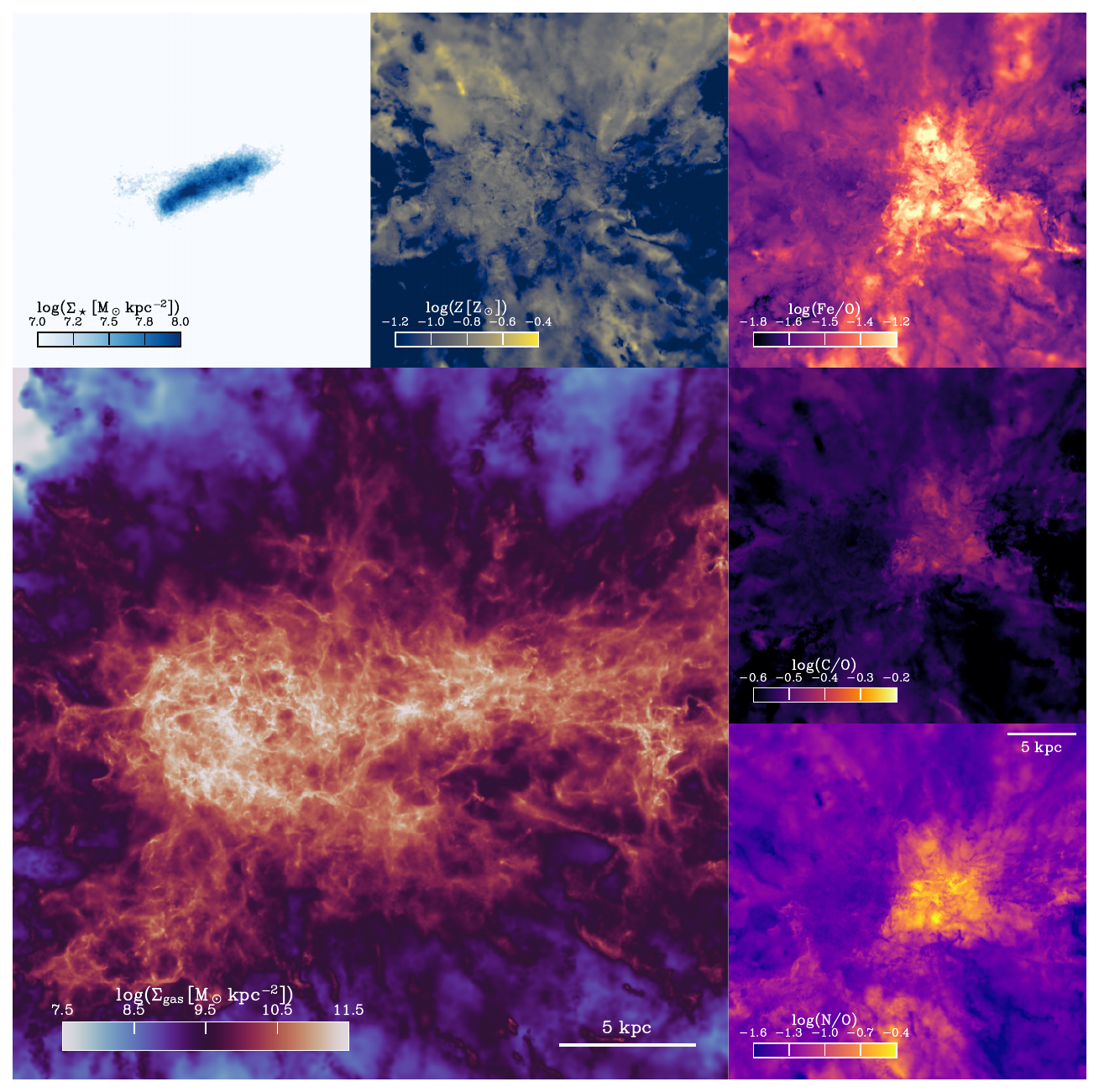}
    \caption{Maps of the central subhalo of m12.6 at $z=6$, with $M_\ast=10^{9.3}\,\mathrm{M_\odot}$. The large panel shows gas surface density and the smaller panels show, clockwise from the upper left, stellar mass surface density, gas-phase metallicity, iron-to-oxygen, carbon-to-oxygen, and nitrogen-to-oxygen. The metallicity and abundance ratios are weighted by gas mass. All maps show the same spatial scales. Considerable variations are seen spatially in the abundance ratios, and in particular the nitrogen-to-oxygen ratio, and such variations are not obviously correlated with the metallicity.}
    \label{fig:metal_maps}
\end{figure*}

\begin{table*}
    \centering
    \begin{tabular}{ccccccc}
        \hline
        \multicolumn{6}{|c|}{Gas-phase redshift-dependent MZR} \\
        \hline
         & $M_0$ & $Z_0$ & $\gamma$ & $\alpha$ & $\beta$\\
        \hline
        $12+\log(\mathrm{O/H})$ & $10.74^{+0.25}_{-0.26}$ & $8.98^{+0.14}_{-0.06}$ & $0.39^{+0.01}_{-0.02}$ & $0.28^{+0.13}_{-0.05}$ & $0.55^{+0.37}_{-0.16}$ \vspace{.1cm} \\
        $Z_\mathrm{gas}\,[\mathrm{Z_\odot}]$ & $10.88^{+0.28}_{-0.15}$ & $0.19^{+0.16}_{-0.08}$ & $0.41^{+0.01}_{-0.00}$ & $0.39^{+0.01}_{-0.05}$ & $0.51^{+0.08}_{-0.14}$\\
        \hline
        \multicolumn{6}{|c|}{Stellar redshift-dependent MZR} \\
        \hline
         & $M_0$ & $Z_0$ & $\gamma$ & $\alpha$ & $\beta$\\
        \hline
        $[\mathrm{Fe/H}]$ & $11.01^{+0.23}_{-0.27}$ & $0.45^{+0.42}_{-0.16}$ & $0.43^{+0.04}_{-0.01}$ & $0.77^{+0.03}_{-0.02}$ & $0.32^{+0.10}_{-0.13}$ \vspace{.1cm} \\
        $Z_\mathrm{\ast}\,[\mathrm{Z_\odot}]$ & $11.27^{+0.75}_{-0.21}$ & $1.63^{+0.09}_{-0.60}$ & $0.56^{+0.01}_{-0.08}$ & $0.43^{+0.03}_{-0.01}$ & $0.10^{+0.06}_{-0.00}$\\
        \hline
    \end{tabular}
    \caption{The best-fit parameters to Eq.~(\ref{eq:mzr_eq}) for the gas-phase oxygen and stellar iron abundances, as well as the gas-phase and stellar total metallicity. Uncertainties were obtained via bootstrap resampling with $n=5000$.}
    \label{tab:mzr_fits}
\end{table*}

We consider only subhalos which are resolved with at least 100 stellar particles, which is effectively a stellar mass cut of $9.09\times10^5\,\mathrm{M}_\odot$, $1.14\times10^5\,\mathrm{M}_\odot$, and $1.42\times10^4\,\mathrm{M}_\odot$ for the 4x, 8x, and 16x resolutions respectively (not accounting for mass loss). Haloes were identified with the friends-of-friends (FOF) algorithm \citep{Davis:1985aa}, and self-gravitating subhalos were identified within the FOF groups using the SUBFIND-HBT algorithm \citep{Springel:2001aa,Springel:2021aa}. We use both central and satellite galaxies (subhaloes) in this analysis, though sometimes exclusively considering centrals where noted. We consider galaxies in the redshift range $3<z<12$. This leaves us with 68137 central subhalos, comprised of 1570 unique merger trees, and 52295 satellite subhaloes, comprised of 1056 unique trees. In total, then, we have a sample of 120\,432 subhaloes, comprised of 1720 unique trees, where the number of trees is less than the sum of centrals and satellites because some galaxies become satellites at later times.

Stellar masses and SFRs for each galaxy are calculated from bound particles within the virial radius, defined by the spherical overdensity criterion $R_\mathrm{vir}=R_{\rm crit,200}$. SFRs are calculated within a given averaging timescale ($t_\mathrm{avg}$) from the stellar particles with
\begin{equation}
  \mathrm{SFR}_{t_\mathrm{avg}} = \frac{\sum_im_{\ast,i}}{t_\mathrm{avg}}\,,
\end{equation}
where $i$ is summing over stellar particles formed within $t_\mathrm{avg}$ and $m_{\ast,i}$ is the stellar particle's initial mass. 

H$\alpha$ luminosities are obtained in post-processing with the Monte Carlo radiative transfer (MCRT) Cosmic Ly$\alpha$ Transfer code \citep[\textsc{colt};][]{Smith:2015aa, Smith:2019aa, Smith:2022aa,McClymont:2025af}. The \textsc{colt} procedure to obtain line and continuum fluxes is exactly as described in \citet{McClymont:2025aa}, which itself largely follows previous works \citep[e.g.][]{Smith:2022aa,Tacchella:2022aa,McClymont:2024aa}. Briefly, \textsc{colt} is used to launch and propagate ionizing photons through the gas, and this procedure is repeated iteratively, updating ionization states until ionization equilibrium is reached, and the ionization states are then used to calculate the H$\alpha$ emission. We calculate the intrinsic H$\beta$ luminosity by dividing the H$\alpha$ luminosities by 2.86 \citep{Osterbrock:2006aa}. This ratio is specifically valid for gas at 10\,000\,K and with an electron density of $10^2~\mathrm{cm^{-3}}$, however it only varies mildly for reasonable values of gas temperature and density, varying by $\sim$4\% from 10\,000\,K to 20\,000\,K and by 2\% from $10^2~\mathrm{cm^{-3}}$ to $10^6~\mathrm{cm^{-3}}$. Although more dramatic offsets have been seen in high-redshift galaxies with \textit{JWST}, this appears to effect a minority of galaxies and the observed ratios imply even for most of this population, the ratio is offset at the $\lesssim20\%$ level \citep{Scarlata:2024aa,Yanagisawa:2024aa,McClymont:2025ad}. H$\beta$ equivalent widths (EWs) are calculated in $f_\nu$ relative to the average continuum in a 5000\AA\,--6000\AA\, window. 

\subsection{Metal enrichment}
\label{sec:Metal enrichment}

Due to the importance of the metal enrichment scheme to this work, we describe the scheme in further detail here. The scheme is largely inherited from \citet{Vogelsberger:2013aa}, particularly with respect to the chemical yields, but see \citet{Marinacci:2019aa} for a more comprehensive overview of the current model, including updates to SNe feedback and stellar winds. The model follows the enrichment of nine chemical elements: H, He, C, N, O, Ne, Mg, Si, and Fe. In Fig.~\ref{fig:metal_maps} we show abundance ratio (Fe/O, C/O, and N/O) and metallicity maps for an example galaxy. Elements are recycled to the ISM through stellar mass loss. CC SNe are responsible for most of the mass loss in massive stars ($>13\,\mathrm{M_\odot}$), whereas less massive stars lose most of their mass via AGB winds. Each stellar particle in the simulation represents a single-age stellar population (SSP) following a \citet{Chabrier:2003aa} IMF with an upper mass cutoff of 100$\mathrm{M_\odot}$, and the mass returned to the ISM of each tracked chemical element is calculated for a given timestep based on elemental mass yields.

The elemental mass yields for CC SNe are from \citet{Portinari:1998aa}, which themselves are calculated from the Padova stellar evolutionary tracks and explosive nucleosynthesis from \citet{Woosley:1995aa}. For AGB stars the elemental mass yields are provided by \citet{Karakas:2010aa} across a range of initial stellar metallicities ($0.0001<Z<0.02$) and masses ($1 < m_\ast\,[\mathrm{M_\odot}] < 6$). The elemental yields are time-dependent for a given SSP, however, the model assumes instantaneous post-main-sequence evolution, which is justified because stars typically spend $\lesssim10\%$ of their lifetimes in a post-main-sequence phase \citep{Vogelsberger:2013aa}. The SNIa mass-loss channel is also included, with the SNIa rate being calculated with a delay time distribution, scaling as $\propto t^{-1.12}$ following the onset of SNIa, 40\,Myr after the stellar particle is formed (see \citealt{Vogelsberger:2013aa} for details). SNIa elemental mass yields are calculated following \citet{Thielemann:2003aa} and \citet{Travaglio:2004aa}

The on-the-fly dust model included in \thzoom largely follows \citet{McKinnon:2016aa,McKinnon:2017aa} and is treated as a scalar property of the gas cells. The model tracks five chemical elements: C, O, Mg, Si, and Fe. However, only the carbon-to-silicon ratio was saved as an output for data storage reasons. Dust is produced in the simulation via AGB stars and through SNe \citep{Dwek:1998aa}. Gas-phase metals can condense onto dust in the ISM, with modified metallicity and temperature-dependent rates introduced in the \thzoom model \citep{Kannan:2025aa}. Dust can be destroyed through shocks \citep{Seab:1983aa,Draine:1993aa,Alarie:2019aa,Flury:2025aa} and sputtering \citep{Draine:1979aa}.

In this work, we generally define gas-phase metallicities using the oxygen abundance ($12+\log(\mathrm{O/H})$) and stellar metallicities with the iron abundance ($[\mathrm{Fe/H}]=\log(\mathrm{Fe/H})-\log(\mathrm{Fe/H})_\odot$), although we do also investigate total metallicity ($Z$). When scaling metallicity to the solar value, we assume $Z_\odot=0.02$. When scaling iron abundances we assume $\log(\mathrm{Fe/H})_\odot+12=7.5$ \citep{Asplund:2009aa}. Metallicities are calculated within twice the stellar half-mass radius. We do not require a minimum number of gas particles in order to avoid biasing our results from excluding low gas fraction galaxies; however, our minimum stellar particle cut ensures all haloes are well-resolved. In the most extreme case, this means that we include galaxies which have zero gas particles within twice the stellar half-mass radius, which comprise 12.5\% of our sample. In reality, these galaxies should of course not have \textit{zero} atoms of gas within their ISM, however, we lack the resolution to resolve such extremely low gas fractions. It is important to include these galaxies in our analysis to accurately report, for example, the correct median gas fraction. However, the gas-phase metallicity for these galaxies is undefined, so we choose to set the metallicity for these galaxies to zero. In general, this may be a strong assumption; however, to minimize the impact of this assumption, we base our fits on the median metallicity values in a given stellar mass and redshift range. This means that our assumption of zero metallicity is actually an assumption that the metallicity of these galaxies is below the median, which is less severe. In any case, the impact on our metallicity fits is small.

\section{Metallicity}
\label{sec:Metallicity}

\begin{figure*} 
\centering
	\includegraphics[width=\textwidth]{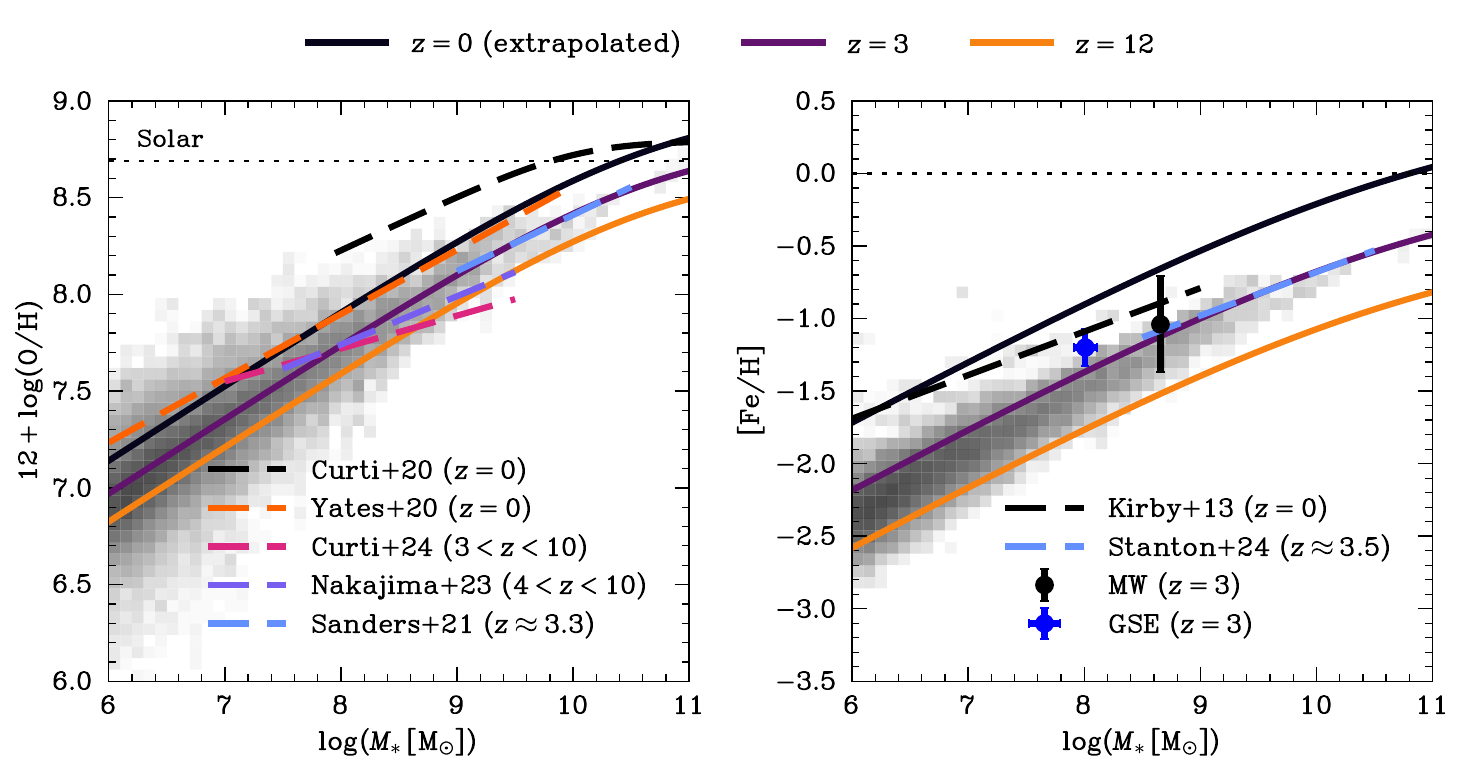}
    \caption{The mass-metallicity relation across redshift, where we show both the gas-phase oxygen abundance (\textit{left panel}) and the stellar iron abundance (\textit{right panel}). The black histograms show log-scaled distribution of all \thzoom galaxies included in this work ($3<z<12$). Curved, coloured lines show our best fits to a redshift-dependent MZR (Eq.~(\ref{eq:mzr_eq})) at $z=12$, $z=3$, and extrapolated down to $z=0$. We compare to observational results at $z=0$ for stellar metallicities of local dwarf galaxies \citep[black;][]{Kirby:2013aa} and at $z\approx3.5$ from the NIRVANDELS survey \citep[light blue;][]{Cullen:2021aa,Stanton:2024aa}. We also show estimates for the MW and GSE at $z=3$ as black and blue points, respectively \citep{Monty:2025aa}. For the gas-phase, we compare with strong line-derived metallicities of SDSS galaxies \citep[black;][]{Curti:2020aa} and electron temperature method-derived metallicities \citep[dark orange;][]{Yates:2020aa}. We also show higher-redshift results from MOSDEF at $z=3.3$ \citep[light blue;][]{Sanders:2021aa}, \textit{JWST} at $3<z<10$ \citep[magenta;][]{Curti:2024ab}, and \textit{JWST} at $4<z<10$ \citep[indigo;][]{Nakajima:2023aa}. Our gas-phase MZR extrapolated to $z=0$ is within $\sim$0.05\,dex of the \citet{Curti:2020aa} MZR at $M_\ast\approx10^{11}~\mathrm{M_\odot}$ and within $\sim$$0.1-0.2$\,dex of the \citet{Yates:2020aa} MZR at $M_\ast\lesssim10^{10}~\mathrm{M_\odot}$. Although we are extrapolating and there is clear tension between different observational MZR fits at $z=0$, this relatively good agreement may suggest a consistent and shallow redshift evolution of the MZR from $z=0$ to $z=12$. In terms of the high-redshift observations, our MZR is in excellent agreement with \citet{Sanders:2021aa} at $z\approx3.3$. The \citet{Curti:2024ab} and \citet{Nakajima:2023aa} MZR fits have extremely shallow slopes, cutting across our relations. This indicates either that there is a dramatic evolution in the MZR from $z=3$ which we do not capture in our simulation, or biases in the observations due to, for example, selection effects.
    }
    \label{fig:mass_met_rel}
\end{figure*}

\subsection{Stellar and gas-phase metallicity}
\label{sec:Stellar and gas-phase metallicity}

\begin{figure*} 
\centering
	\includegraphics[width=\textwidth]{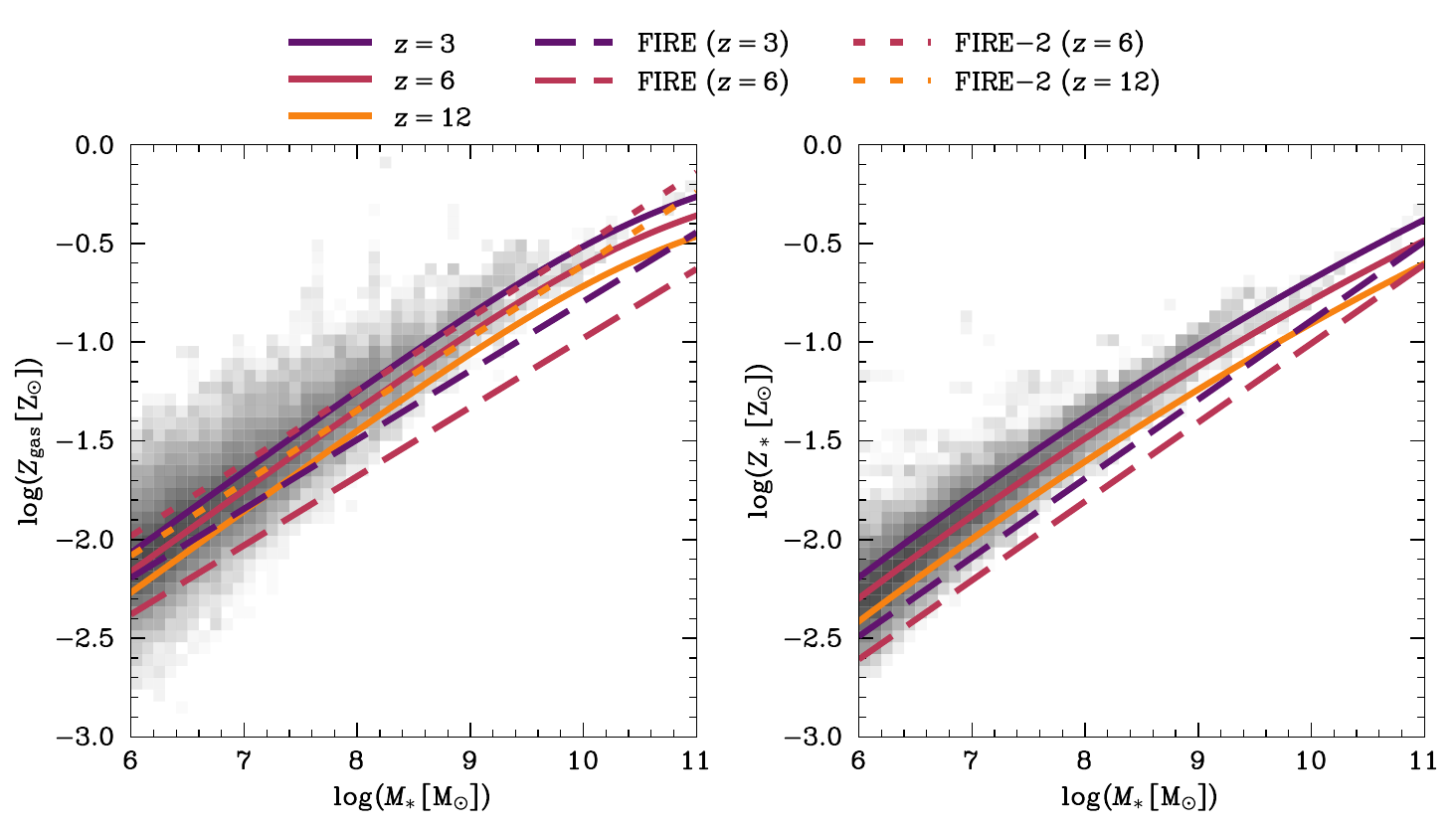}
    \caption{The mass-metallicity relation across redshift, where we show both the gas-phase total metallicity (\textit{left panel}) and the stellar total metallicity (\textit{right panel}). The black histograms show log-scaled distribution of all \thzoom galaxies included in this work. Curved, coloured lines show our best fits to a redshift-dependent MZR at $z=12$, $z=6$, and $z=3$. The MZR from FIRE simulations \citep{Ma:2016aa} are shown at $z=3$ and $z=6$ as dashed lines, and MZR from FIRE-2 simulations are shown as dotted lines at $z=6$ and $z=12$ \citep{Marszewski:2024aa}. The total metallicity-based gas-phase MZR has a steeper redshift dependence compared to the oxygen abundance-based fit, whereas the total metallicity-based gas-phase MZR has a shallower redshift dependence than the iron abundance-based fit. This is due to the characteristic time delay in enrichment for each element. Oxygen-enrichment occurs almost immediately via CC SNe, leading to a flatter redshift evolution, whereas iron is enriched on $\sim$Gyr timescales by SNe Ia, causing a steeper redshift evolution. Even for the total metallicity-based MZR fits, the gas-phase still has a somewhat steeper redshift evolution. Our gas-phase MZR shows a weaker redshift evolution compared to FIRE between $z=3$ and $z=6$, although the stellar MZR evolution is more comparable. The redshift evolution of the FIRE-2 gas-phase MZR is similarly weak to \thzoom. Both FIRE and FIRE-2 find a shallower low-mass slope than \thzoom, but they do not fit a curved mass-dependence, which makes it difficult to compare directly.
    }
    \label{fig:mzr_gas_starZ}
\end{figure*}

\begin{figure*} 
\centering
	\includegraphics[width=\textwidth]{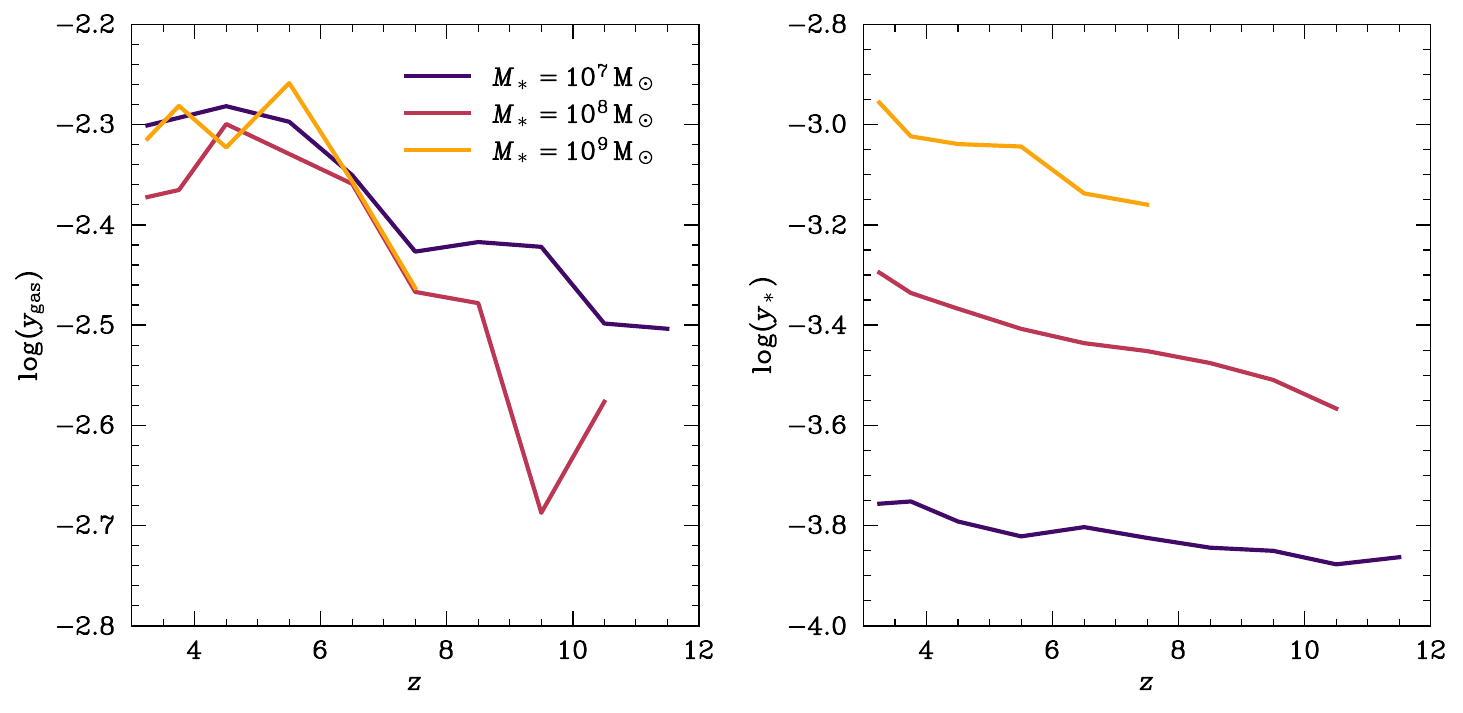}
    \caption{The metal retention efficiency in gas, $y_\mathrm{gas} \equiv M_{Z,\text{gas}} / M_\ast$ (\textit{left}), and stars, $y_\mathrm{\ast} \equiv M_{Z,\mathrm{\ast}} / M_\ast$ (\textit{right}) as a function of redshift for several stellar mass bins. The gas-phase metal retention efficiency remains remarkably constant with stellar mass. $y_\mathrm{gas}$ tends to increase with cosmic time down to $z\approx5$, where it remains approximately constant down to $z=3$. The stellar metal retention efficiency, $y_\ast$, decreases with redshift and increases strongly with stellar mass. In the mass and redshift range we consider here, the stellar retention efficiency is small enough that it does not make up a significant fraction of the metals compared to the gas, however, we expect it to be a more important component in more massive and lower redshift galaxies.
    }
    \label{fig:metal_retention_efficiency}
\end{figure*}
 
At lower masses, the MZR may be fit with a simple power law relation. However, at higher masses, metallicity becomes saturated and therefore the functional form must be adjusted to include a flattening. Although we do not have a large enough sample of galaxies at high masses to identify a saturation metallicity, we do clearly see curvature in our MZR relation, and so fitting a simple power law would be inappropriate. We therefore use the same functional form as \citet{Curti:2020aa}, although with an additional redshift-dependence for the normalization,
\begin{equation}
\label{eq:mzr_eq}
12+\log(\mathrm{O/H}) =  Z_0 -\alpha\log(1+z) - \frac{\gamma}{\beta}\, \log\left( 1 + \left(\frac{M}{M_0}\right)^{-\beta} \right) \,,
\end{equation}
where $Z_0$ is the saturation metallicity, $M_0$ is the turnover mass at which the metallicity approaches saturation, $\gamma$ is the relation slope at lower masses, $\beta$ determines the strength of the turnover, and $\alpha$ is the power-law redshift evolution. Instead of fitting all data directly, we calculate the median metallicity in 0.5\,dex bins of stellar mass in the range  $10^{6}~\mathrm{M_\odot}<M_\ast<10^{11}~\mathrm{M_\odot}$ and bins of 0.5 in redshift in the range $3<z<12$. We also use Eq.~(\ref{eq:mzr_eq}) for the stellar MZR, simply replacing $12+\log(\mathrm{O/H})$ with $[\mathrm{Fe/H}]$. Similarly, for total metallicity gas-phase and stellar MZR fits we replace the oxygen abundance with the bulk gas-phase or stellar metallicity. We note that due to the weak curvature seen in our sample, the saturation mass is less well constrained than in \citet{Curti:2020aa}, however, this form still fits our MZR well. With a larger dataset, particularly with more massive galaxies at higher redshift, we could better constrain the saturation metallicity and its turnover. Indeed, in this case it may be necessary to add a redshift evolution to the turnover mass. Parameters for fits to the gas-phase and stellar MZR are provided in Table~\ref{tab:mzr_fits}.

In Fig.~\ref{fig:mass_met_rel} we show our gas-phase and stellar MZR alongside observational derived fits \citep{Curti:2020aa,Curti:2024ab,Yates:2020aa,Sanders:2021aa,Nakajima:2023aa}. Our $z=3$ gas-phase MZR lies $\sim$0.2\,dex below the locally measured MZR at $M_\ast\approx10^{11}~\mathrm{M_\odot}$, and an extrapolation of our fit down to $z=0$ agrees with the \citet{Curti:2020aa} MZR within $\sim$0.05\,dex. Such agreement is not apparent at lower masses, with our extrapolated $z=0$ MZR $\sim$0.4\,dex below the locally measured MZR for a galaxy with  $M_\ast\approx10^{8}~\mathrm{M_\odot}$. Taken at face value, this implies that the MZR must evolve more rapidly for lower-mass galaxies than for higher-mass galaxies between $z=0$ and $z=3$. However, we are in much closer agreement at low masses with the MZR of \citet{Yates:2020aa}, who fit the MZR using electron temperature-based metallicity measurements for a smaller sample of lower-mass galaxies. We further note that our galaxy sample is on-average lower mass than that of \citet{Curti:2020aa}, leading to further uncertainty on direct comparisons at the high-mass end.

We caution that comparing observed and simulated metallicities can be fraught with issues. One obvious issue is biases due to the observability of galaxies, which can be particularly problematic at the low-mass end due to low SFRs \citep[e.g.][Simmonds et al. in prep.]{Sun:2023aa,McClymont:2025aa}. More fundamentally, the metallicities we measure from the simulations are intrinsic, derived directly from the gas, whereas observed metallicities are derived from emission lines. It has been shown that emission line derived metallicities may be systemically biased \citep[e.g.,][]{Kewley:2008aa}, including due to temperature inhomogeneities \citep[e.g.][]{Peimbert:1967aa,Vale-Asari:2019aa,Cameron:2023ac,Mendez-Delgado:2023aa} and due to the uncertain contributions from diffuse ionised gas \citep{Sanders:2017aa,Sanders:2021aa}, which can make up a large fraction of both collisional and recombination line emission from a galaxies and introduce uncertainties into the derived metallicities partially because the source of its ionisation is debated in the literature \citep{Zhang:2017aa,Belfiore:2022aa,McClymont:2024aa,McCallum:2024aa,McCallum:2024ab}. The choice of metallicity calibration can also have a significant impact \citep[e.g.][]{Curti:2020aa,Sanders:2024aa,Monty:2025aa,Scholte:2025aa}. We anticipate future work directly modeling the metal line emission from \thzoom galaxies in order to compare measured and intrinsic metallicities.

We compare our stellar MZR to the NIRVANDELS survey \citep{Cullen:2021aa} at $z\approx3.5$, finding excellent agreement \citep{Stanton:2024aa}. We also show archaeological estimates for the Milky Way (MW) and Gaia-Sausage-Enceladus \citep[GSE; the progenitor of the last major merger in the MW;][]{Belokurov:2018aa, Helmi:2018aa} determined using abundances from MW field stars and GCs and GC ages \citep{Monty:2025aa}, which also lie close to our MZR. Interestingly, our extrapolated $z=0$ stellar MZR agrees relatively well with the observed metallicities of Local Group dwarfs from \citet{Kirby:2013aa}. This is somewhat puzzling because these dwarf galaxies tend to have old stellar populations, and so we may expect them to agree more with the stellar MZR at $z=3$. However, these Local Group dwarf galaxies may have been enriched via pollution from their environment, given that they tend to show higher metallicities compared to dwarf galaxies in the MATLAS survey \citep{Heesters:2023aa}.

We find a relatively weak evolution of the MZR between $z=3-12$, with the gas-phase metallicity falling by only $\sim$0.2\,dex, although the stellar metallicity falls more appreciably by $\sim$0.4\,dex. In part, this reflects the delayed injection of iron via SNe Ia, on which we base our stellar metallicities, compared to the almost immediate enrichment of oxygen after the onset of star formation via CC SNe. In order to more fairly compare the stellar and gas-phase MZR, we also fit them based on the total metallicity (e.g. the mass fraction of all metals). We show these fits in Fig.~\ref{fig:mzr_gas_starZ}. With the total metallicity-derived MZR, the redshift dependence of the gas-phase fit is steeper and the redshift dependence of the stellar fit is shallower compared to the oxygen and iron abundance-based fits, as expected. Although in closer agreement, the stellar MZR is still somewhat steeper even when total metallicity is used. In effect, this means that although the gas-phase MZR always has a higher normalization because it represents the current state of the ISM, the stellar MZR moves closer in normalization with increasing cosmic time. These results demonstrate that large-volume cosmological simulations should endeavor to explicitly track the evolution of oxygen and iron in order to more fairly compare to observations, especially at high redshift, even if tracking other individual elements (e.g., N, Ne, Si) may be less interesting due to the unresolved ISM in these simulations.

In Fig.~\ref{fig:mzr_gas_starZ} we also show the MZR from FIRE \citep{Ma:2016aa} and FIRE-2 \citep{Marszewski:2024aa} simulations. Both show similar mass dependence, with a slope of $\sim0.35$, which is shallower than the low-mass slope we find in our fits. However, we note that the FIRE and FIRE-2 fits were not carried out with a curved mass-dependence, which may bias their slopes shallower compared to ours. The gas-phase MZR in FIRE shows a steeper redshift evolution compared to \thzoom between $z=3$ and $z=6$, however, FIRE-2 shows similarly shallow redshift evolution as \thzoom. Additionally, the FIRE-2 MZR shows a normalisation somewhat higher than \thzoom ($\sim$0.1\,dex), whereas the FIRE MZR is more appreciably lower ($\sim$0.2\,dex). Although not plotted here, IllustrisTNG shows more significant redshift evolution of the gas-phase MZR than in \thzoom, falling $\sim$0.2\,dex from $z=3$ to $z=6$ \citep{Torrey:2019aa}.

\subsection{Origin of the mass-metallicity relation}
\label{sec:The origin of redshift and metallicity evolution}

\begin{figure} 
\centering
	\includegraphics[width=\columnwidth]{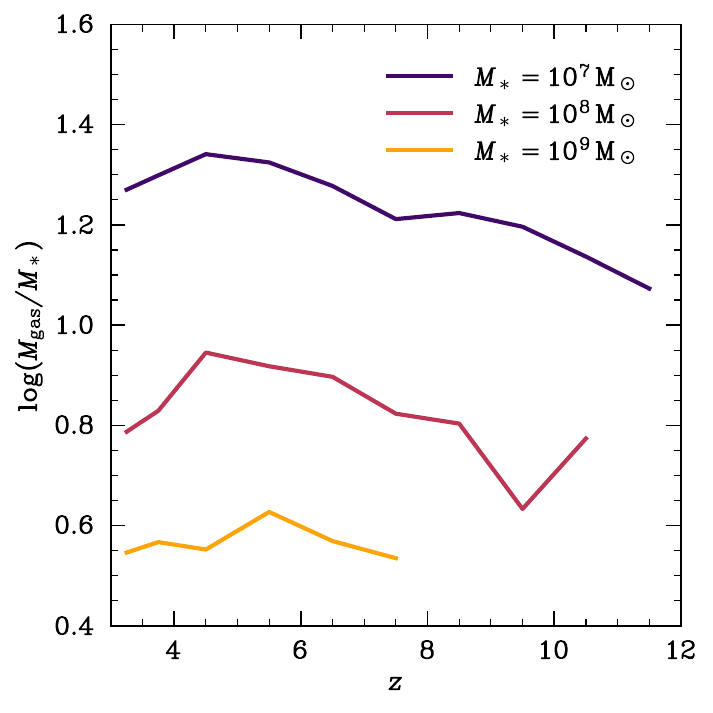}
    \caption{The gas mass to stellar mass ratio, $M_\mathrm{gas}/M_\ast$, as a function of redshift for several stellar mass bins. $M_\mathrm{gas}/M_\ast$ increases with cosmic time down to $z\approx5$, and decreases rapidly down to $z=3$. The behavior of $M_\mathrm{gas}/M_\ast$ at $z\gtrsim5$ is at odds with the generally expected trend of more gas-rich galaxies at higher redshifts. This is because \thzoom galaxies are more bursty at higher redshift and therefore there is a larger population of mini-quenched galaxies with very little gas, dragging down the median. $M_\mathrm{gas}/M_\ast$ shows a strong stellar mass dependence, where more massive galaxies have lower $M_\mathrm{gas}/M_\ast$ because they have converted more of their gas reservoir into stars. 
    }
    \label{fig:gas_fraction}
\end{figure}

The MZR is, in principle, sensitive to a variety of baryonic processes, including star formation, accretion, and outflows, and so understanding the mechanisms which generate it and drive its evolution with redshift can help us understand how these processes change as galaxies build up their mass across cosmic time. In a closed box system, the production of metals is well described by the gas fraction, $M_\mathrm{gas}/(M_\mathrm{gas}+M_\ast)$, or the gas mass to stellar mass ratio, $M_\mathrm{gas}/M_\ast$, which effectively quantifies how evolved a system is. While the closed-box approximation may work reasonably well on the halo scale \citep{Ma:2016aa}, this model does not hold on the galaxy scale due to the significant exchange of metals between the ISM and CGM. We must therefore decompose the metallicity into two components, and define the metal mass within the galaxy, $M_{Z}$, as
\begin{equation}
\label{eq:yield}
Z_i = \frac{y_i}{M_i/M_\ast}  \,,
\end{equation}
where the subscript $i$ represents a particular phase (e.g., gas or stars) and $y_i$ represents the metal retention efficiency $y_i=M_{Z,i}/M_\ast$, which is the same as the effective metal yield used in some analytic models of chemical evolution, such as the leaky and accreting box \citep{Talbot:1971aa,Edmunds:1990aa}. 

The gas-phase metal retention efficiency, $y_\mathrm{gas}$, represents the efficiency with which galaxies are holding onto the metal mass which has been produced by their stars, and it is affected by the production of metals, the locking up of metals in stars or dust, and the outflow of metals from the galaxy. It is notably \textit{not} directly affected by the inflow of pristine gas from the IGM, which is instead reflected only in the gas fraction. For the rest of this section, we are only considering the oxygen abundance-based MZR, so we modify Eq.~(\ref{eq:yield}) to only include the oxygen mass. We note that in this case $y_\ast$ is equivalent to the oxygen mass fraction for stars, but we will refer to it as the stellar metal retention efficiency in this context for consistency with the gas and to emphasize its role in locking up metals.

In principle, we could also use a conversion factor to calculate the total metal mass produced for a given stellar mass, which would allow us to better understand the true fraction of metals which are retained in the ISM. This factor varies depending on the star-formation and enrichment histories of the galaxies, and therefore a global yield may be inappropriate for the mass and redshift range which we consider, and so we do not include a conversion factor. However, we note that because we are only considering oxygen, which is mostly enriched through CCSNe on short timescales, we are less concerned with the evolution of $y_\mathrm{gas}$ being strongly affected by delayed enrichment.

In addition to being locked up in stars, metals can also be deposited into dust and so in theory we could also calculate the dust metal retention efficiency, $y_\mathrm{dust}$. In practice however, while the \thzoom simulations employ a on-the-fly dust model \citep[][Garaldi et al. in prep.]{Kannan:2025aa}, we do not output the oxygen abundance of the dust in our simulation so we cannot self-consistently calculate the $y_\mathrm{dust}$ for oxygen. We expect most of the oxygen mass to be in the gas-phase even at high (solar) metallicity \citep[$\lesssim0.12$\,dex depletion;][]{Peimbert:2010aa}, so we do not expect a significant effect for oxygen abundances, although the impact on the total metallicity may be more important. In this section we focus only on central galaxies.

In Fig.~\ref{fig:metal_retention_efficiency} we show $y_\mathrm{gas}$ and $y_\ast$ as a function of redshift in stellar mass bins. $y_\mathrm{gas}$ shows no significant evolution with stellar mass, although we are only able to probe relatively low-mass bins here ($M_\ast\lesssim10^{9.5}~\mathrm{M_\odot}$) as we need a large sample of galaxies out to high redshift. There are hints of divergence at the high-redshift end but given that the data is fairly noisy, we cannot make strong conclusions. $y_\mathrm{gas}$ does show redshift evolution, increasing with cosmic time down to $z\approx5$ and then stabilizing until the end of our simulations at $z=3$. $y_\mathrm{\ast}$ shows a clear increase with cosmic time, essentially tracking the evolution of the oxygen stellar MZR with redshift, and also increases dramatically with stellar mass. For the mass and redshift range we consider, $y_\mathrm{\ast}$ is never comparable to $y_\mathrm{gas}$ and so does not have an important impact on $y_\mathrm{gas}$. For higher mass galaxies and at lower redshifts, we expect the impact to become more important. 

In Fig.~\ref{fig:gas_fraction} we show $M_\mathrm{gas}/M_\ast$ as a function of redshift in stellar mass bins. $M_\mathrm{gas}/M_\ast$ decreases rapidly with increasing stellar mass, which is a general trend expected as gas is converted to stars. However, the decrease in $M_\mathrm{gas}/M_\ast$ with stellar mass is not consistent with a closed box model, and is instead shallower due to inflowing gas offsetting some of the conversion to stars. $M_\mathrm{gas}/M_\ast$ shows a somewhat similar redshift evolution to $y_\mathrm{gas}$, increasing with cosmic time until $z\approx5$, after which $M_\mathrm{gas}/M_\ast$ decreases with cosmic time down to $z=3$. We note that this decrease may have some mass dependence, potentially being weaker in the highest mass bin, although we do not have sufficient statistics to make a firm conclusion. The increase of $M_\mathrm{gas}/M_\ast$ with cosmic time down to $z\approx5$ is opposite to the general expectation that galaxies become more gas-rich at higher redshift. This is a population-level trend, and reflects an increasing abundance of mini-quenched galaxies with extremely low gas fractions at higher redshift \citep{McClymont:2026aa}. Below $z\approx5$, we find the expected trend of decreasing $M_\mathrm{gas}/M_\ast$ with cosmic time, likely because the consumption of gas into stars becomes more important than the decreasing abundance of mini-quenched galaxies.

The slope of the MZR is dominated by the evolution of $M_\mathrm{gas}/M_\ast$ as a function of $M_\ast$. This is because the gas-phase metal retention efficiency, $y_\mathrm{gas}$, is largely independent of stellar mass, at least where we have good statistics to study this in detail ($M_\ast\lesssim10^{9.5}~\mathrm{M_\odot}$). The lack of $M_\ast$-dependence for $y_\mathrm{gas}$ in the range which we do probe is due to several competing effects. More massive galaxies have a higher fraction of their metals locked into stars, which would act to remove metals from the gas and therefore reduce $y_\mathrm{gas}$. However, this effect is balanced by the fact that lower-mass galaxies are more efficient at ejecting metals from their ISM. As mentioned above, there could be some influence of delayed enrichment, but this effect is likely not strong for oxygen abundances, which we consider here. It is difficult to predict the trend at higher masses because although we may expect $y_\mathrm{gas}$ to increase with stellar mass as higher-mass galaxies are less bursty and their outflows are generally less extreme, the competing effect of metals being increasingly locked into stars (and dust) may be sufficient to offset this trend. We also note that we have only included central galaxies in this analysis. Satellites tend to have both lower $y_\mathrm{gas}$ and lower $M_\mathrm{gas}/M_\ast$, and so a changing satellite fraction as a function of mass would lead to a mass-dependence of $y_\mathrm{gas}$ across the whole galaxy population.

Unlike the $M_\ast$-dependence, the redshift evolution of the MZR is impacted by both $y_\mathrm{gas}$ and $M_\mathrm{gas}/M_\ast$, which vary with redshift. At higher redshift ($z\gtrsim5$), $M_\mathrm{gas}/M_\ast$ increases with cosmic time, which acts to decrease metallicity, but $y_\mathrm{gas}$ increases with cosmic time sufficiently to cause a mild net increase in metallicity. At lower redshift ($z\lesssim5$), the metal retention efficiency plateaus, and subsequent redshift evolution is caused by a decreasing $M_\mathrm{gas}/M_\ast$. 

Overall, the changes in $M_\mathrm{gas}/M_\ast$ are relatively weak as a function of redshift compared to the $M_\ast$ dependence. Additionally, it is important to note that this evolution is primarily driven by the increasing fraction of dramatically mini-quenched galaxies, which have extremely low $M_\mathrm{gas}/M_\ast$, rather than a strong evolution for the average star-forming galaxy. We must, however, caveat our conclusions here by noting that \thzoom simulations are centered around target galaxies and therefore are not representative cosmological volumes.

\subsection{Mass--metallicity scatter}
\label{sec:Mass-metallicity scatter}

\begin{figure} 
\centering
	\includegraphics[width=\columnwidth]{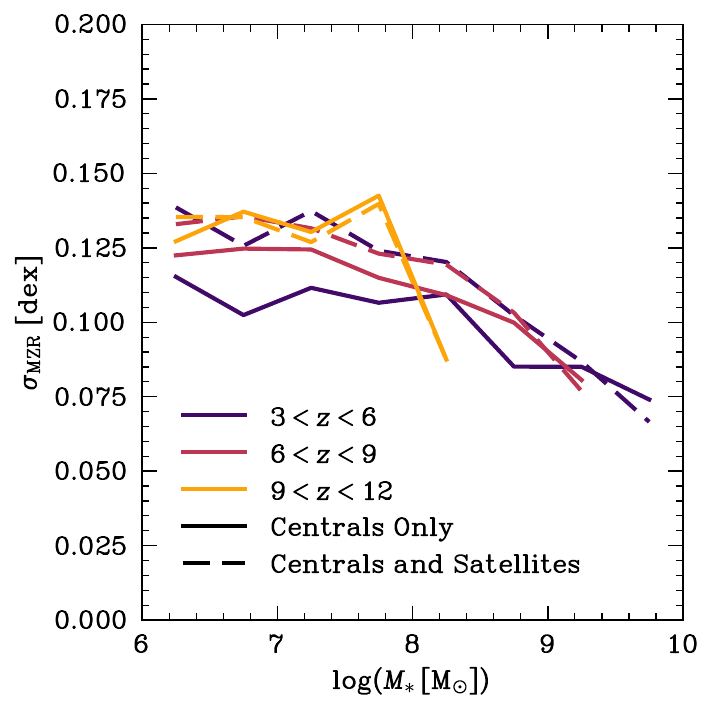}
    \caption{The scatter in the gas-phase MZR, $\sigma_\mathrm{MZR}$, as a function of stellar mass in redshift bins. We show the scatter for central galaxies only (solid lines), as well as for both central and satellite galaxies (dashed lines). When considering central galaxies only, the MZR scatter tends to increase with redshift. However, when satellite galaxies are also included, the MZR scatter shows no strong redshift dependence. Regardless of redshift range or central vs satellite selection, we see that the scatter decreases with increasing stellar mass.
    }
    \label{fig:mzr_scatter}
\end{figure}

In Fig.~\ref{fig:mzr_scatter} we show the scatter of the MZR, $\sigma_\mathrm{MZR}$, as a function of redshift in stellar mass bins. Scatter is measured separately for both central galaxies only and for all (central and satellite) galaxies, and it is calculated as the standard deviation in a given bin. We recenter each bin on its median to avoid artificially high scatter due to any systemic offsets from the median MZR relation. To avoid outliers artificially boosting the scatter, we only consider galaxies within 0.4\,dex of the median, which includes 85.2\% of our sample. Most of those excluded are those with a gas fraction of 0 (12.5\% of the total sample). We visually verified that the measured scatter with this method well described the distribution of metallicity offsets for each bin.

When considering central galaxies only, we see a weak redshift trend, with the scatter at $M_\ast\approx10^{10}~\mathrm{M_\odot}$ falling from $\sigma\approx0.13\,\mathrm{dex}$ to $\sigma\approx0.11\,\mathrm{dex}$ between $z=3-12$. This small redshift trend could plausibly be related to the decreased burstiness of star formation at lower redshifts \citep{McClymont:2025aa}. This could be directly related, as less bursty star formation would mean that metals are produced and ejected stochastically, but also indirectly because it indicates that pristine inflows from the IGM are less powerful and less common. We see no significant redshift trend in scatter for the sample which includes both central and satellite galaxies. One possible explanation of the increased scatter at low redshift when satellites are also included is the increasing importance of metal pollution due to outflows from central galaxies. Independent of redshift or central/satellite selection, we find that scatter decreases with stellar mass, with the scatter falling to $\sigma\approx0.075\,\mathrm{dex}$ at $M_\ast\approx10^{10}~\mathrm{M_\odot}$. Similarly to the redshift trend for central galaxies, the stellar mass trend could be related to the decreasing burstiness of star formation at higher stellar masses.

The MZR scatter measured in the local Universe with SDSS is $\sigma\approx0.075\,\mathrm{dex}$ for $M_\ast\gtrsim10^{9.5}~\mathrm{M_\odot}$ \citep{Curti:2020aa}, which is in remarkably good agreement with the scatter we measure at higher masses, although we should be cautious about drawing strong conclusions due to the difficulty of comparing metallicity between simulations and observations. \citet{Curti:2024ab} measured MZR scatter as $\sigma\approx0.073\,\mathrm{dex}$ for a sample of galaxies at $3<z<10$. However, biases in this measurement may arise from the NIRSpec sample selection, and uncertainties in the high-redshift metallicity calibrations are not yet well understood \citep{Sanders:2024aa}.

\subsection{The fundamental metallicity relation}
\label{sec:Fundamental metallicity relation}

\begin{figure} 
\centering
	\includegraphics[width=\columnwidth]{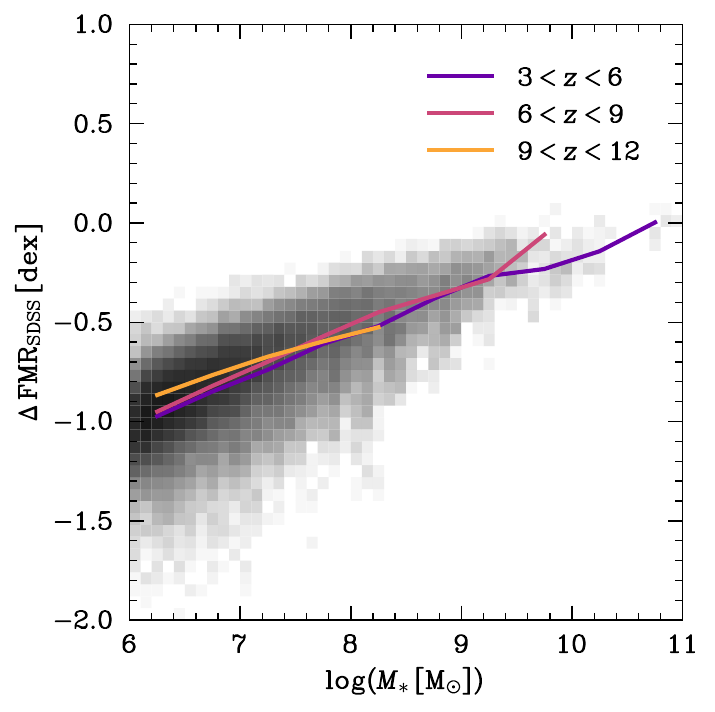}
    \caption{Gas-phase oxygen abundance offset from the locally calibrated FMR \citep{Curti:2020aa} as a function of stellar mass. The black histogram shows the log-scaled distribution of all \thzoom galaxies included in this work. Coloured points show the median offset as a function of stellar mass for redshift bins. At the highest masses ($M_\ast>10^{10.5}\,\mathrm{M_\odot}$), our galaxies align with the low-redshift FMR. However, our galaxies become rapidly inconsistent with the low-redshift FMR with decreasing stellar mass, reaching a $\sim$1\,dex offset at $M_\ast=10^{6}\,\mathrm{M_\odot}$. The redshift evolution is not dramatic, with offsets decreasing only $\sim$0.2\,dex from $z=4$ to $z=14$, although it is notable that the agreement with the local FMR at low masses actually increases at higher redshift. With the caveat that we are comparing simulated to measured metallicity, we suspect that reported offsets is driven in large part by the low-redshift FMR being poorly calibrated at low masses and high sSFRs, rather than any dramatic evolution of the FMR itself.}
    \label{fig:fmr_offset_metallicity_curti}
\end{figure}

\begin{figure*} 
\centering
	\includegraphics[width=\textwidth]{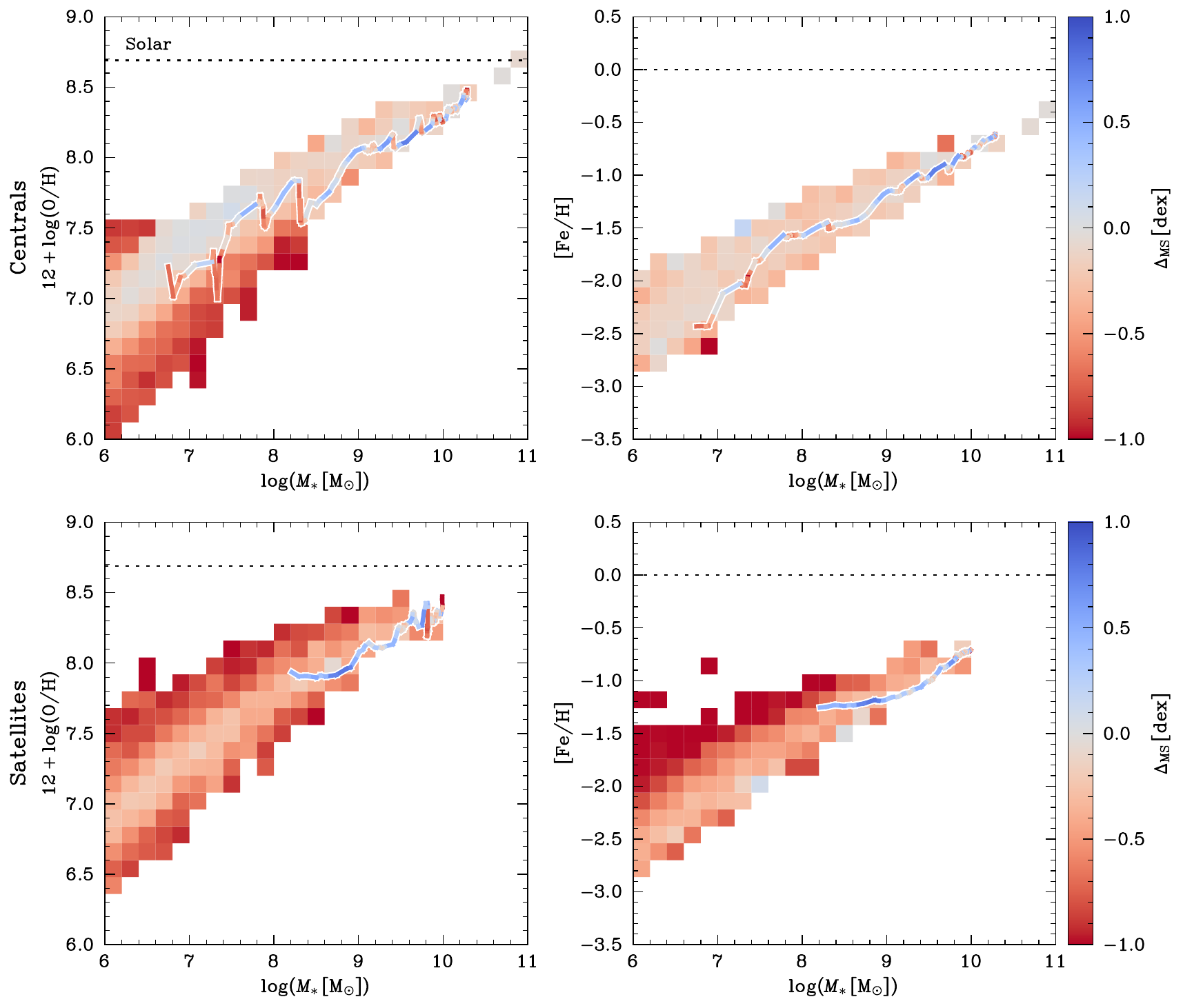}
    \caption{Gas-phase oxygen abundance (\textit{left panels}) and the stellar iron abundance (\textit{right panels}) of galaxies as a function of stellar mass, colored by offset from the \thzoom star-forming main sequence \citep{McClymont:2025aa}. The tracks show the evolution of individual galaxies which are representative of the general population. We show the plots for both central galaxies (\textit{upper panels}) and satellite galaxies (\textit{lower panels}). Only galaxies within 1\,dex of the SFMS are included in this plot to avoid biasing the coloring from galaxies far below the SFMS (Fig.~\ref{fig:sfms_mzr_rel_star} shows across a wider range). The top left panel shows that lower mass ($M_\ast\lesssim10^{9}\,\mathrm{M_\odot}$) central galaxies tend to have high SFRs when they have high metallicities for a given mass, which is the opposite of the expectation from the FMR. The top right panel shows that central galaxies do generally follow the stellar FMR. In the bottom left panel we see that satellites appear to have a complex dependence, with the lowest SFRs at both low and high metallicity for a fixed mass. The bottom right panel shows that satellites follow the stellar metallicity FMR.}
    \label{fig:mass_met_rel_sfms}
\end{figure*}

\begin{figure*} 
\centering
	\includegraphics[width=\textwidth]{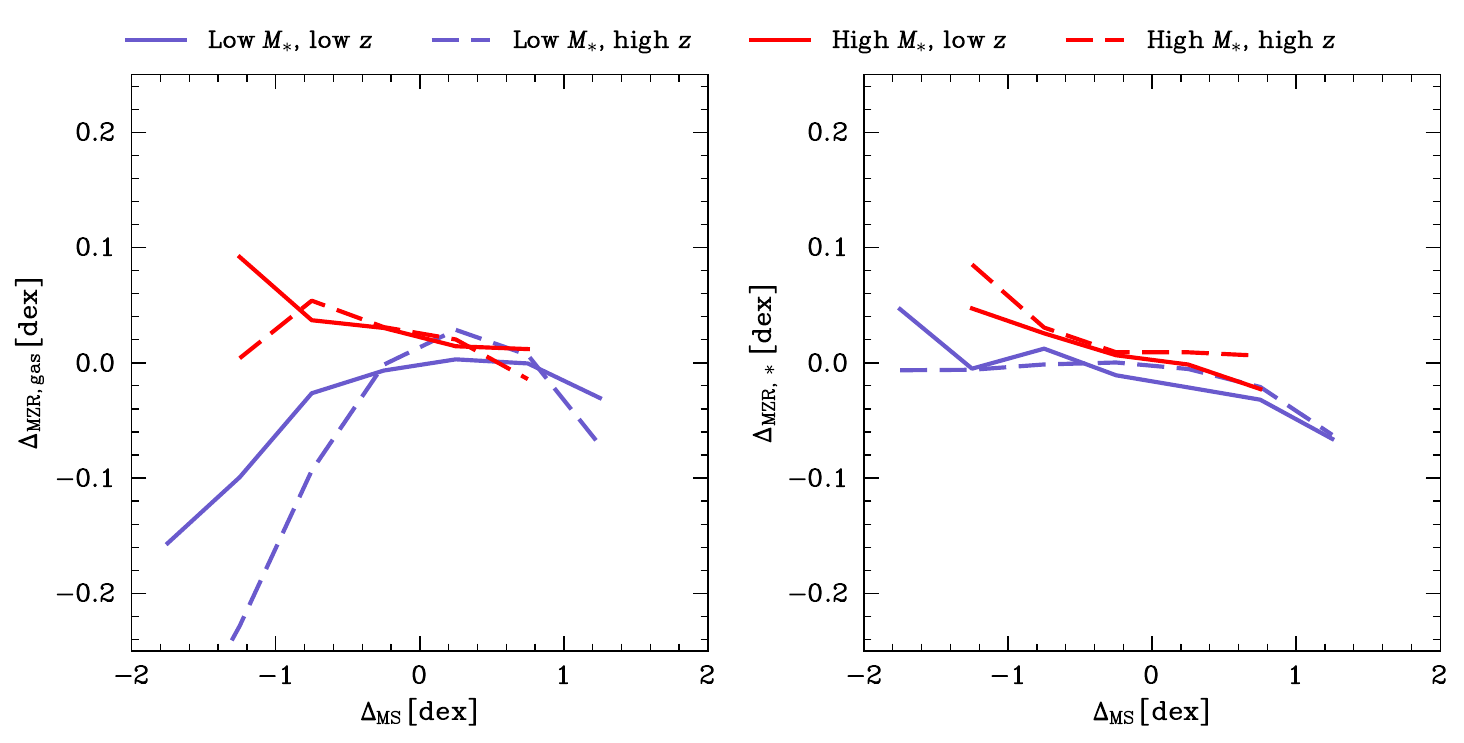}
    \caption{Offset from the gas-phase MZR (\textit{left panel}) and stellar MZR (\textit{right panel}) as a function of offset from the \thzoom star-forming main sequence \citep{McClymont:2025aa}. We only include central galaxies in this plot. The blue lines show median trends for low mass ($M_\ast<10^{9}\,\mathrm{M_\odot}$) galaxies, whereas the red lines show median trends for high mass ($M_\ast>10^{9}\,\mathrm{M_\odot}$) galaxies. Solid lines show the lower redshift sample ($z<5$) and dashed lines show the higher redshift sample ($z>5$). The left panel shows that the higher mass galaxies tend to follow the FMR expectation across redshift, showing increasing metallicity with decreasing SFR at a fixed mass. However, lower mass galaxies show a more complex trend. The most star-forming ($\Delta\mathrm{MS}\gtrsim0.5$) of these galaxies follow the FMR expectation, whereas the rest of this population shows generally decreasing metallicity with decreasing metallicity at a fixed mass. The high-$z$ trend is somewhat more extreme, with the low-$z$ trend appearing flattened. The right panel shows that galaxies across masses and redshifts tend to follow the stellar FMR, where metallicity increases with decreasing SFR at a fixed mass.}
    \label{fig:sfms_mzr_rel_star}
\end{figure*}

The FMR is often interpreted as evidence that galaxies fluctuate around an equilibrium set by gas inflow, outflow, and processing. Recently, there have been indications of offsets from the FMR at $z\gtrsim3$ \citep{Curti:2024ab,Sarkar:2025aa,Pollock:2025aa}, potentially indicating an evolution of the FMR and therefore a fundamental change in the regulation of metallicity in high-redshift galaxies. In Fig.~\ref{fig:fmr_offset_metallicity_curti} we quantify the offset of \thzoom galaxies from the low-$z$ FMR calibration of \citet{Curti:2020aa}. We use $\mathrm{SFR_{10}}$ in the FMR calculation and only include galaxies with $\mathrm{SFR_{10}}>0$.

At $M_{\ast}>10^{10.5}\,\mathrm{M_{\odot}}$ our simulated galaxies lie on the local FMR.  Below this mass the offset grows rapidly, reaching $\sim1$~dex at $M_{\ast}=10^{6}\,\mathrm{M_{\odot}}$. The redshift evolution of the offset is weak (\mbox{$\lesssim0.2$~dex between $z=14$ and $z=4$}), suggesting that the apparent breakdown is driven by the extrapolation of the local relation into a region of parameter space that is poorly constrained, namely low masses and extremely high specific SFRs (sSFRs), rather than by a strong redshift evolution of the FMR. We note that if we instead use the \citet{Andrews:2013aa} FMR parameterization, we find a similar mass dependence but an inverted, weak redshift dependence (see Appendix~\ref{sec:Fundamental metallicity relation parametrization}). Interestingly, \citet{Rowland:2025aa} find no offset to the FMR for massive ($M_{\ast}\sim10^{9.5}\,\mathrm{M_{\odot}}$) galaxies at $z=6-8$, which is qualitatively in line with our trend (and quantitatively in line if we use the \citealt{Andrews:2013aa} FMR as in their work).

It is crucial to note that while the large FMR offsets at high-$z$ may be explained by an overly-extrapolated FMR, this does \textit{not} imply that a perfectly calibrated FMR is universally valid. It may be that a well-calibrated FMR would remove the large systemic offsets with increasing redshift, but is unable to significantly decrease the scatter compared to the MZR due to the evolution of the FMR. 

Fig.~\ref{fig:mass_met_rel_sfms} provides further insight by colouring galaxies in the metallicity vs stellar mass plane according to the offset from the \textsc{thesan-zoom} star-forming main sequence \citep[SFMS;][]{McClymont:2025aa}, $\Delta_\mathrm{MS}$. High-mass centrals ($M_{\ast}>10^{9}\,\mathrm{M_{\odot}}$) follow the classical FMR behaviour: at fixed mass, elevated SFR coincides with lower gas-phase metallicity. In contrast, low-mass centrals display an \textit{inverse} trend whereby metal-rich systems are preferentially \textit{more} star-forming. We interpret this inversion as a signature of stochastic gas accretion in shallow potential wells: once a burst has consumed and/or ejected a large fraction of ISM gas, subsequent pristine gas inflow dilutes the metallicity, temporarily moving the galaxy below the canonical FMR. As the ISM gas is now low metallicity, it is rapidly enriched as star formation ramps up, causing a relatively higher metallicity at higher SFR. The coloured track shows the evolution of an individual galaxy, also colored by $\Delta_\mathrm{MS}$. We can see there are periods where the metallicity drops while $\Delta_\mathrm{MS}$ is low, indicating a low-metallicity inflow. This inflow proceeds to fuel a burst of star formation, moving the galaxy up and to the right of the plot as it gains stellar mass and enriches its ISM, but also increasing $\Delta_\mathrm{MS}$.

Such rapid inflows have been shown to be an important component of bursty star formation in \thzoom galaxies \citep{McClymont:2025aa}, and the impact of these metal-poor inflows on galaxy metallicities and the FMR has been considered in previous observational works \citep[e.g.,][]{Tacchella:2023aa,Langeroodi:2023aa}. Low metallicity inflow can come in the form of pristine gas from the IGM, which itself may mix with expelled gas in the CGM before it reaches the ISM. Additionally, gas-rich mergers can bring inflowing gas of varying metallicity. While mergers play a subdominant role compared to in-situ star formation, the merger rate is found to increase with redshift \citep{Duan:2025aa,Puskas:2025aa}.
Interestingly, \citet{Zier:2025aa} show that spatially inhomogeneous reionisation, which is faithfully modeled in \thzoom, leads to reduced star formation in minihaloes, and therefore to reduced metallicity inflows at later times than would be found using a spatially homogeneous UV-background.

Satellite galaxies exhibit a more complex dependence (lower panels of Fig.~\ref{fig:mass_met_rel_sfms}). Similarly to central galaxies, satellites can undergo metal-poor, low-SFR phases due to dilution by low-metallicity gas. However, they can also experience either metal-rich, low-SFR episodes due to the inflow of metal-rich gas from central galaxies or due to gas stripping, which lowers the SFR without decreasing metallicity. This variety of effects produces a broad ``valley'' in the SFR-metallicity plane. The stellar FMR, however, remains intact for both centrals and satellites.

In Fig.~\ref{fig:sfms_mzr_rel_star} we explore the evolution of the FMR for central galaxies in more detail by considering the offset from the MZR, $\Delta_\mathrm{MZR}$, against $\Delta_\mathrm{MS}$. This figure again shows that the standard gas-phase FMR is followed by higher-mass galaxies across the redshift range, although the trend is quite mild. However, interestingly, at high values of $\Delta_\mathrm{MS}$, the standard FMR reappears, even for the lower-mass galaxies. One plausible explanation for this is that the most extreme starbursts tend to be driven by an extreme inflow of gas from the IGM (or merging galaxies, which would be lower mass, and likely lower metallicity, than the central galaxy) that lasts for longer than typical starbursts \citep{McClymont:2025aa}. In these extreme cases then, SFRs can remain high while low-metallicity gas is still inflowing, leading to simultaneously low metallicity and high $\Delta_\mathrm{MS}$.

Fig.~\ref{fig:sfms_mzr_rel_star} also makes clear that the stellar FMR remains valid across the mass and redshift range, although it is certainly less strong in the low-mass, high-redshift bin. The likely reason for this is that stellar metallicity is not susceptible to such rapid dilution as gas-phase metallicity because stellar metallicity can, by definition, only be altered with the onset of star formation, meaning that most stars form after some level of self-enrichment unless they are part of the very first generation of a new starburst.

\subsubsection{Implications for observations}
\label{sec:Implications for observations}

The prospects for measuring the FMR inversion with observations are unclear. Observable nebular emission is required in order to measure gas-phase metallicity observationally, meaning that metallicity is only practically measurable for a biased subset of observed galaxies, specifically those where star formation has already begun in earnest. Additionally, that not all gas within a galaxy is luminous in emission lines can lead to biased metallicity measurements. 

However, it is possible that the breakdown and inversion of the FMR could have an impact on observed metallicity trends. An inverted FMR means that brighter galaxies are more likely to be metal-rich, flattening the MZR for an observability-limited sample because the observed low-mass galaxies are more likely to be metal-rich than their population as a whole. This biasing effect is further exacerbated by the recovery of the ordinary FMR at $M_{\ast}\gtrsim10^{9}\,\mathrm{M_{\odot}}$. At these higher masses, brighter galaxies are more likely to be metal-poor. In combination, this means that, on average, observed higher-mass galaxies will be biased metal-poor and lower-mass galaxies will be biased metal-rich compared to a complete sample, causing an artificial flattening of the MZR.

\section{Chemical abundances}
\label{sec:Chemical abundances}

Our analysis thus far in this paper has focused exclusively on oxygen abundances in the gas phase and iron abundances in stars, the most commonly used proxies for total metallicity. However, measurements of relative abundances of individual chemical elements also represent rich sources of information on the assembly of galaxies at early times, which is particularly relevant in light of the peculiar chemical enrichment of high-redshift galaxies as seen with \textit{JWST}. In this section, we will study the gas-phase nitrogen, iron, and carbon abundances of high-redshift galaxies.

\subsection{Abundance distribution}
\label{sec:Abundance distribution}

\begin{figure} 
\centering
	\includegraphics[width=\columnwidth]{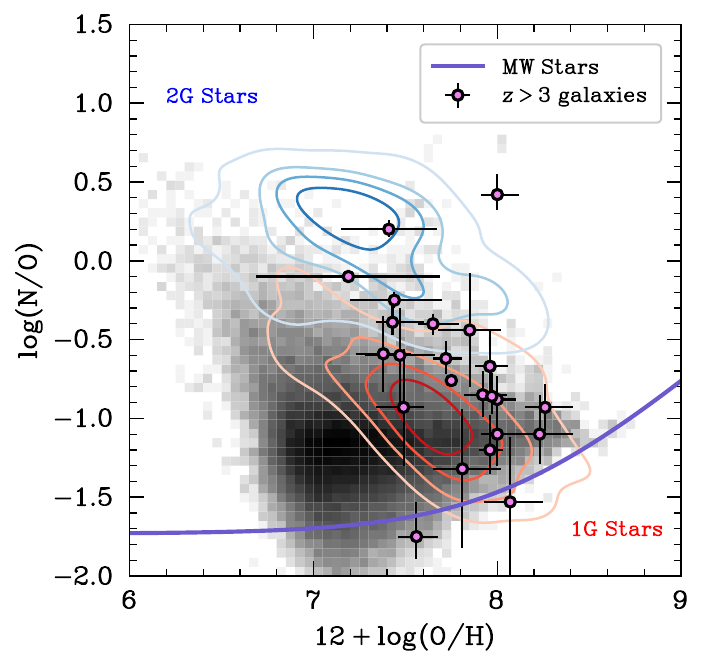}
    \caption{The gas-phase nitrogen to oxygen ratio as a function of the gas-phase oxygen abundance. The black histogram shows the log-scaled distribution of all \thzoom galaxies included in this work. The solid blue line shows an empirical relation to the abundances of Milky Way stars \citep{Nicholls:2017aa}. We also show high-redshift N/O ratios for $z>3$ galaxies (see text for details) as violet points. We also show contours for APOGEE DR17 \citep{Abdurrouf:2022aa} GC stars following \citet{Ji:2026aa}, separately including 1G (red contours) and 2G (blue contours) stars. \thzoom galaxies tend to be somewhat nitrogen-enhanced relative to all Milky Way stars, although the distribution aligns reasonably well with 1G and 2G GC stars. The scatter in N/O at fixed oxygen abundance is large, particularly at lower $12+\log(\mathrm{O/H})$ values. We find a population of NRGs with $\log(\mathrm{N/O})>-0.6\,\mathrm{dex}$, comprising 4\% of the sample. The distribution is consistent with high-redshift galaxy observations, although the observed galaxies lie on the upper end of the distribution, perhaps due to selection effects.}
    \label{fig:no_met_rel}
\end{figure}

\begin{figure} 
\centering
	\includegraphics[width=\columnwidth]{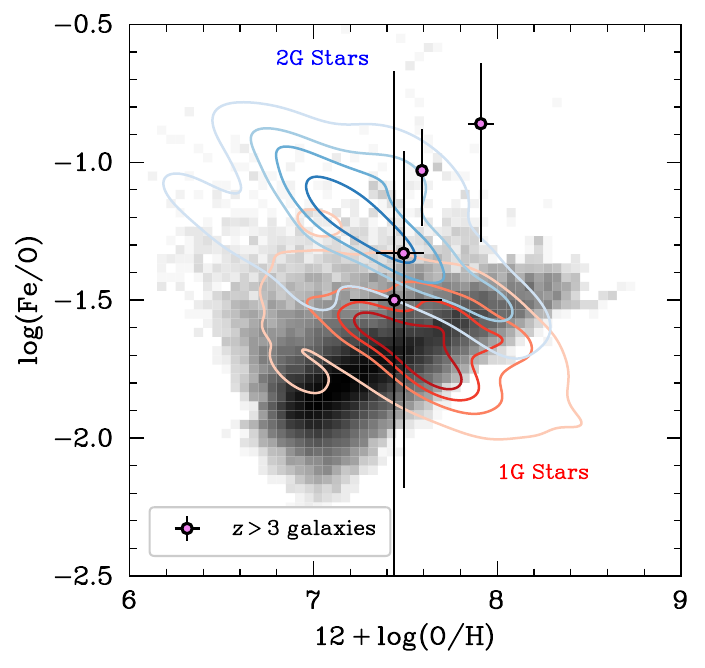}
    \caption{The gas-phase iron to oxygen ratio (Fe/O) as a function of the gas-phase oxygen abundance. The black histogram shows the log-scaled distribution of all \thzoom galaxies included in this work and we compare to 1G and 2G GC stars (red and blue contours, respectively) and high-redshift galaxies \citep[violet points;][]{Nakane:2024aa,Nakane:2025aa,Tacchella:2025aa} as in Fig.~\ref{fig:no_met_rel}. We note that three of the four high-redshift measurements are stellar abundances due to the difficulty of measuring gas-phase iron abundances at high redshift. Fe/O increases with increasing oxygen abundance. The Fe/O distribution shows large scatter relative to the trend, although the scatter is more moderate than for N/O. The simulated distribution is consistent with the limited high-redshift observational constraints, although the observed points lie on the upper end of the distribution. Similarly to N/O, the Fe/O distribution largely overlaps with GC stars in the Milky Way. The trend of increasing Fe/O with increasing oxygen abundance is qualitatively consistent with trends seen in the local Universe once the contribution of iron locked into dust is accounted for \citep{Mendez-Delgado:2024aa}, and with trends seen in MW stars \citep{Nicholls:2017aa}.}
    \label{fig:feo_met_rel}
\end{figure}

\begin{figure} 
\centering
	\includegraphics[width=\columnwidth]{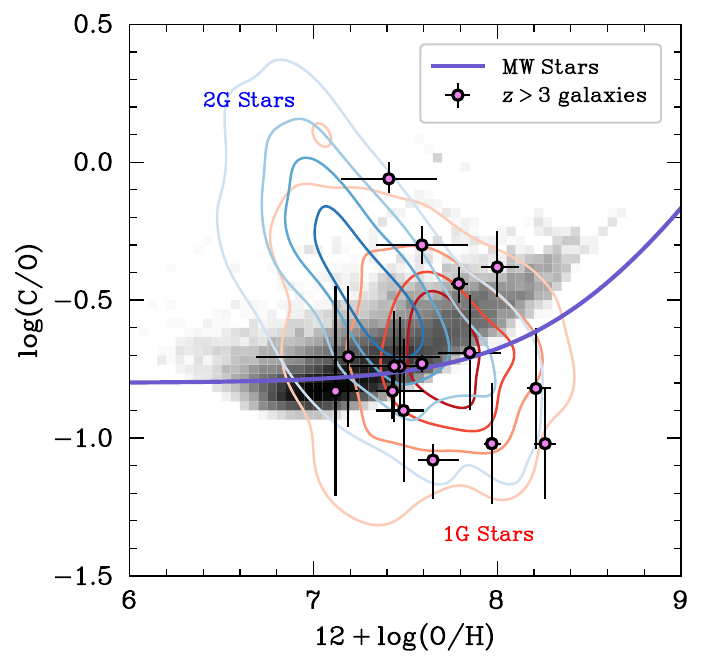}
    \caption{The gas-phase carbon to oxygen ratio (C/O) as a function of the gas-phase oxygen abundance. The black histogram shows the log-scaled distribution of all \thzoom galaxies included in this work and we compare to 1G and 2G GC stars (red and blue contours, respectively), an empirical fit to Milky Way stars \citep[blue line;][]{Nicholls:2017aa}, and high-redshift galaxies (violet points) as in Fig.~\ref{fig:no_met_rel}. The distribution of \thzoom C/O shows much less scatter than N/O or Fe/O, and galaxies are mildly carbon-enhanced relative to Milky Way stars at higher metallicities ($12+\log(\mathrm{O/H})\gtrsim7.6$). The distribution covers the highest density of 1G and 2G GC stars and all but two of the high-redshift galaxy observations are within $1\sigma$. However, it appears that \thzoom galaxies fail to scatter to such low ratios.}
    \label{fig:co_met_rel}
\end{figure}

In order to understand our simulated abundances in context, in this section, we compare them to the abundances of high-redshift galaxies, Milky Way field stars, and Milky Way globular cluster (GC) stars. For GC stars, we use the sample compiled by \citet{Ji:2026aa}, which is comprised of red giant stars observed by the SDSS Apache Point Observatory Galaxy Evolution Experiment (APOGEE) in data release 17 \citep[DR17][]{Abdurrouf:2022aa} cross-matched with GC membership catalogs \citep{Vasiliev:2021aa}. We make a further restriction that the error on the abundance measurement relative to iron (e.g. [O/Fe]) must be less than 0.1\,dex to ensure the distributions are not artificially spread out. GC stars are traditionally split into two generations, with the first generation (1G) showing similar chemical abundances relative to field stars at the same iron abundance, and the second generation (2G) showing enhanced nitrogen, sodium, and depleted oxygen for a given [Fe/H] \citep{Bastian:2018aa}. We also show a fit to the abundances of Milky Way stars derived by \citet{Nicholls:2017aa}.

We use sample of gas-phase galactic chemical abundances compiled by \citet{Ji:2026aa}, including only sources at $z>3$ \citep{Vanzella:2010aa,Cameron:2023ab,Cameron:2024aa,Isobe:2023aa,Larson:2023aa,Pascale:2023aa,Tacchella:2023ab,Ubler:2023aa,Calabro:2024aa,Castellano:2024aa,Ji:2024aa,Labbe:2024aa,Maiolino:2024aa,Schaerer:2024aa,Topping:2024aa,Topping:2025aa,Arellano-Cordova:2025aa,Alvarez-Marquez:2025aa,Curti:2025aa,Stiavelli:2025aa,Tacchella:2025aa,Zhang:2025aa}. This sample is primarily focused on NRGs, so we further augment the sample with other abundance measurements at $z>3$ \citep{Arellano-Cordova:2022aa,Arellano-Cordova:2025aa,DEugenio:2024ab,Hsiao:2025aa,Stiavelli:2025aa}. We take reported abundances at face value, however we note that \citet{Hayes:2025aa} argue that NRGs may be more consistent with local relations if the high electron densities in these galaxies cause an underestimation of abundances.

In Fig.~\ref{fig:no_met_rel} we show N/O as a function of the gas-phase oxygen abundance for all \thzoom galaxies. The bulk of the distribution is moderately nitrogen enhanced relative to the MW relation, with a median across the whole sample of $\log(\mathrm{N/O})\approx -1.17\,\mathrm{dex}$. The spread of N/O is large ($\sigma_{\log(\mathrm{N/O})}=0.3\,\mathrm{dex}$), and we clearly see a population of NRGs with $\log(\mathrm{N/O})>-0.6\,\mathrm{dex}$, which comprise 4\% of the total sample, and the majority of these galaxies have $12+\log(\mathrm{O/H})\lesssim8$. The sample of high-redshift galaxies is well reproduced in our simulation, including the most extreme case of $\log(\mathrm{N/O})\approx 0.5$ at $12+\log(\mathrm{O/H})\approx8$ \citep{Ji:2024aa}. While we do reproduce the abundances of observed high-redshift galaxies, it is clear that the distribution is shifted compared to \thzoom. The lack of lower metallicity galaxies can be easily explained by selection effects because lower metallicity galaxies have lower stellar masses, and are therefore dimmer. Even at higher metallicities, the offset higher N/O ratios seen in observations may too be due to selection effects, where more NRGs have brighter nitrogen lines and therefore have easier to characterize chemical abundances. A much larger observational sample is needed to understand the true distribution of abundances at high redshift. The abundances found in 2G stars correspond to the more extreme upper end of our distribution.

In Fig.~\ref{fig:feo_met_rel} we show Fe/O as a function of the gas-phase oxygen abundance for all \thzoom galaxies. Fe/O increases with oxygen abundance and, while it shows relatively large scatter, this is significantly less than for N/O. There are very few Fe/O measurements for high-redshift galaxies, and the associated uncertainties are large. Additionally, three of the measurements are stellar abundances due to the difficulty of measuring gas-phase iron abundances \citep{Rodriguez:2005aa,Mendez-Delgado:2024aa}, and comparing stellar abundances to the gas-phase could introduce additional biases. Nevertheless, the abundances of \thzoom galaxies are consistent with these measurements within $1\sigma$, although the observed ratios are on the upper end of our distribution. The increase in Fe/O with increasing oxygen abundance is qualitatively in agreement with the trend seen in local nebulae, once the depletion of iron onto dust is accounted for \citep[see the depletion-corrected abundances in][]{Mendez-Delgado:2024aa}.

In Fig.~\ref{fig:co_met_rel} we show C/O as a function of the gas-phase oxygen abundance for all \thzoom galaxies. Our sample is marginally carbon-enriched compared to MW stars at higher metallicities, $12+\log(\mathrm{O/H})\gtrsim7.6$. At lower metallicities, $12+\log(\mathrm{O/H})\lesssim7.6$, our galaxies lie exactly on the relation for MW stars. The scatter of C/O is low relative to both the increasing trend with metallicity and to the scatter in Fe/O and N/O. We generally reproduce the C/O ratios of observed high-redshift galaxies, although there are 3 points with $\log(\mathrm{C/O})<-1\,\mathrm{dex}$ which are more than $1\sigma$ away from our distribution. Additionally, while the distribution of \thzoom galaxies generally well covers the distribution of GC stars, it is clear that 1G stars can scatter to lower C/O ratios. It is plausible that the carbon yields from either primary or secondary enrichment may to too high in our framework, given the highly uncertain and model-dependent nature of calculated yields. Reducing the carbon yield from AGB stars would reduce the slope of C/O with increasing oxygen abundance. On the other hand, reducing the carbon yield from CC SNe would lower the overall abundances somewhat and allow for greater scatter, particularly to lower values of C/O. This solution may be particularly appealing given that other works have shown that these low C/O values are achievable using pure CC SNe yields, such as \citet{Jones:2023aa} using yields from \citet{Nomoto:2013aa}. In any case, the abundances of \thzoom galaxies are within $1\sigma$ of all but two points of the compiled high-redshift observational data, and the distribution shows reasonably strong overlap with the abundances of GC stars. 

\subsection{Evolution of chemical abundances}
\label{sec:Evolution of chemical abundances}

\begin{figure*} 
\centering
	\includegraphics[width=\textwidth]{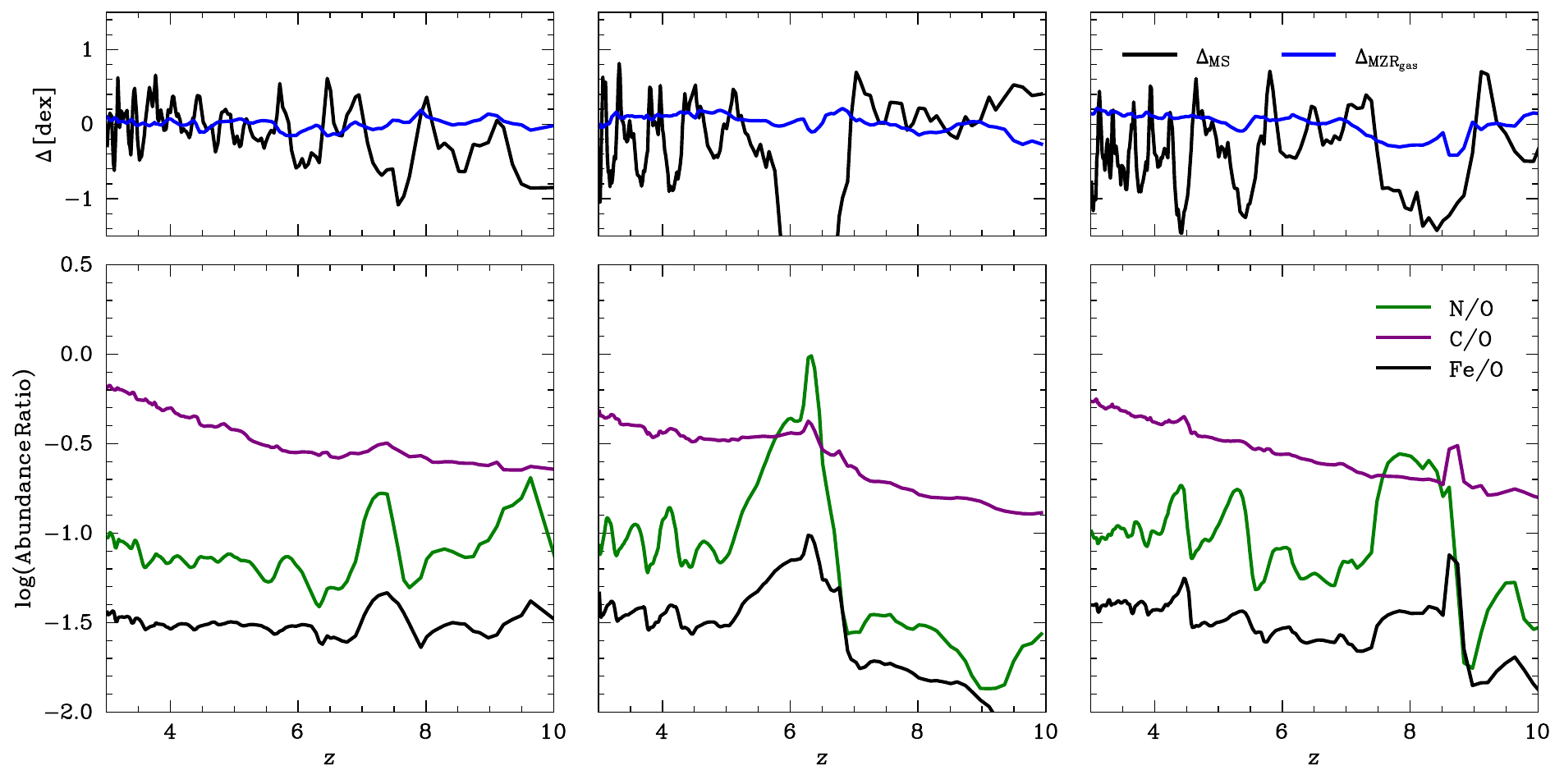}
    \caption{The evolution of three galaxies across cosmic time, with final stellar masses, from left to right, of $10^{11}\,\mathrm{M_\odot}$ (m12.6, subhalo 0), $10^{9.5}\,\mathrm{M_\odot}$ (m11.5, subhalo 0), and $10^{9.3}\,\mathrm{M_\odot}$ (m12.2, subhalo 2). The top panels show the offset from the SFMS, $\Delta\mathrm{MS}$, in black and offset from the redshift-dependent gas-phase MZR, $\Delta_\mathrm{MZR,gas}$, in blue. The lower panels show the evolution of chemical abundance ratios: N/O (red), C/O (blue), and Fe/O (black). The rapid oscillations in $\Delta\mathrm{MS}$ show the bursty nature of \thzoom galaxies \citep{McClymont:2025aa}, and we can clearly see that episodes of peculiar abundances are associated with this bursty behavior. More specifically, we see that NRGs tend to be in a lull of star formation following the end of a starburst.  $\Delta_\mathrm{MZR,gas}$ sometimes decreases following a starburst, likely due to metal-rich outflows, but this is not always the case, and even in the more extreme cases is only by a factor of $\sim0.2\,\mathrm{dex}$. This demonstrates that specifically ejecting CC SNe enriched gas is not the mechanism for generating NRGs, but instead that the total gas mass is reduced following a starburst, leading to a large number of polluting AGB stars relative to the gas mass in the ISM. Nitrogen enhancement following a starburst also tends to be associated with a milder increase in Fe/O. The C/O ratio is sometimes also mildly enhanced, and this tends to be associated with more extreme events.}
    \label{fig:evolution_three}
\end{figure*}

Having established that we reproduce some of the peculiar chemical abundance patterns seen in \textit{JWST} observations, we now aim to understand how they arise in the \thzoom simulations. In Fig.~\ref{fig:evolution_three} we show the evolution of the chemical abundances for three galaxies from $z=10$ to $z=3$. In the upper panel, we also show the offset from the gas-phase MZR, $\Delta_\mathrm{MZR,gas}$, and the offset from the SFMS, $\Delta\mathrm{MS}$.

The N/O ratio shows significant variability as the galaxies evolve, increasing by even an order of magnitude within 10's of Myrs. The periods of N/O enhancement occur during the period of suppressed SFR immediately following the end of a starburst, and decline relatively rapidly with the onset of the next star formation burst. Looking at the offset from the gas-phase MZR, $\Delta_\mathrm{MZR,gas}$, in the top panel, we see that the periods of increased N/O are sometimes associated with the galaxy falling below the MZR, however even the decline in $\Delta_\mathrm{MZR,gas}$ is only $\sim0.2\,\mathrm{dex}$, even in the most extreme cases. In some cases, the $\Delta_\mathrm{MZR,gas}$ actually increases during the onset of the nitrogen-rich period. That the changes in $\Delta_\mathrm{MZR,gas}$ are small aligns with the relatively small scatter around the MZR shown in Fig.~\ref{fig:mzr_scatter}. 

Importantly, we note that we have defined the MZR and $\Delta_\mathrm{MZR,gas}$ purely with oxygen abundance. This means that the nitrogen-rich periods are $\textit{not}$ necessarily associated with the ejection of particularly oxygen-rich gas, which would cause a sharp decrease in metallicity. Instead, a significant amount of gas is ejected and consumed into stars during a starburst period, reducing the amount of gas in the ISM of a galaxy. During the time period when enrichment due to AGB stars is dominant, there is a large population of AGB stars polluting the gas relative to the gas mass. This means that the gas-phase abundances are quickly moved toward the nitrogen-rich abundances of the AGB star winds. Once the next starburst begins, the onset of a fresh generation of CC SNe quickly enriches the gas in oxygen, washing out the nitrogen-rich signature. 

Fig.~\ref{fig:metal_maps} shows abundance maps for a galaxy in such a lull of star formation, corresponding to $z=6$ in the left panel of Fig.~\ref{fig:evolution_three} (m12.6, subhalo 0). We can clearly see that the central regions are enhanced in nitrogen, carbon, and iron relative to oxygen, indicating that oxygen-rich gas has been ejected and that the remaining ISM is dominated by AGB enrichment. This example also shows that the global abundances which we have measured cannot reflect the diversity of abundances within the galaxy, given that the global ratio is $\log(\mathrm{N/O})\approx-1.1$ but we can clearly see ratios reaching $\log(\mathrm{N/O})>-0.4$ on the projected map.

The Fe/O ratio shows similar, albeit more moderate, variability as N/O. This is because SNe Ia, which dominate iron enrichment during this period, do have elevated Fe/O compared to CC SNe. A very rough scaling is that the $\log(\mathrm{Fe/O})$ increases by $\sim$1/3 of the increase in $\log(\mathrm{N/O})$ during these periods.

The C/O ratio is much less variable. During the periods of enhanced N/O and mildly enhanced Fe/O, the C/O ratio is usually not elevated whatsoever. In extreme cases, there are brief increases of C/O by at most $\sim0.2\,\mathrm{dex}$. The C/O ratios increase fairly steadily as the galaxies evolve, which is in line with the population-level trend and low scatter shown in Fig.~\ref{fig:co_met_rel}. This implies that carbon and oxygen are produced on similar timescales, which is indeed the case in our chemical enrichment network.

\subsection{Nitrogen-rich galaxies}
\label{sec:Nitrogen-rich galaxies}

\begin{figure} 
\centering
	\includegraphics[width=\columnwidth]{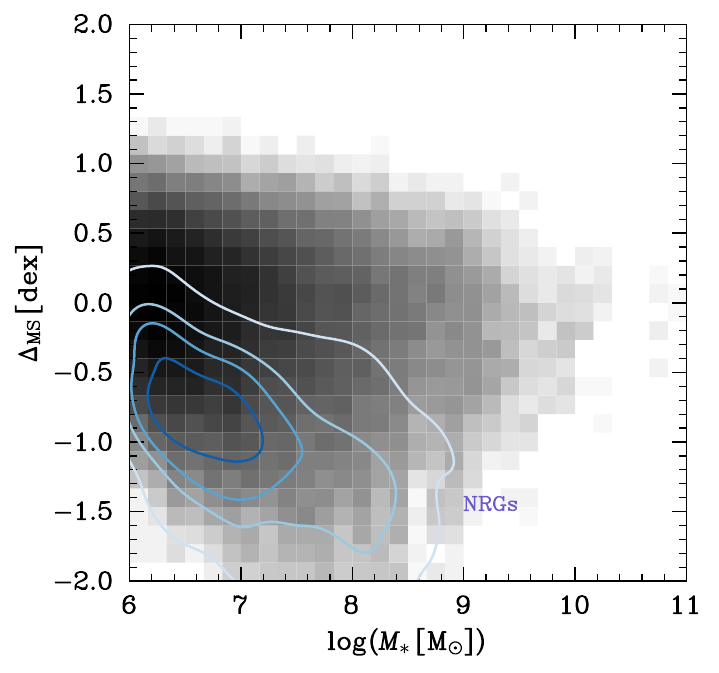}
    \caption{Offset from the SFMS, $\Delta\mathrm{MS}$, as a function of stellar mass. The black histogram shows the log-scaled distribution of all \thzoom galaxies included in this work. Blue contours show the distribution of NRGs ($\log(\mathrm{N/O}>-0.6$) in \thzoom. NRGs tend to lie below the SFMS, their ISMs having been enriched by AGB stars following the end of a starburst. Higher mass galaxies have a larger reservoir of previously enriched gas, have deeper potential wells, and are less bursty, meaning that extreme starbursts are needed to generate a sufficiently high ratio of AGB stars to ISM mass to allow for AGB-dominated enrichment. Therefore, there are fewer NRGs at higher masses, and they generally lie further below the SFMS.}
    \label{fig:nrich_analysis_one}
\end{figure}

\begin{figure} 
\centering
	\includegraphics[width=\columnwidth]{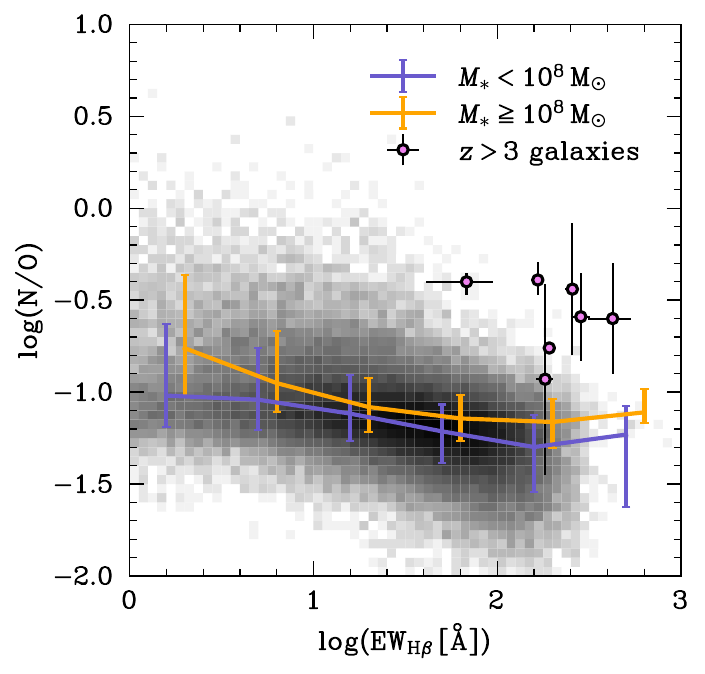}
    \caption{The gas-phase nitrogen to oxygen ratio as a function of EW[H$\beta$]. The black histogram shows the log-scaled distribution of all \thzoom galaxies included in this work. The blue (orange) line shows the median trend and the errorbars show the 16th--84th percentile scatter for galaxies with $M_{\ast}<10^{8}\,\mathrm{M_{\odot}}$ ($M_{\ast}\geq10^{8}\,\mathrm{M_{\odot}}$).
    Across the EW range, the median galaxy is not dramatically nitrogen-rich, with typical values of $\log(\mathrm{N/O})\sim-1$. NRGs generally show lower EW[H$\beta$], with the majority below $\mathrm{EW[H\beta]}<40$\,\AA, although the most extreme, rare cases can reach $\mathrm{EW[H\beta]}\gtrsim100$\,\AA.}
    \label{fig:hb_ew_no}
\end{figure}

While we do produce the elevated nitrogen abundances seen in \textit{JWST} observations, it is important to understand whether our simulated NRGs have similar properties to the observed galaxies. One notable feature of the nitrogen-rich phases explored in the previous section is that they have relatively low SFRs. We explore this on a population level in Fig.~\ref{fig:nrich_analysis_one}, where we show the distribution of our \thzoom galaxies for $\Delta\mathrm{MS}$ against stellar mass as a histogram. NRGs, defined as those with $\log(\mathrm{N/O})>-0.6$, are plotted on top as a contour. This plot confirms our conclusion from the previous section, where elevated N/O is primarily seen in galaxies with suppressed SFRs.

This is important because observed NRGs tend to have high Balmer line equivalent widths, which is generally associated with starbursts. To examine this directly, in Fig.~\ref{fig:hb_ew_no} we show N/O as a function of EW[H$\beta$] for \thzoom galaxies compared to \textit{JWST}-observed NRGs. For the \thzoom galaxies, we show separate median lines for two stellar mass bins, $M_{\ast}<10^{8}\,\mathrm{M_{\odot}}$ and $M_{\ast}\geq10^{8}\,\mathrm{M_{\odot}}$. The distribution of \textit{JWST}-observed NRGs is shifted compared to our sample of galaxies, generally being exclusively found at high EW[H$\beta$].

Galaxies observed with NIRSpec represent a heavily biased sample, with brighter, heavily star-forming galaxies being more likely to be observed. Of the NIRSpec observed sample, only objects with extremely bright emission lines have even mildly constraining measurements of N/O, making the sample of N/O measured galaxies even more heavily biased. We should therefore be careful in making strong conclusions about the relationship between EW and N/O from individual nitrogen-rich observations. One method to help alleviate the signal-to-noise bias is to stack galaxy spectra in order to produce averaged spectra with higher S/N, although we note that this method is still subject to NIRSpec selection function biases. \citet{Hayes:2025aa} use stacks of high-redshift galaxies observed with \textit{JWST}/NIRSpec, where they stack based on $\mathrm{EW[\OIII]}$. The uncertainties on the N/O ratios of their stacks are large, however the high $\mathrm{EW[\OIII]}$ stacks do not show signs of being more nitrogen-rich than the lower EW stacks. In fact, the highest N/O ratios are found for the lowest EW stack for which they show, $500<\mathrm{EW[\OIII]}<1000$\,\AA.

Regardless of the average N/O value at a given EW[H$\beta$], which is not clearly constrained observationally, the fact that individual NRGs are observed with high EW[H$\beta$] means that our model should produce such galaxies, even if they are rare. In our scenario, we would expect high-EW NRGs to be found when a lulling, AGB-enriched galaxy undergoes a subsequent starburst, which would illuminate the pre-enriched gas. The primary reason this is not seen is because we are measuring galaxy-scale abundances, which are quickly diluted by inflowing gas or via CC SNe at the onset of the second burst of star formation. We explore this effect in the following section, where we consider abundances on the GMC scale. 

Another mechanism to generate high-EW NRGs are AGN, which are not included in the \thzoom model. An AGN that is rapidly accreting gas at the onset of the second starburst, when the gas is nitrogen-rich, would be able to produce high EW emission without also enriching the ISM and diluting the nitrogen, unlike star formation. This potentially also explains the association between AGN and nitrogen-rich abundances found by \citet{Isobe:2025aa}. There are also changes which could be made to the chemical enrichment network that would help produce galaxies with both high N/O and high EW[H$\beta$], such as the inclusion of enrichment from Wolf-Rayet stars. Wolf-Rayet enrichment provides an additional source of nitrogen-rich metals which is delayed from the CC SNe \citep{Kobayashi:2024aa}. This would increase the amount of nitrogen injected into the ISM following the quenching of the first starburst, meaning that it would take longer for the second starburst to dilute the nitrogen with CC SNe enrichment.

We also note that the observed gas-phase chemical abundances which we are comparing to are measured using emission line ratios. Even without considering the biases which can be introduced in these measurements (see Section~\ref{sec:Stellar and gas-phase metallicity}), they are certainly not measuring mass-weighted abundances within twice the stellar half-mass radius as we are. We anticipate future work generating mock emission lines for more direct comparison with observations.

Nitrogen-rich \thzoom galaxies are generally moderately iron-enriched and not particularly C/O rich compared to the general high-redshift galaxy population. There is not yet sufficient data to study these potential associations observationally at this stage; however measurement of Fe/O for a large number of high-redshift galaxies may help test proposed scenarios for creating NRGs. 

\subsection{Nitrogen-rich giant molecular clouds}
\label{sec:Nitrogen-rich giant molecular clouds}

\begin{figure} 
\centering
	\includegraphics[width=\columnwidth]{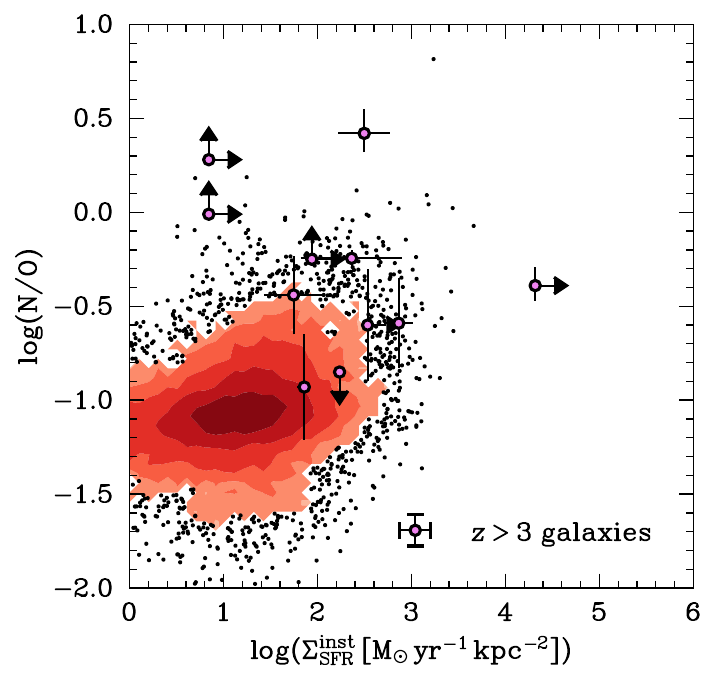}
    \caption{The instantaneous SFR-weighted gas-phase nitrogen to oxygen ratio as a function of instantaneous SFR surface density in GMCs. Despite nitrogen-rich abundances being generated during lulls of star formation, nitrogen-rich GMCs can be found at the onset of a fresh starburst, as the pre-polluted gas collapses. Nitrogen-rich GMCs are found even once the galaxy's global abundances become typical for the galaxy's metallicity, due to some initial scattered star formation or the re-accretion of CC SNe-enriched gas, although they are quickly enriched by the CC SNe of their young stars. These pre-polluted, nitrogen-rich GMCs have $\Sigma_{\mathrm{SFR}}^\mathrm{inst}$ comparable to observed NRGs, and these GMCs may dominate the line emission, and in particular the UV line emission, resolving the apparent tension seen when comparing the EW[H$\beta$] of galaxies and their global abundance ratios. }
    \label{fig:gmc_sfr_density_no}
\end{figure}

While the forward modeling of metal line emission is outside of the scope of this work, we can gain insight into the abundances of star-forming regions by analyzing the chemical abundances of giant molecular clouds (GMCs). The gas in GMCs is in close proximity to a large population of young stars, which dominate the LyC output of galaxies. The abundances in these regions, then, may dominate the global metal line emission, particularly in the UV lines which tend to arise from highly ionized species with relatively small ionization zones.

The method for identifying GMCs in the \thzoom simulations was presented in \citet{Wang:2025aa}, however, we briefly summarize the key points here. GMCs were identified with \textsc{CloudPhinder}\footnote{https://github.com/mikegrudic/CloudPhinder} \citep{Guszejnov:2020aa}, which was applied to search for self-gravitating structures of cold (<1000\,K) dense ($n>100\,\mathrm{cm^{-3}}$ gas and young stellar objects (<3\,Myr, YSOs). In this section, we use the instantaneous SFR as measured from the gas associated with the GMCs in order to characterize actively star-forming GMCs. Additionally, instead of using the mass-weighted abundances, we use the SFR-weighted abundances of the gas cells to demonstrate better the chemical makeup of newly forming stars in the GMC. The SFR surface density is then calculated as $\Sigma_{\mathrm{SFR}}^\mathrm{inst} = \mathrm{SFR}_\mathrm{inst} / (\pi R_\mathrm{e}^2)$, where $R_\mathrm{e}^2$ is the effective radius of the GMC.
Using mass-weighted abundances and SFRs measured from associated YSOs quantitatively changes our results, but they remain qualitatively unchanged, and we show this comparison in Appendix~\ref{sec:Giant molecular cloud chemical abundance definitions}.

In Fig.~\ref{fig:gmc_sfr_density_no} we show the gas-phase N/O of GMCs as a function of $\Sigma_{\mathrm{SFR}}^\mathrm{inst}$. We can immediately see that we do indeed form nitrogen-rich GMCs. Nitrogen-rich GMCs tend to form at the onset of a starburst from the collapsing gas, which was polluted by AGB stars from the previous burst.

We also plot \textit{JWST}-observed NRGs. The SFR surface density for many of these NRGs is a lower limit as they are photometrically unresolved. While the observed SFR surface density is not identical to that calculated for GMCs because it uses the effective radius of the whole galaxy, we expect $\Sigma_{\mathrm{SFR}}^\mathrm{inst}$ to be comparable due to the cluster-dominated nature of high-redshift star formation. Such cluster-dominated star formation has been discussed in the context of simulations \citep[e.g.][e.g.,]{Belokurov:2022aa,Belokurov:2023aa}, but observations have also provided direct evidence \citep[e.g.,][]{Vanzella:2022aa,Vanzella:2023aa,Adamo:2024aa,Mowla:2024aa,Fujimoto:2025aa}. That we find nitrogen-rich GMCs with $\Sigma_{\mathrm{SFR}}^\mathrm{inst}$ comparable to those found in observed NRGs indicates that we are able to reproduce the extreme properties of observed NRGs. Line emission may be dominated by these bright, nitrogen-rich clumps, which explains the apparent discrepancy between our results and observed NRGs discussed in the previous section.

Notably, nitrogen-rich GMCs can sometimes be found in galaxies that have ordinary global N/O ratios. This indicates that while the wider ISM has been enriched by CC SNe, either through some initial distributed star formation or through the re-accretion of ejected gas, nitrogen-rich GMCs are still able to collapse and form. We also note some instances of nitrogen-rich GMCs forming in more massive galaxies, even though the global abundances never become nitrogen-rich. This indicates that the global bursty behavior leading to nitrogen-enrichment seen in Fig.~\ref{fig:evolution_three} can also occur on spatially resolved scales in more massive galaxies, where parts of the ISM are evacuated without its complete ejection. It may be possible to observe such events with spatially resolved spectroscopy, and indeed \citet{Scholtz:2025ab} found indications of a nitrogen-rich clump within a larger galaxy.

\section{Discussion}
\label{sec:Discussion}

\begin{figure*} 
\centering
	\includegraphics[width=\textwidth]{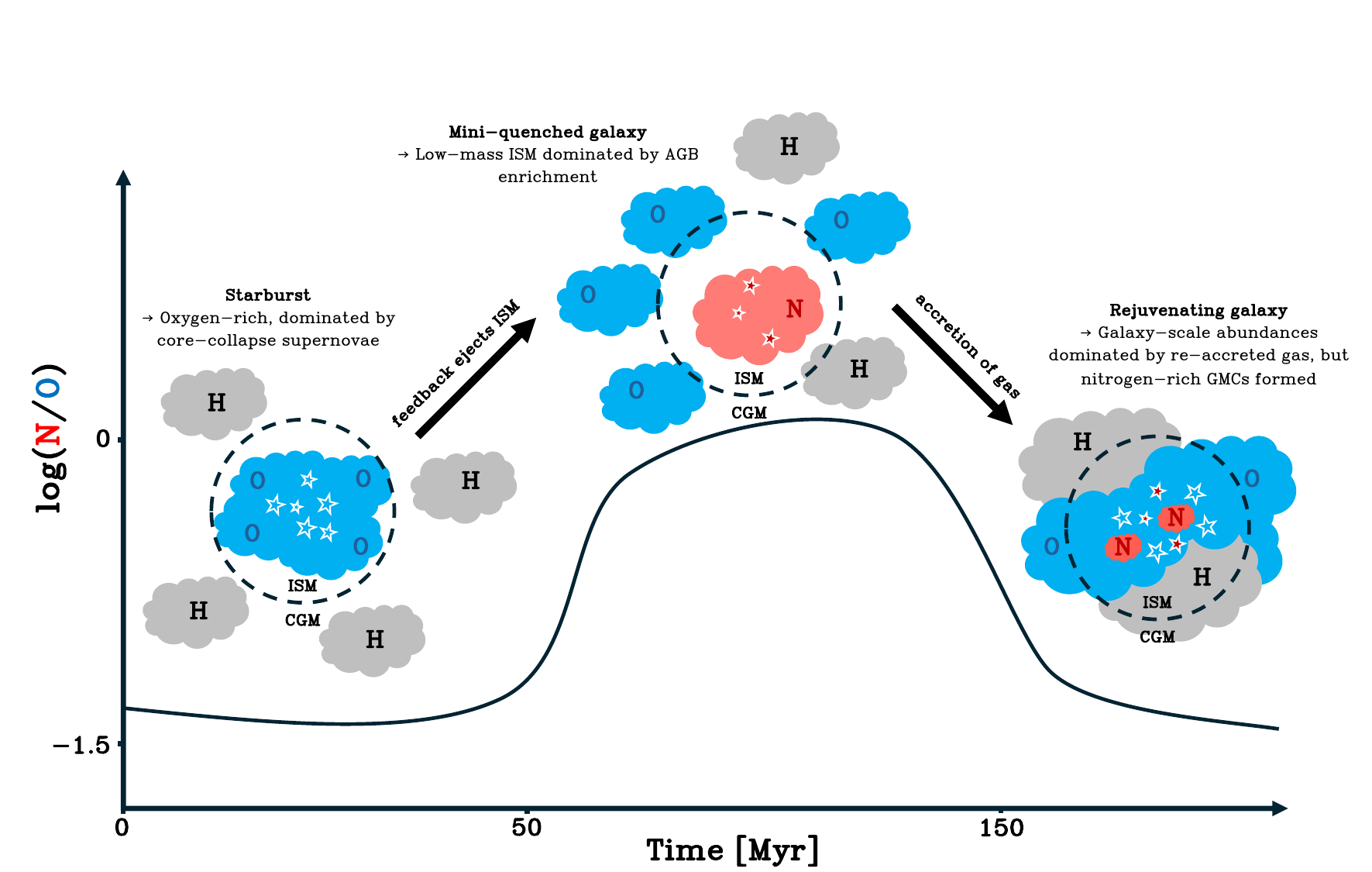}
    \caption{Diagram showing how NRGs arise in the \thzoom simulations. Galaxies undergoing starbursts are oxygen-rich, being dominated by enrichment due to CC SNe. Star formation ceases as gas is consumed into stars and ejected from the galaxy, leaving behind a low mass, diffuse ISM. This low mass ISM is subsequently dominated by enrichment from the winds of the relatively large number of AGB stars, causing the gas to become nitrogen-rich. Subsequently, pristine and previously ejected gas inflow to the galaxy, causing the galaxy-scale abundances to once again be dominated by CC SNe ejecta. However, the diffuse nitrogen-rich gas can collapse down to form nitrogen-rich GMCs with extreme SFR surface densities. These GMCs may dominate the line emission of the galaxy, and in particular the UV line emission. }
    \label{fig:metal_scheme_one}
\end{figure*}

\subsection{Out of equilibrium metallicity in the high-redshift regime}

In the high-redshift regime, simple models of chemical enrichment, such as closed, leaky, and accreting box models, may still perform well on the halo scale, provided that metal-loaded outflows to the IGM are not common \citep{Ma:2016aa}. However, these models struggle on the galaxy scale because bursty star formation in low-mass and high-redshift galaxies causes rapid fluctuations in the gas fraction due to rapid inflow, outflows, and the consumption of gas into stars. In the \thzoom simulations, the slope of the MZR is set by the gas fraction. This is because the metal retention efficiency does not appear to vary with stellar mass, at least where we have sufficient statistics to test this across a wide redshift range, $M_{\ast}\lesssim10^{9.5}\,\mathrm{M_{\odot}}$. The redshift evolution of the MZR is weak, and we find that it is set by a trend of increasing metal retention efficiency down to $z\approx5$, and then by decreasing gas fractions.

Another clear break with the low redshift Universe is the inversion of the FMR seen in \thzoom galaxies (see Fig.~\ref{fig:sfms_mzr_rel_star}). Such a breakdown is not seen in lower resolution, large volume cosmological simulations \citep{Garcia:2025aa,Garcia:2025ab}. \citet{Torrey:2018aa} analyzed the origin of the FMR at lower redshifts ($z\lesssim4$) in the IllustrisTNG simulation \citep{Pillepich:2018aa} and found that the FMR arises due to the similarity of variability timescales for SFRs and metallicity. They note that for models with burstier star-formation histories, the SFR can easily vary on more rapid timescales than metallicity, which would weaken the FMR. This effect is certainly present in \thzoom galaxies, where we see SFRs fall rapidly compared to changes in metallicity, as in Fig.~\ref{fig:evolution_three}. However, more extreme than simply a weaker FMR, we find evidence for an inverted FMR in low-mass galaxies. This is due to the additional effect of dilution, which is much more extreme at both lower masses and higher redshifts due to the smaller ratio of enriched gas to pristine inflow.

While we do find a breakdown in the FMR, we do not claim that this is responsible for the large offsets from the FMR recently seen in \textit{JWST} observations \citep[e.g.][]{Curti:2024ab}. Instead, these offsets are likely due to the low-redshift FMR being poorly calibrated at low masses and high SFRs. When we apply a low-redshift FMR to our galaxies, we find a median offset of $\sim$1\,dex at $M_{\ast}=10^{6}\,\mathrm{M_{\odot}}$, whereas there is relative agreement at $M_{\ast}=10^{11}\,\mathrm{M_{\odot}}$ (see Fig.~\ref{fig:mzr_scatter}). A recalibrated FMR could be calculated which would fit the median metallicities at high redshift more effectively, although we would predict that this recalibrated FMR would be ineffective at reducing scatter relative to the MZR for lower mass ($M_{\ast}\lesssim10^{9}\,\mathrm{M_{\odot}}$) galaxies at $z>3$, unless a more complicated function is used to account for and inversion of the correlation between metallicity and SFR. However, the prospects for measuring the inverted FMR as predicted in our simulations are uncertain, as it would require measuring gas-phase metallicities for galaxies with low gas masses and SFRs.

Given the chaotic nature of metal enrichment discussed here, we may expect a strong evolution of MZR scatter with redshift, however this is not the case (see Fig.~\ref{fig:mzr_scatter}). In fact, we only see a redshift evolution of the scatter when considering exclusively central galaxies, and even then only for lower masses ($M_{\ast}\lesssim10^{9}\,\mathrm{M_{\odot}}$). On the other hand, the mass dependence of the MZR scatter is clear. This picture aligns somewhat with the burstiness of \thzoom galaxies, which is a function of mass, and which does increase with redshift, albeit relatively weakly \citep{McClymont:2025aa}. It is interesting to note that the overall normalization of the MZR is also only weakly dependent on redshift compared to stellar mass. At least in the redshift range studied here, $3<z<12$, stellar mass is clearly the dominant factor in driving both variability and normalization of galaxy-scale metallicity, with redshift having a much weaker effect.

\subsection{Disentangling high-redshift chemical enrichment models}

A plethora of models have recently been proposed to explain the puzzling chemical abundances of high-redshift galaxies, and in particular, NRGs. These include fast-rotating massive stars \citep{Vink:2023aa,Nandal:2024ab,Tsiatsiou:2024aa}, very, extremely, or supermassive stars \citep[$M_\star>100~M_\odot$, $10^3~M_\odot<M_\star<10^4~M_\odot$, and $M_\star>10^4~M_\odot$, respectively;][]{Charbonnel:2023aa,Nagele:2023aa,Nandal:2024aa,Nandal:2025aa,Gieles:2025aa}, Wolf-Rayet stars \citep{Kobayashi:2024aa,Watanabe:2024aa}, tidal disruption events \citep{Cameron:2023ab}, and AGB star enrichment combined with differential outflows \citep{Rizzuti:2025aa} or pristine gas inflows \citep{DAntona:2023aa}.

The scenario seen in \thzoom is most similar to the differential winds discussed in \citet{Rizzuti:2025aa}, who similarly invoke differential winds to preferentially eject gas enriched by CC SNe, however, the models are not identical. \citet{Rizzuti:2025aa} explore models where differential winds are achieved spatially, where star formation can be constantly ongoing and CC SNe-enriched gas is constantly ejected. On the other hand, as we have shown, differential winds are achieved temporally in the \thzoom simulations, where a large fraction of the ISM is ejected following a starburst, leaving the remaining gas to be dominated by AGB wind enrichment. A schematic diagram illustrating this scenario is shown in Fig.~\ref{fig:metal_scheme_one}, including the recollapse of this diffuse, nitrogen-rich gas into nitrogen-rich GMCs. We also note that SFH variations have been invoked to explain differences in the N/O plateau between galaxies at low redshift \citet{Berg:2020aa}, although this effect arises due to the longer-term SFH, rather than individual burst-quench cycles.

We emphasize that nothing in the \thzoom simulations was designed to produce the peculiar abundances examined in this work, and in fact, the employed chemical enrichment scheme was largely implemented over a decade ago \citep{Vogelsberger:2013aa}. That such abundances appear in our simulation with this established chemical enrichment model, which does not invoke any non-standard enrichment sources, is remarkable. 

\section{Conclusions}
\label{sec:Conclusions}

We have used the \thzoom radiation-hydrodynamic zoom-in simulations to investigate how early galaxies acquire, lose, and recycle their heavy elements during the first two billion years of cosmic time ($3<z<12$). The combination of parsec-scale resolution, multi-frequency radiation transport, on-the-fly dust, and non-equilibrium chemistry allows us to connect global metallicity trends with the bursty, small-scale physics that drives them. Additionally, our chemical network, which tracks the abundances of nine chemical elements (H, He, C, N, O, Ne, Mg, Si, and Fe), enables us to explore the chemical enrichment of the first galaxies. Our key findings are as follows:

\begin{itemize} 
\item \textbf{A slowly evolving mass-metallicity relation:} We fit redshift-dependent gas-phase and stellar MZR, finding general agreement with observations across a wide range of redshifts, including extrapolation to $z=0$. Although we do not have a sufficient sample of high-mass galaxies to see a turnover in the MZR, we do clearly see curvature toward higher masses.
\item \textbf{Distinct redshift evolution of total metallicity versus individual elements:} The choice of chemical tracer has a dramatic impact on the redshift evolution of the MZR. Using their traditional tracers of oxygen and iron abundance for gas-phase and stellar metallicity, respectively, we find $12+\log(\mathrm{O/H})\propto(1+z)^{0.28}$ and $\mathrm{[Fe/H]}\propto(1+z)^{0.77}$. If we instead fit the total mass fraction of metals, we find scalings of $Z_\mathrm{gas}\propto(1+z)^{0.39}$ and $Z_\ast\propto(1+z)^{0.43}$. This reflects the differing enrichment timescales of elements, with oxygen being rapidly enriched by CC SNe, while iron is enriched on Gyr timescales by SNe Ia. The remaining steepness of the stellar MZR relative to the gas-phase MZR represents the stellar metallicity ‘‘catching-up’’ to the ISM metallicity with cosmic time.
\item \textbf{The combined role of metal retention and gas fractions in driving the mass-metallicity relation:} The slope of the MZR is governed by the decrease of $M_\mathrm{gas}/M_\ast$ with increasing stellar mass, with the metal retention efficiency, $y_\mathrm{gas}$, showing no clear mass dependence for central galaxies up to at least $M_\ast\lesssim10^{9.5}~\mathrm{M_\odot}$. On the other hand, the weak redshift dependence has a more complicated origin. Down to $z\approx5$, the evolution is driven by an increasing $y_\mathrm{gas}$ with cosmic time, which offsets a surprisingly \textit{increasing} $M_\mathrm{gas}/M_\ast$. From $z\approx5$ to the end of our simulations at $z=3$, $y_\mathrm{gas}$ stabilizes, and the subsequent evolution is driven by $M_\mathrm{gas}/M_\ast$, which inverts at $z\approx5$ and now decreases with cosmic time.
\item \textbf{Breakdown of the fundamental metallicity relation in low-mass galaxies:} At $M_\ast\lesssim10^{9}\,{\rm M_\odot}$ the classical anti-correlation between SFR and gas metallicity not only weakens but \textit{reverses} for central galaxies, except for the most extremely star-forming galaxies ($\Delta_\mathrm{MS}\gtrsim0.5$\,dex). Such an inversion arises due to the prevalence of pristine gas diluting the ISM of low-SFR galaxies, which then enrich their gas as their SFR ramps up. Satellite galaxies show a more complex, valley-like dependence due to the additional impact of metal pollution from their central. The stellar FMR, in contrast, remains as expected for all galaxies because most stars form once the galaxy has already been somewhat pre-enriched.
\item \textbf{Generation of NRGs without exotic yields via temporally differential winds:} Our standard chemical enrichment network combined with bursty star formation generates order-of-magnitude excursions in N/O on $\lesssim100$\,Myr time-scales, reproducing extreme values reported by recent \textit{JWST} studies \citep[e.g.][]{Isobe:2023aa,Ji:2026aa}. These variations are achieved via temporally differential winds, where outflows rapidly follow starbursts and eject CC SNe-enriched gas. This leaves a relatively low mass ISM, which is then dominated by the nitrogen-heavy enrichment from AGB stars.
\item \textbf{Most NRGs are lulling galaxies, but nitrogen-rich GMCs formed from pre-enriched gas match observations:} NRGs show a milder enhancement in Fe/O, whereas C/O is generally not enhanced, although it can be in the more extreme nitrogen-rich episodes. NRGs are typically found below the SFMS and usually have $\mathrm{EW[H\beta]}<40$\,\AA, although there are some cases with $\mathrm{EW[H\beta]}\gtrsim100$\,\AA. Our proposed scenario may be reconciled with observations of high-EW ultra-NRGs by considering either an AGN or young stars preferentially illuminating the pre-enriched gas at the beginning of a new starburst. This reconciliation is supported by the fact that we find nitrogen-rich GMCs with high SFR surface densities at the onset of starbursts.
\end{itemize}

The analysis presented in this paper has focused on mass-weighted metallicities within galaxies, which are not directly comparable with observations. In the future, we plan to forward model metal emission lines from the \thzoom simulations.

\section*{Acknowledgements}

The authors are grateful to the referee for their helpful comments, which improved the manuscript. The authors gratefully acknowledge the Gauss Centre for Supercomputing e.V. (\url{www.gauss-centre.eu}) for funding this project by providing computing time on the GCS Supercomputer SuperMUC-NG at Leibniz Supercomputing Centre (\url{www.lrz.de}), under project pn29we. WM thanks the Science and Technology Facilities Council (STFC) Center for Doctoral Training (CDT) in Data Intensive Science at the University of Cambridge (STFC grant number 2742968) for a PhD studentship. WM and ST acknowledge support by the Royal Society Research Grant G125142. RK acknowledges support of the Natural Sciences and Engineering Research Council of Canada (NSERC) through a Discovery Grant and a Discovery Launch Supplement (funding reference numbers RGPIN-2024-06222 and DGECR-2024-00144) and York University's Global Research Excellence Initiative. XS acknowledges the support from the NASA theory grant JWST-AR-04814. XS acknowledges the support from the National Aeronautics and Space Administration (NASA) theory grant JWST-AR-04814.

\section*{Data Availability}

All simulation data, including snapshots, group, and subhalo catalogs and merger trees will be made publicly available in the near future. Data will be distributed via \url{www.thesan-project.com}. Before the public data release, data underlying this article will be shared on reasonable request to the corresponding author(s).



\bibliographystyle{mnras}
\bibliography{mainbib} 

@article{Shen:2025aa,
	adsurl = {https://ui.adsabs.harvard.edu/abs/2025MNRAS.tmp.2012S},
	archiveprefix = {arXiv},
	author = {{Shen}, Xuejian and {Kannan}, Rahul and {Puchwein}, Ewald and {Smith}, Aaron and {Vogelsberger}, Mark and {Borrow}, Josh and {Garaldi}, Enrico and {Keating}, Laura and {Zier}, Oliver and {McClymont}, William and {Tacchella}, Sandro and {Wang}, Zihao and {Hernquist}, Lars},
	doi = {10.1093/mnras/staf2119},
	eprint = {2503.01949},
	journal = {\mnras},
	month = dec,
	primaryclass = {astro-ph.GA},
	title = {{The THESAN-ZOOM project: Star-formation efficiencies in high-redshift galaxies}},
	year = 2025}

@article{Ji:2026aa,
	adsurl = {https://ui.adsabs.harvard.edu/abs/2026MNRAS.545f2110J},
	author = {{Ji}, Xihan and {Belokurov}, Vasily and {Maiolino}, Roberto and {Monty}, Stephanie and {Isobe}, Yuki and {Kravtsov}, Andrey and {McClymont}, William and {{\"U}bler}, Hannah},
	doi = {10.1093/mnras/staf2110},
	eid = {staf2110},
	journal = {\mnras},
	month = jan,
	number = {3},
	pages = {staf2110},
	title = {{Connecting JWST discovered N/O-enhanced galaxies to globular clusters: evidence from chemical imprints}},
	volume = {545},
	year = 2026}

@article{McClymont:2026aa,
	adsurl = {https://ui.adsabs.harvard.edu/abs/2026MNRAS.545f2092M},
	archiveprefix = {arXiv},
	author = {{McClymont}, William and {Tacchella}, Sandro and {Ji}, Xihan and {Kannan}, Rahul and {Maiolino}, Roberto and {Simmonds}, Charlotte and {Smith}, Aaron and {Puchwein}, Ewald and {Garaldi}, Enrico and {Vogelsberger}, Mark and {D'Eugenio}, Francesco and {Keating}, Laura and {Shen}, Xuejian and {Trefoloni}, Bartolomeo and {Zier}, Oliver},
	doi = {10.1093/mnras/staf2092},
	eid = {staf2092},
	eprint = {2506.13852},
	journal = {\mnras},
	month = jan,
	number = {1},
	pages = {staf2092},
	primaryclass = {astro-ph.GA},
	title = {{Overmassive black holes in the early Universe can be explained by gas-rich, dark matter-dominated galaxies}},
	volume = {545},
	year = 2026}

@article{McClymont:2025ab,
	adsurl = {https://ui.adsabs.harvard.edu/abs/2025MNRAS.544.1732M},
	archiveprefix = {arXiv},
	author = {{McClymont}, William and {Tacchella}, Sandro and {Smith}, Aaron and {Kannan}, Rahul and {Puchwein}, Ewald and {Borrow}, Josh and {Garaldi}, Enrico and {Keating}, Laura and {Vogelsberger}, Mark and {Zier}, Oliver and {Shen}, Xuejian and {Popovic}, Filip},
	doi = {10.1093/mnras/staf1861},
	eprint = {2503.04894},
	journal = {\mnras},
	month = dec,
	number = {2},
	pages = {1732-1747},
	primaryclass = {astro-ph.GA},
	title = {{The THESAN-ZOOM project: central starbursts and inside-out quenching govern galaxy sizes in the early Universe}},
	volume = {544},
	year = 2025}

@article{Wang:2025aa,
	adsurl = {https://ui.adsabs.harvard.edu/abs/2025MNRAS.544.2675W},
	archiveprefix = {arXiv},
	author = {{Wang}, Zihao and {Shen}, Xuejian and {Vogelsberger}, Mark and {Li}, Hui and {Kannan}, Rahul and {Puchwein}, Ewald and {Smith}, Aaron and {Borrow}, Josh and {Garaldi}, Enrico and {Keating}, Laura and {Zier}, Oliver and {McClymont}, William and {Tacchella}, Sandro and {Ni}, Yang and {Hernquist}, Lars},
	doi = {10.1093/mnras/staf1677},
	eprint = {2505.05554},
	journal = {\mnras},
	month = dec,
	number = {3},
	pages = {2675-2697},
	primaryclass = {astro-ph.GA},
	title = {{The THESAN-ZOOM project: star formation efficiency from giant molecular clouds to galactic scale in high-redshift starbursts}},
	volume = {544},
	year = 2025}

@article{Peimbert:2010aa,
	adsurl = {https://ui.adsabs.harvard.edu/abs/2010ApJ...724..791P},
	archiveprefix = {arXiv},
	author = {{Peimbert}, Antonio and {Peimbert}, Manuel},
	doi = {10.1088/0004-637X/724/1/791},
	eprint = {1006.0692},
	journal = {\apj},
	month = nov,
	number = {1},
	pages = {791-798},
	primaryclass = {astro-ph.GA},
	title = {{On the O/H, Mg/H, Si/H, and Fe/H Gas and Dust Abundance Ratios in Galactic and Extragalactic H II Regions}},
	volume = {724},
	year = 2010}

@article{Hayes:2025aa,
	adsurl = {https://ui.adsabs.harvard.edu/abs/2025ApJ...982...14H},
	archiveprefix = {arXiv},
	author = {{Hayes}, Matthew J. and {Saldana-Lopez}, Alberto and {Citro}, Annalisa and {James}, Bethan L. and {Mingozzi}, Matilde and {Scarlata}, Claudia and {Martinez}, Zorayda and {Berg}, Danielle A.},
	doi = {10.3847/1538-4357/adaea1},
	eid = {14},
	eprint = {2411.09262},
	journal = {\apj},
	month = mar,
	number = {1},
	pages = {14},
	primaryclass = {astro-ph.GA},
	title = {{On the Average Ultraviolet Emission-line Spectra of High-redshift Galaxies: Hot and Cold, Carbon-poor, Nitrogen Modest, and Oozing Ionizing Photons}},
	volume = {982},
	year = 2025}

@article{Morel:2025aa,
	adsurl = {https://ui.adsabs.harvard.edu/abs/2025arXiv251120484M},
	archiveprefix = {arXiv},
	author = {{Morel}, I. and {Schaerer}, D. and {Marques-Chaves}, R. and {Prantzos}, N. and {Charbonnel}, C. and {Brammer}, G. and {Xiao}, M. and {Dessauges-Zavadsky}, M.},
	doi = {10.48550/arXiv.2511.20484},
	eid = {arXiv:2511.20484},
	eprint = {2511.20484},
	journal = {arXiv e-prints},
	month = nov,
	pages = {arXiv:2511.20484},
	primaryclass = {astro-ph.GA},
	title = {{Discovery of new N-emitters over a wide redshift range}},
	year = 2025}

@article{Topping:2025aa,
	adsurl = {https://ui.adsabs.harvard.edu/abs/2025ApJ...980..225T},
	archiveprefix = {arXiv},
	author = {{Topping}, Michael W. and {Stark}, Daniel P. and {Senchyna}, Peter and {Chen}, Zuyi and {Zitrin}, Adi and {Endsley}, Ryan and {Charlot}, St{\'e}phane and {Furtak}, Lukas J. and {Maseda}, Michael V. and {Plat}, Adele and {Smit}, Renske and {Mainali}, Ramesh and {Chevallard}, Jacopo and {Molyneux}, Stephen and {Rigby}, Jane R.},
	doi = {10.3847/1538-4357/ada95c},
	eid = {225},
	eprint = {2407.19009},
	journal = {\apj},
	month = feb,
	number = {2},
	pages = {225},
	primaryclass = {astro-ph.GA},
	title = {{Deep Rest-UV JWST/NIRSpec Spectroscopy of Early Galaxies: The Demographics of C IV and N-emitters in the Reionization Era}},
	volume = {980},
	year = 2025}

@article{Nomoto:2013aa,
	adsurl = {https://ui.adsabs.harvard.edu/abs/2013ARA&A..51..457N},
	author = {{Nomoto}, Ken'ichi and {Kobayashi}, Chiaki and {Tominaga}, Nozomu},
	doi = {10.1146/annurev-astro-082812-140956},
	journal = {\araa},
	month = aug,
	number = {1},
	pages = {457-509},
	title = {{Nucleosynthesis in Stars and the Chemical Enrichment of Galaxies}},
	volume = {51},
	year = 2013}

@article{Jones:2023aa,
	adsurl = {https://ui.adsabs.harvard.edu/abs/2023ApJ...951L..17J},
	archiveprefix = {arXiv},
	author = {{Jones}, Tucker and {Sanders}, Ryan and {Chen}, Yuguang and {Wang}, Xin and {Morishita}, Takahiro and {Roberts-Borsani}, Guido and {Treu}, Tommaso and {Dressler}, Alan and {Merlin}, Emiliano and {Paris}, Diego and {Santini}, Paola and {Bergamini}, Pietro and {Henry}, A. and {Huntzinger}, Erin and {Nanayakkara}, Themiya and {Boyett}, Kristan and {Bradac}, Marusa and {Brammer}, Gabriel and {Calabr{\'o}}, Antonello and {Glazebrook}, Karl and {Grasha}, Kathryn and {Mascia}, Sara and {Pentericci}, Laura and {Trenti}, Michele and {Vulcani}, Benedetta},
	doi = {10.3847/2041-8213/acd938},
	eid = {L17},
	eprint = {2301.07126},
	journal = {\apjl},
	month = jul,
	number = {1},
	pages = {L17},
	primaryclass = {astro-ph.GA},
	title = {{Early Results from GLASS-JWST. XXI. Rapid Asembly of a Galaxy at z = 6.23 Revealed by Its C/O Abundance}},
	volume = {951},
	year = 2023}

@article{Mendez-Delgado:2024aa,
	adsurl = {https://ui.adsabs.harvard.edu/abs/2024A&A...690A.248M},
	archiveprefix = {arXiv},
	author = {{M{\'e}ndez-Delgado}, J.~E. and {Kreckel}, K. and {Esteban}, C. and {Garc{\'\i}a-Rojas}, J. and {Carigi}, L. and {Sander}, A.~A.~C. and {Palla}, M. and {Chru{\'s}li{\'n}ska}, M. and {De Looze}, I. and {Rela{\~n}o}, M. and {van der Giessen}, S.~A. and {Reyes-Rodr{\'\i}guez}, E. and {S{\'a}nchez}, S.~F.},
	doi = {10.1051/0004-6361/202450928},
	eid = {A248},
	eprint = {2408.06215},
	journal = {\aap},
	month = oct,
	pages = {A248},
	primaryclass = {astro-ph.GA},
	title = {{Gas-phase Fe/O and Fe/N abundances in star-forming regions: Relations between nucleosynthesis, metallicity, and dust}},
	volume = {690},
	year = 2024}

@article{Rodriguez:2005aa,
	adsurl = {https://ui.adsabs.harvard.edu/abs/2005ApJ...626..900R},
	archiveprefix = {arXiv},
	author = {{Rodr{\'\i}guez}, M{\'o}nica and {Rubin}, Robert H.},
	doi = {10.1086/429958},
	eprint = {astro-ph/0504131},
	journal = {\apj},
	month = jun,
	number = {2},
	pages = {900-908},
	primaryclass = {astro-ph},
	title = {{The [Fe IV] Discrepancy: Constraining the Iron Abundances in Nebulae}},
	volume = {626},
	year = 2005}

@article{Nakane:2025aa,
	adsurl = {https://ui.adsabs.harvard.edu/abs/2025ApJ...994...65N},
	archiveprefix = {arXiv},
	author = {{Nakane}, Minami and {Ouchi}, Masami and {Nakajima}, Kimihiko and {Ono}, Yoshiaki and {Harikane}, Yuichi and {Isobe}, Yuki and {Nomoto}, Ken'ichi and {Ishigaki}, Miho N. and {Yanagisawa}, Hiroto and {Kashino}, Daichi and {Tominaga}, Nozomu and {Takahashi}, Koh and {Nishigaki}, Moka and {Takeda}, Yui and {Watanabe}, Kuria},
	doi = {10.3847/1538-4357/ae04e6},
	eid = {65},
	eprint = {2503.11457},
	journal = {\apj},
	month = nov,
	number = {1},
	pages = {65},
	primaryclass = {astro-ph.GA},
	title = {{Fe Abundances of Early Galaxies at z = 9─12 Derived with Deep JWST Spectra}},
	volume = {994},
	year = 2025}

@article{Nakane:2024aa,
	adsurl = {https://ui.adsabs.harvard.edu/abs/2024ApJ...976..122N},
	archiveprefix = {arXiv},
	author = {{Nakane}, Minami and {Ouchi}, Masami and {Nakajima}, Kimihiko and {Harikane}, Yuichi and {Tominaga}, Nozomu and {Takahashi}, Koh and {Kashino}, Daichi and {Yanagisawa}, Hiroto and {Watanabe}, Kuria and {Nomoto}, Ken'ichi and {Isobe}, Yuki and {Nishigaki}, Moka and {Ishigaki}, Miho N. and {Ono}, Yoshiaki and {Takeda}, Yui},
	doi = {10.3847/1538-4357/ad84e8},
	eid = {122},
	eprint = {2407.14470},
	journal = {\apj},
	month = nov,
	number = {1},
	pages = {122},
	primaryclass = {astro-ph.GA},
	title = {{Low [O/Fe] Ratio in a Luminous Galaxy at the Early Cosmic Epoch (z > 10): Signature of Short Delay Time or Bright Hypernovae/Pair-instability Supernovae?}},
	volume = {976},
	year = 2024}

@article{Vale-Asari:2019aa,
	adsurl = {https://ui.adsabs.harvard.edu/abs/2019MNRAS.489.4721V},
	archiveprefix = {arXiv},
	author = {{Vale Asari}, N. and {Couto}, G.~S. and {Cid Fernandes}, R. and {Stasi{\'n}ska}, G. and {de Amorim}, A.~L. and {Ruschel-Dutra}, D. and {Werle}, A. and {Florido}, T.~Z.},
	doi = {10.1093/mnras/stz2470},
	eprint = {1907.08635},
	journal = {\mnras},
	month = nov,
	number = {4},
	pages = {4721-4733},
	primaryclass = {astro-ph.GA},
	title = {{Diffuse ionized gas and its effects on nebular metallicity estimates of star-forming galaxies}},
	volume = {489},
	year = 2019}

@article{Kewley:2008aa,
	adsurl = {https://ui.adsabs.harvard.edu/abs/2008ApJ...681.1183K},
	archiveprefix = {arXiv},
	author = {{Kewley}, Lisa J. and {Ellison}, Sara L.},
	doi = {10.1086/587500},
	eprint = {0801.1849},
	journal = {\apj},
	month = jul,
	number = {2},
	pages = {1183-1204},
	primaryclass = {astro-ph},
	title = {{Metallicity Calibrations and the Mass-Metallicity Relation for Star-forming Galaxies}},
	volume = {681},
	year = 2008}

@article{Flury:2025aa,
	adsurl = {https://ui.adsabs.harvard.edu/abs/2025MNRAS.543.3367F},
	archiveprefix = {arXiv},
	author = {{Flury}, Sophia R. and {Arellano-C{\'o}rdova}, Karla Z. and {Moran}, Edward C. and {Einsig}, Alaina},
	doi = {10.1093/mnras/staf1615},
	eprint = {2412.06763},
	journal = {\mnras},
	month = nov,
	number = {4},
	pages = {3367-3381},
	primaryclass = {astro-ph.GA},
	title = {{New ionization models and the shocking nitrogen excess at z > 5}},
	volume = {543},
	year = 2025}

@article{Alarie:2019aa,
	adsurl = {https://ui.adsabs.harvard.edu/abs/2019RMxAA..55..377A},
	archiveprefix = {arXiv},
	author = {{Alarie}, A. and {Morisset}, C.},
	doi = {10.22201/ia.01851101p.2019.55.02.21},
	eprint = {1908.08579},
	journal = {\rmxaa},
	month = oct,
	pages = {377-394},
	primaryclass = {astro-ph.GA},
	title = {{Extensive Online Shock Model Database}},
	volume = {55},
	year = 2019}

@article{Draine:1993aa,
	adsurl = {https://ui.adsabs.harvard.edu/abs/1993ARA&A..31..373D},
	author = {{Draine}, Bruce T. and {McKee}, Christopher F.},
	doi = {10.1146/annurev.aa.31.090193.002105},
	journal = {\araa},
	month = jan,
	pages = {373-432},
	title = {{Theory of interstellar shocks.}},
	volume = {31},
	year = 1993}

@article{Senchyna:2024aa,
	adsurl = {https://ui.adsabs.harvard.edu/abs/2024ApJ...966...92S},
	archiveprefix = {arXiv},
	author = {{Senchyna}, Peter and {Plat}, Adele and {Stark}, Daniel P. and {Rudie}, Gwen C. and {Berg}, Danielle and {Charlot}, St{\'e}phane and {James}, Bethan L. and {Mingozzi}, Matilde},
	doi = {10.3847/1538-4357/ad235e},
	eid = {92},
	eprint = {2303.04179},
	journal = {\apj},
	month = may,
	number = {1},
	pages = {92},
	primaryclass = {astro-ph.GA},
	title = {{GN-z11 in Context: Possible Signatures of Globular Cluster Precursors at Redshift 10}},
	volume = {966},
	year = 2024}

@article{Monty:2025aa,
	adsurl = {https://ui.adsabs.harvard.edu/abs/2025MNRAS.542.1443M},
	archiveprefix = {arXiv},
	author = {{Monty}, Stephanie and {Strom}, Allison L. and {Stanton}, Thomas M. and {Chru{\'s}li{\'n}ska}, Martyna and {Cullen}, Fergus and {Kobayashi}, Chiaki and {Starkenburg}, Tjitske and {Bhattacharya}, Souradeep and {Sanders}, Jason L. and {Gieles}, Mark},
	doi = {10.1093/mnras/staf1213},
	eprint = {2507.14094},
	journal = {\mnras},
	month = sep,
	number = {2},
	pages = {1443-1464},
	primaryclass = {astro-ph.GA},
	title = {{ChemZz I: comparing oxygen and iron abundance patterns in the Milky Way, the Local Group, and Cosmic Noon}},
	volume = {542},
	year = 2025}

@article{Mendez-Delgado:2023aa,
	adsurl = {https://ui.adsabs.harvard.edu/abs/2023Natur.618..249M},
	archiveprefix = {arXiv},
	author = {{M{\'e}ndez-Delgado}, J. Eduardo and {Esteban}, C{\'e}sar and {Garc{\'\i}a-Rojas}, Jorge and {Kreckel}, Kathryn and {Peimbert}, Manuel},
	doi = {10.1038/s41586-023-05956-2},
	eprint = {2305.11578},
	journal = {\nat},
	month = jun,
	number = {7964},
	pages = {249-251},
	primaryclass = {astro-ph.GA},
	title = {{Temperature inhomogeneities cause the abundance discrepancy in H II regions}},
	volume = {618},
	year = 2023}

@article{Peimbert:1967aa,
	adsurl = {https://ui.adsabs.harvard.edu/abs/1967ApJ...150..825P},
	author = {{Peimbert}, Manuel},
	doi = {10.1086/149385},
	journal = {\apj},
	month = dec,
	pages = {825},
	title = {{Temperature Determinations of H II Regions}},
	volume = {150},
	year = 1967}

@article{Nandal:2025aa,
	adsurl = {https://ui.adsabs.harvard.edu/abs/2025ApJ...994L..11N},
	archiveprefix = {arXiv},
	author = {{Nandal}, Devesh and {Whalen}, Daniel J. and {Latif}, Muhammad A. and {Heger}, Alexander},
	doi = {10.3847/2041-8213/ae1a63},
	eid = {L11},
	eprint = {2502.04435},
	journal = {\apjl},
	month = nov,
	number = {1},
	pages = {L11},
	primaryclass = {astro-ph.GA},
	title = {{1000─10,000 M$_{{\ensuremath{\odot}}}$ Primordial Stars Created the Nitrogen Excess in GS 3073 at z = 5.55}},
	volume = {994},
	year = 2025}

@article{McClymont:2025aa,
	adsurl = {https://ui.adsabs.harvard.edu/abs/2025MNRAS.544..513M},
	archiveprefix = {arXiv},
	author = {{McClymont}, William and {Tacchella}, Sandro and {Smith}, Aaron and {Kannan}, Rahul and {Puchwein}, Ewald and {Borrow}, Josh and {Garaldi}, Enrico and {Keating}, Laura and {Vogelsberger}, Mark and {Zier}, Oliver and {Shen}, Xuejian and {Popovic}, Filip and {Simmonds}, Charlotte},
	doi = {10.1093/mnras/staf1660},
	eprint = {2503.00106},
	journal = {\mnras},
	month = nov,
	number = {1},
	pages = {513-534},
	primaryclass = {astro-ph.GA},
	title = {{The THESAN-ZOOM project: burst, quench, repeat {\textendash} unveiling the evolution of high-redshift galaxies along the star-forming main sequence}},
	volume = {544},
	year = 2025}

@article{Zier:2025aa,
	adsurl = {https://ui.adsabs.harvard.edu/abs/2025MNRAS.544..391Z},
	archiveprefix = {arXiv},
	author = {{Zier}, Oliver and {Kannan}, Rahul and {Smith}, Aaron and {Puchwein}, Ewald and {Vogelsberger}, Mark and {Borrow}, Josh and {Garaldi}, Enrico and {Keating}, Laura and {McClymont}, William and {Shen}, Xuejian and {Hernquist}, Lars},
	doi = {10.1093/mnras/staf1052},
	eprint = {2503.02927},
	journal = {\mnras},
	month = nov,
	number = {1},
	pages = {391-409},
	primaryclass = {astro-ph.GA},
	title = {{The THESAN{\textendash}ZOOM project: long-term imprints of external reionization on galaxy evolution}},
	volume = {544},
	year = 2025}

@article{Zier:2025ab,
	adsurl = {https://ui.adsabs.harvard.edu/abs/2025MNRAS.544..410Z},
	archiveprefix = {arXiv},
	author = {{Zier}, Oliver and {Kannan}, Rahul and {Smith}, Aaron and {Puchwein}, Ewald and {Vogelsberger}, Mark and {Borrow}, Josh and {Garaldi}, Enrico and {Keating}, Laura and {McClymont}, William and {Shen}, Xuejian and {Hernquist}, Lars},
	doi = {10.1093/mnras/staf1053},
	eprint = {2503.03806},
	journal = {\mnras},
	month = nov,
	number = {1},
	pages = {410-429},
	primaryclass = {astro-ph.GA},
	title = {{The THESAN-ZOOM project: Population III star formation continues until the end of reionization}},
	volume = {544},
	year = 2025}

@article{Hsiao:2025aa,
	adsurl = {https://ui.adsabs.harvard.edu/abs/2025ApJ...993...70H},
	archiveprefix = {arXiv},
	author = {{Hsiao}, Tiger Yu-Yang and {Topping}, Michael W. and {Coe}, Dan and {Chisholm}, John and {Berg}, Danielle A. and {Abdurro'uf} and {{\'A}lvarez-M{\'a}rquez}, Javier and {Maiolino}, Roberto and {Dayal}, Pratika and {Furtak}, Lukas J.},
	doi = {10.3847/1538-4357/ae07d7},
	eid = {70},
	eprint = {2409.04625},
	journal = {\apj},
	month = nov,
	number = {1},
	pages = {70},
	primaryclass = {astro-ph.GA},
	title = {{First Direct Carbon Abundance Measured at z > 10 in the Lensed Galaxy MACS0647-JD}},
	volume = {993},
	year = 2025}

@article{Kannan:2025aa,
	adsurl = {https://ui.adsabs.harvard.edu/abs/2025OJAp....8E.153K},
	archiveprefix = {arXiv},
	author = {{Kannan}, Rahul and {Puchwein}, Ewald and {Smith}, Aaron and {Borrow}, Josh and {Garaldi}, Enrico and {Keating}, Laura and {Vogelsberger}, Mark and {Zier}, Oliver and {McClymont}, William and {Shen}, Xuejian and {Popovic}, Filip and {Tacchella}, Sandro and {Hernquist}, Lars and {Springel}, Volker},
	doi = {10.33232/001c.145804},
	eid = {153},
	eprint = {2502.20437},
	journal = {The Open Journal of Astrophysics},
	month = oct,
	pages = {153},
	primaryclass = {astro-ph.GA},
	title = {{Introducing the THESAN-ZOOM project: radiation-hydrodynamic simulations of high-redshift galaxies with a multi-phase interstellar medium}},
	volume = {8},
	year = 2025}

@article{McClymont:2025af,
	adsurl = {https://ui.adsabs.harvard.edu/abs/2025arXiv251013952M},
	archiveprefix = {arXiv},
	author = {{McClymont}, William and {Smith}, Aaron and {Tacchella}, Sandro},
	eid = {arXiv:2510.13952},
	eprint = {2510.13952},
	journal = {arXiv e-prints},
	month = oct,
	pages = {arXiv:2510.13952},
	primaryclass = {astro-ph.GA},
	title = {{Modelling the nebular emission of galaxies across cosmic time with COLT}},
	year = 2025}

@article{Tacchella:2025aa,
	adsurl = {https://ui.adsabs.harvard.edu/abs/2025MNRAS.540..851T},
	archiveprefix = {arXiv},
	author = {{Tacchella}, Sandro and {McClymont}, William and {Scholtz}, Jan and {Maiolino}, Roberto and {Ji}, Xihan and {Villanueva}, Natalia C. and {Charlot}, St{\'e}phane and {D'Eugenio}, Francesco and {Helton}, Jakob M. and {Williams}, Christina C. and {Witstok}, Joris and {Bhatawdekar}, Rachana and {Carniani}, Stefano and {Chevallard}, Jacopo and {Curti}, Mirko and {Hainline}, Kevin and {Ji}, Zhiyuan and {Johnson}, Benjamin D. and {Leja}, Joel and {Li}, Yijia and {Maseda}, Michael V. and {Pusk{\'a}s}, D{\'a}vid and {Rieke}, Marcia and {Robertson}, Brant and {Shivaei}, Irene and {Silcock}, Maddie S. and {Simmonds}, Charlotte and {{\"U}bler}, Hannah and {Willmer}, Christopher N.~A. and {Willott}, Chris},
	doi = {10.1093/mnras/staf718},
	eprint = {2404.02194},
	journal = {\mnras},
	month = jun,
	number = {1},
	pages = {851-870},
	primaryclass = {astro-ph.GA},
	title = {{Resolving the nature and putative nebular emission of GS9422: an obscured AGN without exotic stars}},
	volume = {540},
	year = 2025}

@article{Duan:2025aa,
	adsurl = {https://ui.adsabs.harvard.edu/abs/2025MNRAS.540..774D},
	archiveprefix = {arXiv},
	author = {{Duan}, Qiao and {Conselice}, Christopher J. and {Li}, Qiong and {Austin}, Duncan and {Harvey}, Thomas and {Adams}, Nathan J. and {Duncan}, Kenneth J. and {Trussler}, James and {Ferreira}, Leonardo and {Westcott}, Lewi and {Harris}, Honor and {Windhorst}, Rogier A. and {Holwerda}, Benne W. and {Broadhurst}, Thomas J. and {Coe}, Dan and {Cohen}, Seth H. and {Du}, Xiaojing and {Driver}, Simon P. and {Frye}, Brenda and {Grogin}, Norman A. and {Hathi}, Nimish P. and {Jansen}, Rolf A. and {Koekemoer}, Anton M. and {Marshall}, Madeline A. and {Nonino}, Mario and {Ortiz}, III, Rafael and {Pirzkal}, Nor and {Robotham}, Aaron and {Ryan}, Russell E. and {Summers}, Jake and {D'Silva}, Jordan C.~J. and {Willmer}, Christopher N.~A. and {Yan}, Haojing},
	doi = {10.1093/mnras/staf638},
	eprint = {2407.09472},
	journal = {\mnras},
	month = jun,
	number = {1},
	pages = {774-805},
	primaryclass = {astro-ph.GA},
	title = {{Galaxy mergers in the epoch of reionization {\textendash} I. A JWST study of pair fractions, merger rates, and stellar mass accretion rates at z = 4.5{\textendash}11.5}},
	volume = {540},
	year = 2025}

@article{Jamieson:2025aa,
	adsurl = {https://ui.adsabs.harvard.edu/abs/2025MNRAS.541.1088J},
	archiveprefix = {arXiv},
	author = {{Jamieson}, Nathan and {Smith}, Aaron and {Neyer}, Meredith and {Kannan}, Rahul and {Garaldi}, Enrico and {Vogelsberger}, Mark and {Hernquist}, Lars and {Zier}, Oliver and {Shen}, Xuejian and {Kakiichi}, Koki},
	doi = {10.1093/mnras/staf996},
	eprint = {2411.08943},
	journal = {\mnras},
	month = aug,
	number = {2},
	pages = {1088-1105},
	primaryclass = {astro-ph.GA},
	title = {{The THESAN project: tracking the expansion and merger histories of ionized bubbles during the Epoch of Reionization}},
	volume = {541},
	year = 2025}

@article{Garcia:2025aa,
	adsurl = {https://ui.adsabs.harvard.edu/abs/2025ApJ...989..147G},
	archiveprefix = {arXiv},
	author = {{Garcia}, Alex M. and {Torrey}, Paul and {Bhagwat}, Aniket and {Wright}, Ruby J. and {Chen}, Qian-Hui and {Grasha}, Kathryn and {Ridolfo}, Sophia and {Hemler}, Z.~S. and {Sarkar}, Arnab and {Chakraborty}, Priyanka and {Nelson}, Erica J. and {Sanders}, Ryan L. and {Costa}, Tiago and {Vogelsberger}, Mark and {Kewley}, Lisa J. and {Ellison}, Sara L. and {Hernquist}, Lars},
	doi = {10.3847/1538-4357/adea51},
	eid = {147},
	eprint = {2503.03804},
	journal = {\apj},
	month = aug,
	number = {2},
	pages = {147},
	primaryclass = {astro-ph.GA},
	title = {{Metallicity Gradients in Modern Cosmological Simulations. I. Tension between Smooth Stellar Feedback Models and Observations}},
	volume = {989},
	year = 2025}

@article{Marszewski:2025aa,
	adsurl = {https://ui.adsabs.harvard.edu/abs/2025ApJ...991L...4M},
	archiveprefix = {arXiv},
	author = {{Marszewski}, Andrew and {Faucher-Gigu{\`e}re}, Claude-Andr{\'e} and {Feldmann}, Robert and {Sun}, Guochao},
	doi = {10.3847/2041-8213/adf74b},
	eid = {L4},
	eprint = {2505.22712},
	journal = {\apjl},
	month = sep,
	number = {1},
	pages = {L4},
	primaryclass = {astro-ph.GA},
	title = {{Explaining the Weak Evolution of the High-redshift Mass{\textendash}Metallicity Relation with Galaxy Burst Cycles}},
	volume = {991},
	year = 2025}

@article{Gieles:2025aa,
	adsurl = {https://ui.adsabs.harvard.edu/abs/2025MNRAS.tmp.1257G},
	archiveprefix = {arXiv},
	author = {{Gieles}, Mark and {Padoan}, Paolo and {Charbonnel}, Corinne and {Vink}, Jorick S. and {Ram{\'\i}rez-Galeano}, Laura},
	doi = {10.1093/mnras/staf1314},
	eprint = {2501.12138},
	journal = {\mnras},
	month = aug,
	primaryclass = {astro-ph.GA},
	title = {{Globular cluster formation from inertial inflows: accreting extremely massive stars as the origin of abundance anomalies}},
	year = 2025}

@article{Fujimoto:2025aa,
	adsurl = {https://ui.adsabs.harvard.edu/abs/2025NatAs.tmp..157F},
	archiveprefix = {arXiv},
	author = {{Fujimoto}, S. and {Ouchi}, M. and {Kohno}, K. and {Valentino}, F. and {Gim{\'e}nez-Arteaga}, C. and {Brammer}, G.~B. and {Furtak}, L.~J. and {Kohandel}, M. and {Oguri}, M. and {Pallottini}, A. and {Richard}, J. and {Zitrin}, A. and {Bauer}, F.~E. and {Boylan-Kolchin}, M. and {Dessauges-Zavadsky}, M. and {Egami}, E. and {Finkelstein}, S.~L. and {Ma}, Z. and {Smail}, I. and {Watson}, D. and {Hutchison}, T.~A. and {Rigby}, J.~R. and {Welch}, B.~D. and {Ao}, Y. and {Bradley}, L.~D. and {Caminha}, G.~B. and {Caputi}, K.~I. and {Espada}, D. and {Endsley}, R. and {Fudamoto}, Y. and {Gonz{\'a}lez-L{\'o}pez}, J. and {Hatsukade}, B. and {Koekemoer}, A.~M. and {Kokorev}, V. and {Laporte}, N. and {Lee}, M. and {Magdis}, G.~E. and {Ono}, Y. and {Rizzo}, F. and {Shibuya}, T. and {Shimasaku}, K. and {Sun}, F. and {Toft}, S. and {Umehata}, H. and {Wang}, T. and {Yajima}, H.},
	doi = {10.1038/s41550-025-02592-w},
	eprint = {2402.18543},
	journal = {Nature Astronomy},
	month = aug,
	primaryclass = {astro-ph.GA},
	title = {{Primordial rotating disk composed of at least 15 dense star-forming clumps at cosmic dawn}},
	year = 2025}

@article{Belokurov:2018aa,
	adsurl = {https://ui.adsabs.harvard.edu/abs/2018MNRAS.478..611B},
	archiveprefix = {arXiv},
	author = {{Belokurov}, V. and {Erkal}, D. and {Evans}, N.~W. and {Koposov}, S.~E. and {Deason}, A.~J.},
	doi = {10.1093/mnras/sty982},
	eprint = {1802.03414},
	journal = {\mnras},
	month = jul,
	number = {1},
	pages = {611-619},
	primaryclass = {astro-ph.GA},
	title = {{Co-formation of the disc and the stellar halo}},
	volume = {478},
	year = 2018}

@article{Helmi:2018aa,
	adsurl = {https://ui.adsabs.harvard.edu/abs/2018Natur.563...85H},
	archiveprefix = {arXiv},
	author = {{Helmi}, Amina and {Babusiaux}, Carine and {Koppelman}, Helmer H. and {Massari}, Davide and {Veljanoski}, Jovan and {Brown}, Anthony G.~A.},
	doi = {10.1038/s41586-018-0625-x},
	eprint = {1806.06038},
	journal = {\nat},
	month = oct,
	number = {7729},
	pages = {85-88},
	primaryclass = {astro-ph.GA},
	title = {{The merger that led to the formation of the Milky Way's inner stellar halo and thick disk}},
	volume = {563},
	year = 2018}

@article{Rowland:2025aa,
	adsurl = {https://ui.adsabs.harvard.edu/abs/2025arXiv250110559R},
	archiveprefix = {arXiv},
	author = {{Rowland}, Lucie E. and {Stefanon}, Mauro and {Bouwens}, Rychard and {Hodge}, Jacqueline and {Algera}, Hiddo and {Fisher}, Rebecca and {Dayal}, Pratika and {Pallottini}, Andrea and {Stark}, Daniel P. and {Heintz}, Kasper E. and {Aravena}, Manuel and {Bowler}, Rebecca and {Cescon}, Karin and {Endsley}, Ryan and {Ferrara}, Andrea and {Gonzalez}, Valentino and {Graziani}, Luca and {Gulis}, Cindy and {Herard-Demanche}, Thomas and {Inami}, Hanae and {Laza-Ramos}, Andr{\`e}s and {van Leeuwen}, Ivana and {de Looze}, Ilse and {Nanayakkara}, Themiya and {Oesch}, Pascal and {Ormerod}, Katherine and {Sartorio}, Nina S. and {Schouws}, Sander and {Smit}, Renske and {Sommovigo}, Laura and {Toft}, Sune and {Weaver}, John R. and {van der Werf}, Paul},
	doi = {10.48550/arXiv.2501.10559},
	eid = {arXiv:2501.10559},
	eprint = {2501.10559},
	journal = {arXiv e-prints},
	month = jan,
	pages = {arXiv:2501.10559},
	primaryclass = {astro-ph.GA},
	title = {{REBELS-IFU: Evidence for metal-rich massive galaxies at z\raisebox{-0.5ex}\textasciitilde6-8}},
	year = 2025}

@article{Scholtz:2025ab,
	adsurl = {https://ui.adsabs.harvard.edu/abs/2025MNRAS.539.2463S},
	archiveprefix = {arXiv},
	author = {{Scholtz}, J. and {Curti}, M. and {D'Eugenio}, F. and {{\"U}bler}, H. and {Maiolino}, R. and {Marconcini}, C. and {Smit}, R. and {Perna}, M. and {Witstok}, J. and {Arribas}, S. and {B{\"o}ker}, T. and {Bunker}, A.~J. and {Carniani}, S. and {Charlot}, S. and {Cresci}, G. and {Lamperti}, I. and {Parlanti}, E. and {P{\'e}rez-Gonz{\'a}lez}, P.~G. and {Rodr{\'\i}guez Del Pino}, B. and {Venturi}, G.},
	doi = {10.1093/mnras/staf518},
	eprint = {2411.07695},
	journal = {\mnras},
	month = may,
	number = {3},
	pages = {2463-2484},
	primaryclass = {astro-ph.GA},
	title = {{GA-NIFS: ISM properties and metal enrichment in a merger-driven starburst during the epoch of reionization probed with JWST and ALMA}},
	volume = {539},
	year = 2025}

@article{Berg:2020aa,
	adsurl = {https://ui.adsabs.harvard.edu/abs/2020ApJ...893...96B},
	archiveprefix = {arXiv},
	author = {{Berg}, Danielle A. and {Pogge}, Richard W. and {Skillman}, Evan D. and {Croxall}, Kevin V. and {Moustakas}, John and {Rogers}, Noah S.~J. and {Sun}, Jiayi},
	doi = {10.3847/1538-4357/ab7eab},
	eid = {96},
	eprint = {2001.05002},
	journal = {\apj},
	month = apr,
	number = {2},
	pages = {96},
	primaryclass = {astro-ph.GA},
	title = {{CHAOS IV: Gas-phase Abundance Trends from the First Four CHAOS Galaxies}},
	volume = {893},
	year = 2020}

@article{Scholte:2025aa,
	adsurl = {https://ui.adsabs.harvard.edu/abs/2025MNRAS.540.1800S},
	archiveprefix = {arXiv},
	author = {{Scholte}, D. and {Cullen}, F. and {Carnall}, A.~C. and {Arellano-C{\'o}rdova}, K.~Z. and {Stanton}, T.~M. and {Barrufet}, L. and {Begley}, R. and {Bondestam}, C. and {Donnan}, C.~T. and {Dunlop}, J.~S. and {Leung}, H. -H. and {McLeod}, D.~J. and {McLure}, R.~J. and {Moustakas}, J.~M. and {Pollock}, C.~L. and {Shapley}, A.~E. and {Stevenson}, S. and {Zou}, H.},
	doi = {10.1093/mnras/staf834},
	eprint = {2502.10499},
	journal = {\mnras},
	month = jun,
	number = {2},
	pages = {1800-1826},
	primaryclass = {astro-ph.GA},
	title = {{The JWST EXCELS survey: probing strong-line diagnostics and the chemical evolution of galaxies over cosmic time using T$_{e}$-metallicities}},
	volume = {540},
	year = 2025}

@article{Tacchella:2023aa,
	adsurl = {https://ui.adsabs.harvard.edu/abs/2023MNRAS.522.6236T},
	archiveprefix = {arXiv},
	author = {{Tacchella}, Sandro and {Johnson}, Benjamin D. and {Robertson}, Brant E. and {Carniani}, Stefano and {D'Eugenio}, Francesco and {Kumari}, Nimisha and {Maiolino}, Roberto and {Nelson}, Erica J. and {Suess}, Katherine A. and {{\"U}bler}, Hannah and {Williams}, Christina C. and {Adebusola}, Alabi and {Alberts}, Stacey and {Arribas}, Santiago and {Bhatawdekar}, Rachana and {Bonaventura}, Nina and {Bowler}, Rebecca A.~A. and {Bunker}, Andrew J. and {Cameron}, Alex J. and {Curti}, Mirko and {Egami}, Eiichi and {Eisenstein}, Daniel J. and {Frye}, Brenda and {Hainline}, Kevin and {Helton}, Jakob M. and {Ji}, Zhiyuan and {Looser}, Tobias J. and {Lyu}, Jianwei and {Perna}, Michele and {Rawle}, Timothy and {Rieke}, George and {Rieke}, Marcia and {Saxena}, Aayush and {Sandles}, Lester and {Shivaei}, Irene and {Simmonds}, Charlotte and {Sun}, Fengwu and {Willmer}, Christopher N.~A. and {Willott}, Chris J. and {Witstok}, Joris},
	doi = {10.1093/mnras/stad1408},
	eprint = {2208.03281},
	journal = {\mnras},
	month = jul,
	number = {4},
	pages = {6236-6249},
	primaryclass = {astro-ph.GA},
	title = {{JWST NIRCam + NIRSpec: interstellar medium and stellar populations of young galaxies with rising star formation and evolving gas reservoirs}},
	volume = {522},
	year = 2023}

@article{Arellano-Cordova:2022aa,
	adsurl = {https://ui.adsabs.harvard.edu/abs/2022ApJ...940L..23A},
	archiveprefix = {arXiv},
	author = {{Arellano-C{\'o}rdova}, Karla Z. and {Berg}, Danielle A. and {Chisholm}, John and {Arrabal Haro}, Pablo and {Dickinson}, Mark and {Finkelstein}, Steven L. and {Leclercq}, Floriane and {Rogers}, Noah S.~J. and {Simons}, Raymond C. and {Skillman}, Evan D. and {Trump}, Jonathan R. and {Kartaltepe}, Jeyhan S.},
	doi = {10.3847/2041-8213/ac9ab2},
	eid = {L23},
	eprint = {2208.02562},
	journal = {\apjl},
	month = nov,
	number = {1},
	pages = {L23},
	primaryclass = {astro-ph.GA},
	title = {{A First Look at the Abundance Pattern-O/H, C/O, and Ne/O-in z > 7 Galaxies with JWST/NIRSpec}},
	volume = {940},
	year = 2022}

@article{Stiavelli:2025aa,
	adsurl = {https://ui.adsabs.harvard.edu/abs/2025ApJ...981..136S},
	archiveprefix = {arXiv},
	author = {{Stiavelli}, Massimo and {Morishita}, Takahiro and {Chiaberge}, Marco and {Leethochawalit}, Nicha and {Norman}, Colin and {Ricotti}, Massimo and {Roberts-Borsani}, Guido and {Treu}, Tommaso and {Vanzella}, Eros and {Wyse}, Rosemary F.~G. and {Zhang}, Yechi and {Boyett}, Kit},
	doi = {10.3847/1538-4357/adb5f3},
	eid = {136},
	eprint = {2412.06517},
	journal = {\apj},
	month = mar,
	number = {2},
	pages = {136},
	primaryclass = {astro-ph.GA},
	title = {{What Can We Learn from the Nitrogen Abundance of High-z Galaxies?}},
	volume = {981},
	year = 2025}

@article{Belokurov:2023aa,
	adsurl = {https://ui.adsabs.harvard.edu/abs/2023MNRAS.525.4456B},
	archiveprefix = {arXiv},
	author = {{Belokurov}, Vasily and {Kravtsov}, Andrey},
	doi = {10.1093/mnras/stad2241},
	eprint = {2306.00060},
	journal = {\mnras},
	month = nov,
	number = {3},
	pages = {4456-4473},
	primaryclass = {astro-ph.GA},
	title = {{Nitrogen enrichment and clustered star formation at the dawn of the Galaxy}},
	volume = {525},
	year = 2023}

@article{Belokurov:2022aa,
	adsurl = {https://ui.adsabs.harvard.edu/abs/2022MNRAS.514..689B},
	archiveprefix = {arXiv},
	author = {{Belokurov}, Vasily and {Kravtsov}, Andrey},
	doi = {10.1093/mnras/stac1267},
	eprint = {2203.04980},
	journal = {\mnras},
	month = jul,
	number = {1},
	pages = {689-714},
	primaryclass = {astro-ph.GA},
	title = {{From dawn till disc: Milky Way's turbulent youth revealed by the APOGEE+Gaia data}},
	volume = {514},
	year = 2022}

@article{Adamo:2024aa,
	adsurl = {https://ui.adsabs.harvard.edu/abs/2024Natur.632..513A},
	archiveprefix = {arXiv},
	author = {{Adamo}, Angela and {Bradley}, Larry D. and {Vanzella}, Eros and {Claeyssens}, Ad{\'e}la{\"\i}de and {Welch}, Brian and {Diego}, Jose M. and {Mahler}, Guillaume and {Oguri}, Masamune and {Sharon}, Keren and {Abdurro'uf} and {Hsiao}, Tiger Yu-Yang and {Xu}, Xinfeng and {Messa}, Matteo and {Lassen}, Augusto E. and {Zackrisson}, Erik and {Brammer}, Gabriel and {Coe}, Dan and {Kokorev}, Vasily and {Ricotti}, Massimo and {Zitrin}, Adi and {Fujimoto}, Seiji and {Inoue}, Akio K. and {Resseguier}, Tom and {Rigby}, Jane R. and {Jim{\'e}nez-Teja}, Yolanda and {Windhorst}, Rogier A. and {Hashimoto}, Takuya and {Tamura}, Yoichi},
	doi = {10.1038/s41586-024-07703-7},
	eprint = {2401.03224},
	journal = {\nat},
	month = aug,
	number = {8025},
	pages = {513-516},
	primaryclass = {astro-ph.GA},
	title = {{Bound star clusters observed in a lensed galaxy 460 Myr after the Big Bang}},
	volume = {632},
	year = 2024}

@article{Vanzella:2022aa,
	adsurl = {https://ui.adsabs.harvard.edu/abs/2022A&A...659A...2V},
	archiveprefix = {arXiv},
	author = {{Vanzella}, E. and {Castellano}, M. and {Bergamini}, P. and {Meneghetti}, M. and {Zanella}, A. and {Calura}, F. and {Caminha}, G.~B. and {Rosati}, P. and {Cupani}, G. and {Me{\v{s}}tri{\'c}}, U. and {Brammer}, G. and {Tozzi}, P. and {Mercurio}, A. and {Grillo}, C. and {Sani}, E. and {Cristiani}, S. and {Nonino}, M. and {Merlin}, E. and {Pignataro}, G.~V.},
	doi = {10.1051/0004-6361/202141590},
	eid = {A2},
	eprint = {2106.10280},
	journal = {\aap},
	month = mar,
	pages = {A2},
	primaryclass = {astro-ph.GA},
	title = {{High star cluster formation efficiency in the strongly lensed Sunburst Lyman-continuum galaxy at z = 2.37}},
	volume = {659},
	year = 2022}

@article{Mowla:2024aa,
	adsurl = {https://ui.adsabs.harvard.edu/abs/2024Natur.636..332M},
	archiveprefix = {arXiv},
	author = {{Mowla}, Lamiya and {Iyer}, Kartheik and {Asada}, Yoshihisa and {Desprez}, Guillaume and {Tan}, Vivian Yun Yan and {Martis}, Nicholas and {Sarrouh}, Ghassan and {Strait}, Victoria and {Abraham}, Roberto and {Brada{\v{c}}}, Maru{\v{s}}a and {Brammer}, Gabriel and {Muzzin}, Adam and {Pacifici}, Camilla and {Ravindranath}, Swara and {Sawicki}, Marcin and {Willott}, Chris and {Estrada-Carpenter}, Vince and {Jahan}, Nusrath and {Noirot}, Ga{\"e}l and {Matharu}, Jasleen and {Rihtar{\v{s}}i{\v{c}}}, Gregor and {Zabl}, Johannes},
	doi = {10.1038/s41586-024-08293-0},
	eprint = {2402.08696},
	journal = {\nat},
	month = dec,
	number = {8042},
	pages = {332-336},
	primaryclass = {astro-ph.GA},
	title = {{Formation of a low-mass galaxy from star clusters in a 600-million-year-old Universe}},
	volume = {636},
	year = 2024}

@article{Vanzella:2023aa,
	adsurl = {https://ui.adsabs.harvard.edu/abs/2023ApJ...945...53V},
	archiveprefix = {arXiv},
	author = {{Vanzella}, Eros and {Claeyssens}, Ad{\'e}la{\"\i}de and {Welch}, Brian and {Adamo}, Angela and {Coe}, Dan and {Diego}, Jose M. and {Mahler}, Guillaume and {Khullar}, Gourav and {Kokorev}, Vasily and {Oguri}, Masamune and {Ravindranath}, Swara and {Furtak}, Lukas J. and {Hsiao}, Tiger Yu-Yang and {Abdurro'uf} and {Mandelker}, Nir and {Brammer}, Gabriel and {Bradley}, Larry D. and {Brada{\v{c}}}, Maru{\v{s}}a and {Conselice}, Christopher J. and {Dayal}, Pratika and {Nonino}, Mario and {Andrade-Santos}, Felipe and {Windhorst}, Rogier A. and {Pirzkal}, Nor and {Sharon}, Keren and {de Mink}, S.~E. and {Fujimoto}, Seiji and {Zitrin}, Adi and {Eldridge}, Jan J. and {Norman}, Colin},
	doi = {10.3847/1538-4357/acb59a},
	eid = {53},
	eprint = {2211.09839},
	journal = {\apj},
	month = mar,
	number = {1},
	pages = {53},
	primaryclass = {astro-ph.GA},
	title = {{JWST/NIRCam Probes Young Star Clusters in the Reionization Era Sunrise Arc}},
	volume = {945},
	year = 2023}

@article{DEugenio:2024ab,
	adsurl = {https://ui.adsabs.harvard.edu/abs/2024A&A...689A.152D},
	archiveprefix = {arXiv},
	author = {{D'Eugenio}, Francesco and {Maiolino}, Roberto and {Carniani}, Stefano and {Chevallard}, Jacopo and {Curtis-Lake}, Emma and {Witstok}, Joris and {Charlot}, Stephane and {Baker}, William M. and {Arribas}, Santiago and {Boyett}, Kristan and {Bunker}, Andrew J. and {Curti}, Mirko and {Eisenstein}, Daniel J. and {Hainline}, Kevin and {Ji}, Zhiyuan and {Johnson}, Benjamin D. and {Kumari}, Nimisha and {Looser}, Tobias J. and {Nakajima}, Kimihiko and {Nelson}, Erica and {Rieke}, Marcia and {Robertson}, Brant and {Scholtz}, Jan and {Smit}, Renske and {Sun}, Fengwu and {Venturi}, Giacomo and {Tacchella}, Sandro and {{\"U}bler}, Hannah and {Willmer}, Christopher N.~A. and {Willott}, Chris},
	doi = {10.1051/0004-6361/202348636},
	eid = {A152},
	eprint = {2311.09908},
	journal = {\aap},
	month = sep,
	pages = {A152},
	primaryclass = {astro-ph.GA},
	title = {{JADES: Carbon enrichment 350 Myr after the Big Bang}},
	volume = {689},
	year = 2024}

@article{McClymont:2025ad,
	adsurl = {https://ui.adsabs.harvard.edu/abs/2025MNRAS.540..190M},
	archiveprefix = {arXiv},
	author = {{McClymont}, William and {Tacchella}, Sandro and {D'Eugenio}, Francesco and {Witten}, Callum and {Ji}, Xihan and {Smith}, Aaron and {Maiolino}, Roberto and {Arribas}, Santiago and {Scholtz}, Jan and {Simmonds}, Charlotte and {Witstok}, Joris},
	doi = {10.1093/mnras/staf745},
	eprint = {2405.15859},
	journal = {\mnras},
	month = jun,
	number = {1},
	pages = {190-203},
	primaryclass = {astro-ph.GA},
	title = {{The density-bounded twilight of starbursts in the early Universe}},
	volume = {540},
	year = 2025}

@article{Puskas:2025aa,
	adsurl = {https://ui.adsabs.harvard.edu/abs/2025MNRAS.540.2146P},
	archiveprefix = {arXiv},
	author = {{Pusk{\'a}s}, D{\'a}vid and {Tacchella}, Sandro and {Simmonds}, Charlotte and {Hainline}, Kevin and {D'Eugenio}, Francesco and {Alberts}, Stacey and {Arribas}, Santiago and {Baker}, William M. and {Bunker}, Andrew J. and {Carniani}, Stefano and {Charlot}, St{\'e}phane and {Duan}, Qiao and {Eisenstein}, Daniel J. and {Ji}, Zhiyuan and {Johnson}, Benjamin D. and {Jones}, Gareth C. and {Maiolino}, Roberto and {McClymont}, William and {Rieke}, Marcia and {Rinaldi}, Pierluigi and {Robertson}, Brant and {{\"U}bler}, Hannah and {Williams}, Christina C. and {Willmer}, Christopher N.~A. and {Willott}, Chris and {Witstok}, Joris},
	doi = {10.1093/mnras/staf813},
	eprint = {2502.01721},
	journal = {\mnras},
	month = jul,
	number = {3},
	pages = {2146-2175},
	primaryclass = {astro-ph.GA},
	title = {{Constraining the major merger history of z \raisebox{-0.5ex}\textasciitilde 3-9 galaxies using JADES: dominant in situ star formation}},
	volume = {540},
	year = 2025}

@article{Isobe:2025aa,
	adsurl = {https://ui.adsabs.harvard.edu/abs/2025MNRAS.541L..71I},
	archiveprefix = {arXiv},
	author = {{Isobe}, Yuki and {Maiolino}, Roberto and {D'Eugenio}, Francesco and {Curti}, Mirko and {Ji}, Xihan and {Juod{\v{z}}balis}, Ignas and {Scholtz}, Jan and {Feltre}, Anne and {Charlot}, St{\'e}phane and {{\"U}bler}, Hannah and {J. Bunker}, Andrew and {Carniani}, Stefano and {Curtis-Lake}, Emma and {Ji}, Zhiyuan and {Kumari}, Nimisha and {Rinaldi}, Pierluigi and {Robertson}, Brant and {Willott}, Chris and {Witstok}, Joris},
	doi = {10.1093/mnrasl/slaf056},
	eprint = {2502.12091},
	journal = {\mnras},
	month = jul,
	number = {1},
	pages = {L71-L79},
	primaryclass = {astro-ph.GA},
	title = {{JADES: nitrogen enhancement in high-redshift broad-line active galactic nuclei}},
	volume = {541},
	year = 2025}

@article{Arellano-Cordova:2025aa,
	adsurl = {https://ui.adsabs.harvard.edu/abs/2025MNRAS.540.2991A},
	archiveprefix = {arXiv},
	author = {{Arellano-C{\'o}rdova}, K.~Z. and {Cullen}, F. and {Carnall}, A.~C. and {Scholte}, D. and {Stanton}, T.~M. and {Kobayashi}, C. and {Martinez}, Z. and {Berg}, D.~A. and {Barrufet}, L. and {Begley}, R. and {Donnan}, C.~T. and {Dunlop}, J.~S. and {Hamadouche}, M.~L. and {McLeod}, D.~J. and {McLure}, R.~J. and {Rowlands}, K. and {Shapley}, A.~E.},
	doi = {10.1093/mnras/staf855},
	eprint = {2412.10557},
	journal = {\mnras},
	month = jul,
	number = {4},
	pages = {2991-3007},
	primaryclass = {astro-ph.GA},
	title = {{The JWST EXCELS survey: direct estimates of C, N, and O abundances in two relatively metal-rich galaxies at z ≃ 5}},
	volume = {540},
	year = 2025}

@article{Sarkar:2025aa,
	adsurl = {https://ui.adsabs.harvard.edu/abs/2025ApJ...978..136S},
	archiveprefix = {arXiv},
	author = {{Sarkar}, Arnab and {Chakraborty}, Priyanka and {Vogelsberger}, Mark and {McDonald}, Michael and {Torrey}, Paul and {Garcia}, Alex M. and {Khullar}, Gourav and {Ferland}, Gary J. and {Forman}, William and {Wolk}, Scott and {Schneider}, Benjamin and {Bautz}, Mark and {Miller}, Eric and {Grant}, Catherine and {ZuHone}, John},
	doi = {10.3847/1538-4357/ad8f32},
	eid = {136},
	eprint = {2408.07974},
	journal = {\apj},
	month = jan,
	number = {2},
	pages = {136},
	primaryclass = {astro-ph.GA},
	title = {{Unveiling the Cosmic Chemistry: Revisiting the Mass{\textendash}Metallicity Relation with JWST/NIRSpec at 4 < z < 10}},
	volume = {978},
	year = 2025}

@article{Cullen:2021aa,
	adsurl = {https://ui.adsabs.harvard.edu/abs/2021MNRAS.505..903C},
	archiveprefix = {arXiv},
	author = {{Cullen}, F. and {Shapley}, A.~E. and {McLure}, R.~J. and {Dunlop}, J.~S. and {Sanders}, R.~L. and {Topping}, M.~W. and {Reddy}, N.~A. and {Amor{\'\i}n}, R. and {Begley}, R. and {Bolzonella}, M. and {Calabr{\`o}}, A. and {Carnall}, A.~C. and {Castellano}, M. and {Cimatti}, A. and {Cirasuolo}, M. and {Cresci}, G. and {Fontana}, A. and {Fontanot}, F. and {Garilli}, B. and {Guaita}, L. and {Hamadouche}, M. and {Hathi}, N.~P. and {Mannucci}, F. and {McLeod}, D.~J. and {Pentericci}, L. and {Saxena}, A. and {Talia}, M. and {Zamorani}, G.},
	doi = {10.1093/mnras/stab1340},
	eprint = {2103.06300},
	journal = {\mnras},
	month = jul,
	number = {1},
	pages = {903-920},
	primaryclass = {astro-ph.GA},
	title = {{The NIRVANDELS Survey: a robust detection of {\ensuremath{\alpha}}-enhancement in star-forming galaxies at z ≃ 3.4}},
	volume = {505},
	year = 2021}

@article{Pascale:2023aa,
	adsurl = {https://ui.adsabs.harvard.edu/abs/2023ApJ...957...77P},
	archiveprefix = {arXiv},
	author = {{Pascale}, Massimo and {Dai}, Liang and {McKee}, Christopher F. and {Tsang}, Benny T. -H.},
	doi = {10.3847/1538-4357/acf75c},
	eid = {77},
	eprint = {2301.10790},
	journal = {\apj},
	month = nov,
	number = {2},
	pages = {77},
	primaryclass = {astro-ph.GA},
	title = {{Nitrogen-enriched, Highly Pressurized Nebular Clouds Surrounding a Super Star Cluster at Cosmic Noon}},
	volume = {957},
	year = 2023}

@article{Wagoner:1967aa,
	adsurl = {https://ui.adsabs.harvard.edu/abs/1967ApJ...148....3W},
	author = {{Wagoner}, Robert V. and {Fowler}, William A. and {Hoyle}, F.},
	doi = {10.1086/149126},
	journal = {\apj},
	month = apr,
	pages = {3},
	title = {{On the Synthesis of Elements at Very High Temperatures}},
	volume = {148},
	year = 1967}

@article{Peebles:1966aa,
	adsurl = {https://ui.adsabs.harvard.edu/abs/1966ApJ...146..542P},
	author = {{Peebles}, P.~J.~E.},
	doi = {10.1086/148918},
	journal = {\apj},
	month = nov,
	pages = {542},
	title = {{Primordial Helium Abundance and the Primordial Fireball. II}},
	volume = {146},
	year = 1966}

@article{Hoyle:1964aa,
	adsurl = {https://ui.adsabs.harvard.edu/abs/1964Natur.203.1108H},
	author = {{Hoyle}, F. and {Tayler}, R.~J.},
	doi = {10.1038/2031108a0},
	journal = {\nat},
	month = sep,
	number = {4950},
	pages = {1108-1110},
	title = {{The Mystery of the Cosmic Helium Abundance}},
	volume = {203},
	year = 1964}

@article{Klessen:2023aa,
	adsurl = {https://ui.adsabs.harvard.edu/abs/2023ARA&A..61...65K},
	archiveprefix = {arXiv},
	author = {{Klessen}, Ralf S. and {Glover}, Simon C.~O.},
	doi = {10.1146/annurev-astro-071221-053453},
	eprint = {2303.12500},
	journal = {\araa},
	month = aug,
	pages = {65-130},
	primaryclass = {astro-ph.CO},
	title = {{The First Stars: Formation, Properties, and Impact}},
	volume = {61},
	year = 2023}

@article{Karakas:2014aa,
	adsurl = {https://ui.adsabs.harvard.edu/abs/2014PASA...31...30K},
	archiveprefix = {arXiv},
	author = {{Karakas}, Amanda I. and {Lattanzio}, John C.},
	doi = {10.1017/pasa.2014.21},
	eid = {e030},
	eprint = {1405.0062},
	journal = {\pasa},
	month = jul,
	pages = {e030},
	primaryclass = {astro-ph.SR},
	title = {{The Dawes Review 2: Nucleosynthesis and Stellar Yields of Low- and Intermediate-Mass Single Stars}},
	volume = {31},
	year = 2014}

@article{Stanton:2024aa,
	adsurl = {https://ui.adsabs.harvard.edu/abs/2024MNRAS.532.3102S},
	archiveprefix = {arXiv},
	author = {{Stanton}, T.~M. and {Cullen}, F. and {McLure}, R.~J. and {Shapley}, A.~E. and {Arellano-C{\'o}rdova}, K.~Z. and {Begley}, R. and {Amor{\'\i}n}, R. and {Barrufet}, L. and {Calabr{\`o}}, A. and {Carnall}, A.~C. and {Cirasuolo}, M. and {Dunlop}, J.~S. and {Donnan}, C.~T. and {Hamadouche}, M.~L. and {Liu}, F.~Y. and {McLeod}, D.~J. and {Pentericci}, L. and {Pozzetti}, L. and {Sanders}, R.~L. and {Scholte}, D. and {Topping}, M.~W.},
	doi = {10.1093/mnras/stae1705},
	eprint = {2405.00774},
	journal = {\mnras},
	month = aug,
	number = {3},
	pages = {3102-3119},
	primaryclass = {astro-ph.GA},
	title = {{The NIRVANDELS survey: the stellar and gas-phase mass-metallicity relations of star-forming galaxies at z = 3.5}},
	volume = {532},
	year = 2024}

@article{Kirby:2013aa,
	adsurl = {https://ui.adsabs.harvard.edu/abs/2013ApJ...779..102K},
	archiveprefix = {arXiv},
	author = {{Kirby}, Evan N. and {Cohen}, Judith G. and {Guhathakurta}, Puragra and {Cheng}, Lucy and {Bullock}, James S. and {Gallazzi}, Anna},
	doi = {10.1088/0004-637X/779/2/102},
	eid = {102},
	eprint = {1310.0814},
	journal = {\apj},
	month = dec,
	number = {2},
	pages = {102},
	primaryclass = {astro-ph.GA},
	title = {{The Universal Stellar Mass-Stellar Metallicity Relation for Dwarf Galaxies}},
	volume = {779},
	year = 2013}

@article{Heesters:2023aa,
	adsurl = {https://ui.adsabs.harvard.edu/abs/2023A&A...676A..33H},
	archiveprefix = {arXiv},
	author = {{Heesters}, Nick and {M{\"u}ller}, Oliver and {Marleau}, Francine R. and {Duc}, Pierre-Alain and {S{\'a}nchez-Janssen}, Rub{\'e}n and {Poulain}, M{\'e}lina and {Habas}, Rebecca and {Lim}, Sungsoon and {Durrell}, Patrick R.},
	doi = {10.1051/0004-6361/202346441},
	eid = {A33},
	eprint = {2305.04593},
	journal = {\aap},
	month = aug,
	pages = {A33},
	primaryclass = {astro-ph.GA},
	title = {{Radial velocities and stellar population properties of 56 MATLAS dwarf galaxies observed with MUSE}},
	volume = {676},
	year = 2023}

@article{Pollock:2025aa,
	adsurl = {https://ui.adsabs.harvard.edu/abs/2025arXiv250615779P},
	archiveprefix = {arXiv},
	author = {{Pollock}, Clara L. and {Gottumukkala}, Rashmi and {Heintz}, Kasper E. and {Brammer}, Gabriel B. and {Roberts-Borsani}, Guido and {Oesch}, Pascal A. and {Witstok}, Joris and {Arellano-C{\'o}rdova}, Karla Z. and {Cullen}, Fergus and {Scholte}, Dirk and {Terp}, Chamilla and {Rowland}, Lucie and {Sneppen}, Albert and {Ito}, Kei and {Valentino}, Francesco and {Matthee}, Jorryt and {Watson}, Darach and {Toft}, Sune},
	doi = {10.48550/arXiv.2506.15779},
	eid = {arXiv:2506.15779},
	eprint = {2506.15779},
	journal = {arXiv e-prints},
	month = jun,
	pages = {arXiv:2506.15779},
	primaryclass = {astro-ph.GA},
	title = {{Novel $z\sim~10$ auroral line measurements extend the gradual offset of the FMR deep into the first Gyr of cosmic time}},
	year = 2025}

@article{Naidu:2025ab,
	adsurl = {https://ui.adsabs.harvard.edu/abs/2025arXiv250511263N},
	archiveprefix = {arXiv},
	author = {{Naidu}, Rohan P. and {Oesch}, Pascal A. and {Brammer}, Gabriel and {Weibel}, Andrea and {Li}, Yijia and {Matthee}, Jorryt and {Chisholm}, John and {Pollock}, Clara L. and {Heintz}, Kasper E. and {Johnson}, Benjamin D. and {Shen}, Xuejian and {Hviding}, Raphael E. and {Leja}, Joel and {Tacchella}, Sandro and {Ganguly}, Arpita and {Witten}, Callum and {Atek}, Hakim and {Belli}, Sirio and {Bose}, Sownak and {Bouwens}, Rychard and {Dayal}, Pratika and {Decarli}, Roberto and {de Graaff}, Anna and {Fudamoto}, Yoshinobu and {Giovinazzo}, Emma and {Greene}, Jenny E. and {Illingworth}, Garth and {Inoue}, Akio K. and {Kane}, Sarah G. and {Labbe}, Ivo and {Leonova}, Ecaterina and {Marques-Chaves}, Rui and {Meyer}, Romain A. and {Nelson}, Erica J. and {Roberts-Borsani}, Guido and {Schaerer}, Daniel and {Simcoe}, Robert A. and {Stefanon}, Mauro and {Sugahara}, Yuma and {Toft}, Sune and {van der Wel}, Arjen and {van Dokkum}, Pieter and {Walter}, Fabian and {Watson}, Darach and {Weaver}, John R. and {Whitaker}, Katherine E.},
	doi = {10.48550/arXiv.2505.11263},
	eid = {arXiv:2505.11263},
	eprint = {2505.11263},
	journal = {arXiv e-prints},
	month = may,
	pages = {arXiv:2505.11263},
	primaryclass = {astro-ph.GA},
	title = {{A Cosmic Miracle: A Remarkably Luminous Galaxy at $z_{\rm{spec}}=14.44$ Confirmed with JWST}},
	year = 2025}

@article{Guszejnov:2020aa,
	adsurl = {https://ui.adsabs.harvard.edu/abs/2020MNRAS.492..488G},
	archiveprefix = {arXiv},
	author = {{Guszejnov}, D{\'a}vid and {Grudi{\'c}}, Michael Y. and {Offner}, Stella S.~R. and {Boylan-Kolchin}, Michael and {Faucher-Gigu{\`e}re}, Claude-Andr{\'e} and {Wetzel}, Andrew and {Benincasa}, Samantha M. and {Loebman}, Sarah},
	doi = {10.1093/mnras/stz3527},
	eprint = {1910.01163},
	journal = {\mnras},
	month = feb,
	number = {1},
	pages = {488-502},
	primaryclass = {astro-ph.GA},
	title = {{Evolution of giant molecular clouds across cosmic time}},
	volume = {492},
	year = 2020}

@article{Garcia:2025ab,
	adsurl = {https://ui.adsabs.harvard.edu/abs/2025MNRAS.536..119G},
	archiveprefix = {arXiv},
	author = {{Garcia}, Alex M. and {Torrey}, Paul and {Ellison}, Sara L. and {Grasha}, Kathryn and {Chen}, Qian-Hui and {Hemler}, Z.~S. and {Zimmerman}, Dhruv T. and {Wright}, Ruby J. and {Zovaro}, Henry R.~M. and {Nelson}, Erica J. and {Sanders}, Ryan L. and {Kewley}, Lisa J. and {Hernquist}, Lars},
	doi = {10.1093/mnras/stae2587},
	eprint = {2407.06254},
	journal = {\mnras},
	month = jan,
	number = {1},
	pages = {119-144},
	primaryclass = {astro-ph.GA},
	title = {{Does the fundamental metallicity relation evolve with redshift? - II. The evolution in normalization of the mass-metallicity relation}},
	volume = {536},
	year = 2025}

@article{Finlator:2008aa,
	adsurl = {https://ui.adsabs.harvard.edu/abs/2008MNRAS.385.2181F},
	archiveprefix = {arXiv},
	author = {{Finlator}, Kristian and {Dav{\'e}}, Romeel},
	doi = {10.1111/j.1365-2966.2008.12991.x},
	eprint = {0704.3100},
	journal = {\mnras},
	month = apr,
	number = {4},
	pages = {2181-2204},
	primaryclass = {astro-ph},
	title = {{The origin of the galaxy mass-metallicity relation and implications for galactic outflows}},
	volume = {385},
	year = 2008}

@article{Larson:1974aa,
	adsurl = {https://ui.adsabs.harvard.edu/abs/1974MNRAS.169..229L},
	author = {{Larson}, Richard B.},
	doi = {10.1093/mnras/169.2.229},
	journal = {\mnras},
	month = nov,
	pages = {229-246},
	title = {{Effects of supernovae on the early evolution of galaxies}},
	volume = {169},
	year = 1974}

@article{Larson:1972aa,
	adsurl = {https://ui.adsabs.harvard.edu/abs/1972NPhS..236....7L},
	author = {{Larson}, Richard B.},
	doi = {10.1038/physci236007a0},
	journal = {Nature Physical Science},
	month = mar,
	number = {62},
	pages = {7-8},
	title = {{Effect of Infalling Matter on the Heavy Element Content of a Galaxy}},
	volume = {236},
	year = 1972}

@article{Tinsley:1980aa,
	adsurl = {https://ui.adsabs.harvard.edu/abs/1980FCPh....5..287T},
	archiveprefix = {arXiv},
	author = {{Tinsley}, B.~M.},
	doi = {10.48550/arXiv.2203.02041},
	eprint = {2203.02041},
	journal = {\fcp},
	month = jan,
	pages = {287-388},
	primaryclass = {astro-ph.GA},
	title = {{Evolution of the Stars and Gas in Galaxies}},
	volume = {5},
	year = 1980}

@article{Wilkins:2023aa,
	adsurl = {https://ui.adsabs.harvard.edu/abs/2023MNRAS.519.3118W},
	archiveprefix = {arXiv},
	author = {{Wilkins}, Stephen M. and {Vijayan}, Aswin P. and {Lovell}, Christopher C. and {Roper}, William J. and {Irodotou}, Dimitrios and {Caruana}, Joseph and {Seeyave}, Louise T.~C. and {Kuusisto}, Jussi K. and {Thomas}, Peter A. and {Parris}, Shedeur A.~K.},
	doi = {10.1093/mnras/stac3280},
	eprint = {2204.09431},
	journal = {\mnras},
	month = feb,
	number = {2},
	pages = {3118-3128},
	primaryclass = {astro-ph.GA},
	title = {{First light and reionization epoch simulations (FLARES) V: the redshift frontier}},
	volume = {519},
	year = 2023}

@article{Langan:2020aa,
	adsurl = {https://ui.adsabs.harvard.edu/abs/2020MNRAS.494.1988L},
	archiveprefix = {arXiv},
	author = {{Langan}, Ivanna and {Ceverino}, Daniel and {Finlator}, Kristian},
	doi = {10.1093/mnras/staa880},
	eprint = {1910.11729},
	journal = {\mnras},
	month = may,
	number = {2},
	pages = {1988-1993},
	primaryclass = {astro-ph.GA},
	title = {{Weak evolution of the mass-metallicity relation at cosmic dawn in the FirstLight simulations}},
	volume = {494},
	year = 2020}

@article{Bassini:2024aa,
	adsurl = {https://ui.adsabs.harvard.edu/abs/2024MNRAS.532L..14B},
	archiveprefix = {arXiv},
	author = {{Bassini}, Luigi and {Feldmann}, Robert and {Gensior}, Jindra and {Faucher-Gigu{\`e}re}, Claude-Andr{\'e} and {Cenci}, Elia and {Moreno}, Jorge and {Bernardini}, Mauro and {Liang}, Lichen},
	doi = {10.1093/mnrasl/slae036},
	eprint = {2401.13824},
	journal = {\mnras},
	month = jul,
	number = {1},
	pages = {L14-L20},
	primaryclass = {astro-ph.GA},
	title = {{Inflow and outflow properties, not total gas fractions, drive the evolution of the mass-metallicity relation}},
	volume = {532},
	year = 2024}

@article{Rhoads:2023aa,
	adsurl = {https://ui.adsabs.harvard.edu/abs/2023ApJ...942L..14R},
	archiveprefix = {arXiv},
	author = {{Rhoads}, James E. and {Wold}, Isak G.~B. and {Harish}, Santosh and {Kim}, Keunho J. and {Pharo}, John and {Malhotra}, Sangeeta and {Gabrielpillai}, Austen and {Jiang}, Tianxing and {Yang}, Huan},
	doi = {10.3847/2041-8213/acaaaf},
	eid = {L14},
	eprint = {2207.13020},
	journal = {\apjl},
	month = jan,
	number = {1},
	pages = {L14},
	primaryclass = {astro-ph.GA},
	title = {{Finding Peas in the Early Universe with JWST}},
	volume = {942},
	year = 2023}

@article{Topping:2021aa,
	adsurl = {https://ui.adsabs.harvard.edu/abs/2021MNRAS.506.1237T},
	archiveprefix = {arXiv},
	author = {{Topping}, Michael W. and {Shapley}, Alice E. and {Sanders}, Ryan L. and {Kriek}, Mariska and {Reddy}, Naveen A. and {Coil}, Alison L. and {Mobasher}, Bahram and {Siana}, Brian and {Freeman}, William R. and {Shivaei}, Irene and {Azadi}, Mojegan and {Price}, Sedona H. and {Leung}, Gene C.~K. and {Fetherolf}, Tara and {de Groot}, Laura and {Zick}, Tom and {Fornasini}, Francesca M. and {Barro}, Guillermo and {Runco}, Jordan N.},
	doi = {10.1093/mnras/stab1793},
	eprint = {2103.09245},
	journal = {\mnras},
	month = sep,
	number = {1},
	pages = {1237-1249},
	primaryclass = {astro-ph.GA},
	title = {{The MOSDEF survey: the mass-metallicity relationship and the existence of the FMR at z 1.5}},
	volume = {506},
	year = 2021}

@article{Heintz:2023aa,
	adsurl = {https://ui.adsabs.harvard.edu/abs/2023NatAs...7.1517H},
	archiveprefix = {arXiv},
	author = {{Heintz}, Kasper E. and {Brammer}, Gabriel B. and {Gim{\'e}nez-Arteaga}, Clara and {Strait}, Victoria B. and {Lagos}, Claudia del P. and {Vijayan}, Aswin P. and {Matthee}, Jorryt and {Watson}, Darach and {Mason}, Charlotte A. and {Hutter}, Anne and {Toft}, Sune and {Fynbo}, Johan P.~U. and {Oesch}, Pascal A.},
	doi = {10.1038/s41550-023-02078-7},
	eprint = {2212.02890},
	journal = {Nature Astronomy},
	month = dec,
	pages = {1517-1524},
	primaryclass = {astro-ph.GA},
	title = {{Dilution of chemical enrichment in galaxies 600 Myr after the Big Bang}},
	volume = {7},
	year = 2023}

@article{Shapley:2023aa,
	adsurl = {https://ui.adsabs.harvard.edu/abs/2023ApJ...950L...1S},
	archiveprefix = {arXiv},
	author = {{Shapley}, Alice E. and {Reddy}, Naveen A. and {Sanders}, Ryan L. and {Topping}, Michael W. and {Brammer}, Gabriel B.},
	doi = {10.3847/2041-8213/acd939},
	eid = {L1},
	eprint = {2303.00410},
	journal = {\apjl},
	month = jun,
	number = {1},
	pages = {L1},
	primaryclass = {astro-ph.GA},
	title = {{JWST/NIRSpec Measurements of the Relationships between Nebular Emission-line Ratios and Stellar Mass at z 3-6}},
	volume = {950},
	year = 2023}

@article{Langeroodi:2023aa,
	adsurl = {https://ui.adsabs.harvard.edu/abs/2023ApJ...957...39L},
	archiveprefix = {arXiv},
	author = {{Langeroodi}, Danial and {Hjorth}, Jens and {Chen}, Wenlei and {Kelly}, Patrick L. and {Williams}, Hayley and {Lin}, Yu-Heng and {Scarlata}, Claudia and {Zitrin}, Adi and {Broadhurst}, Tom and {Diego}, Jose M. and {Huang}, Xiaosheng and {Filippenko}, Alexei V. and {Foley}, Ryan J. and {Jha}, Saurabh and {Koekemoer}, Anton M. and {Oguri}, Masamune and {Perez-Fournon}, Ismael and {Pierel}, Justin and {Poidevin}, Frederick and {Strolger}, Lou},
	doi = {10.3847/1538-4357/acdbc1},
	eid = {39},
	eprint = {2212.02491},
	journal = {\apj},
	month = nov,
	number = {1},
	pages = {39},
	primaryclass = {astro-ph.GA},
	title = {{Evolution of the Mass-Metallicity Relation from Redshift z {\ensuremath{\approx}} 8 to the Local Universe}},
	volume = {957},
	year = 2023}

@article{Matthee:2023aa,
	adsurl = {https://ui.adsabs.harvard.edu/abs/2023ApJ...950...67M},
	archiveprefix = {arXiv},
	author = {{Matthee}, Jorryt and {Mackenzie}, Ruari and {Simcoe}, Robert A. and {Kashino}, Daichi and {Lilly}, Simon J. and {Bordoloi}, Rongmon and {Eilers}, Anna-Christina},
	doi = {10.3847/1538-4357/acc846},
	eid = {67},
	eprint = {2211.08255},
	journal = {\apj},
	month = jun,
	number = {1},
	pages = {67},
	primaryclass = {astro-ph.GA},
	title = {{EIGER. II. First Spectroscopic Characterization of the Young Stars and Ionized Gas Associated with Strong H{\ensuremath{\beta}} and [O III] Line Emission in Galaxies at z = 5-7 with JWST}},
	volume = {950},
	year = 2023}

@article{Carniani:2025aa,
	adsurl = {https://ui.adsabs.harvard.edu/abs/2025A&A...696A..87C},
	archiveprefix = {arXiv},
	author = {{Carniani}, Stefano and {D'Eugenio}, Francesco and {Ji}, Xihan and {Parlanti}, Eleonora and {Scholtz}, Jan and {Sun}, Fengwu and {Venturi}, Giacomo and {Bakx}, Tom J.~L.~C. and {Curti}, Mirko and {Maiolino}, Roberto and {Tacchella}, Sandro and {Zavala}, Jorge A. and {Hainline}, Kevin and {Witstok}, Joris and {Johnson}, Benjamin D. and {Alberts}, Stacey and {Bunker}, Andrew J. and {Charlot}, St{\'e}phane and {Eisenstein}, Daniel J. and {Helton}, Jakob M. and {Jakobsen}, Peter and {Kumari}, Nimisha and {Robertson}, Brant and {Saxena}, Aayush and {{\"U}bler}, Hannah and {Williams}, Christina C. and {Willmer}, Christopher N.~A. and {Willott}, Chris},
	doi = {10.1051/0004-6361/202452451},
	eid = {A87},
	eprint = {2409.20533},
	journal = {\aap},
	month = apr,
	pages = {A87},
	primaryclass = {astro-ph.GA},
	title = {{The eventful life of a luminous galaxy at z = 14: metal enrichment, feedback, and low gas fraction?}},
	volume = {696},
	year = 2025}

@article{Sanders:2018aa,
	adsurl = {https://ui.adsabs.harvard.edu/abs/2018ApJ...858...99S},
	archiveprefix = {arXiv},
	author = {{Sanders}, Ryan L. and {Shapley}, Alice E. and {Kriek}, Mariska and {Freeman}, William R. and {Reddy}, Naveen A. and {Siana}, Brian and {Coil}, Alison L. and {Mobasher}, Bahram and {Dav{\'e}}, Romeel and {Shivaei}, Irene and {Azadi}, Mojegan and {Price}, Sedona H. and {Leung}, Gene and {Fetherolf}, Tara and {de Groot}, Laura and {Zick}, Tom and {Fornasini}, Francesca M. and {Barro}, Guillermo},
	doi = {10.3847/1538-4357/aabcbd},
	eid = {99},
	eprint = {1711.00224},
	journal = {\apj},
	month = may,
	number = {2},
	pages = {99},
	primaryclass = {astro-ph.GA},
	title = {{The MOSDEF Survey: A Stellar Mass-SFR-Metallicity Relation Exists at z {\ensuremath{\sim}} 2.3}},
	volume = {858},
	year = 2018}

@article{Sanders:2024aa,
	adsurl = {https://ui.adsabs.harvard.edu/abs/2024ApJ...962...24S},
	archiveprefix = {arXiv},
	author = {{Sanders}, Ryan L. and {Shapley}, Alice E. and {Topping}, Michael W. and {Reddy}, Naveen A. and {Brammer}, Gabriel B.},
	doi = {10.3847/1538-4357/ad15fc},
	eid = {24},
	eprint = {2303.08149},
	journal = {\apj},
	month = feb,
	number = {1},
	pages = {24},
	primaryclass = {astro-ph.GA},
	title = {{Direct T $_{e}$-based Metallicities of z = 2{\textendash}9 Galaxies with JWST/NIRSpec: Empirical Metallicity Calibrations Applicable from Reionization to Cosmic Noon}},
	volume = {962},
	year = 2024}

@article{Telford:2016aa,
	adsurl = {https://ui.adsabs.harvard.edu/abs/2016ApJ...827...35T},
	archiveprefix = {arXiv},
	author = {{Telford}, O. Grace and {Dalcanton}, Julianne J. and {Skillman}, Evan D. and {Conroy}, Charlie},
	doi = {10.3847/0004-637X/827/1/35},
	eid = {35},
	eprint = {1606.08850},
	journal = {\apj},
	month = aug,
	number = {1},
	pages = {35},
	primaryclass = {astro-ph.GA},
	title = {{Exploring Systematic Effects in the Relation Between Stellar Mass, Gas Phase Metallicity, and Star Formation Rate}},
	volume = {827},
	year = 2016}

@article{Salim:2015aa,
	adsurl = {https://ui.adsabs.harvard.edu/abs/2015ApJ...808...25S},
	archiveprefix = {arXiv},
	author = {{Salim}, Samir and {Lee}, Janice C. and {Dav{\'e}}, Romeel and {Dickinson}, Mark},
	doi = {10.1088/0004-637X/808/1/25},
	eid = {25},
	eprint = {1506.03080},
	journal = {\apj},
	month = jul,
	number = {1},
	pages = {25},
	primaryclass = {astro-ph.GA},
	title = {{On the Mass-Metallicity-Star Formation Rate Relation for Galaxies at z{\ensuremath{\sim}}2}},
	volume = {808},
	year = 2015}

@article{Yates:2012aa,
	adsurl = {https://ui.adsabs.harvard.edu/abs/2012MNRAS.422..215Y},
	archiveprefix = {arXiv},
	author = {{Yates}, Robert M. and {Kauffmann}, Guinevere and {Guo}, Qi},
	doi = {10.1111/j.1365-2966.2012.20595.x},
	eprint = {1107.3145},
	journal = {\mnras},
	month = may,
	number = {1},
	pages = {215-231},
	primaryclass = {astro-ph.CO},
	title = {{The relation between metallicity, stellar mass and star formation in galaxies: an analysis of observational and model data}},
	volume = {422},
	year = 2012}

@article{Lee:2006aa,
	adsurl = {https://ui.adsabs.harvard.edu/abs/2006ApJ...647..970L},
	archiveprefix = {arXiv},
	author = {{Lee}, Henry and {Skillman}, Evan D. and {Cannon}, John M. and {Jackson}, Dale C. and {Gehrz}, Robert D. and {Polomski}, Elisha F. and {Woodward}, Charles E.},
	doi = {10.1086/505573},
	eprint = {astro-ph/0605036},
	journal = {\apj},
	month = aug,
	number = {2},
	pages = {970-983},
	primaryclass = {astro-ph},
	title = {{On Extending the Mass-Metallicity Relation of Galaxies by 2.5 Decades in Stellar Mass}},
	volume = {647},
	year = 2006}

@article{Baker:2023ac,
	adsurl = {https://ui.adsabs.harvard.edu/abs/2023MNRAS.521.4173B},
	archiveprefix = {arXiv},
	author = {{Baker}, William M. and {Maiolino}, Roberto},
	doi = {10.1093/mnras/stad802},
	eprint = {2303.08145},
	journal = {\mnras},
	month = may,
	number = {3},
	pages = {4173-4179},
	primaryclass = {astro-ph.GA},
	title = {{Stellar mass, not dynamical mass nor gravitational potential, drives the mass-metallicity relationship}},
	volume = {521},
	year = 2023}

@article{Edmunds:1990aa,
	adsurl = {https://ui.adsabs.harvard.edu/abs/1990MNRAS.246..678E},
	author = {{Edmunds}, M.~G.},
	journal = {\mnras},
	month = oct,
	pages = {678},
	title = {{General Constraints on the Effect of Gas Flows in the Chemical Evolution of Galaxies}},
	volume = {246},
	year = 1990}

@article{Searle:1972aa,
	adsurl = {https://ui.adsabs.harvard.edu/abs/1972ApJ...173...25S},
	author = {{Searle}, Leonard and {Sargent}, Wallace L.~W.},
	doi = {10.1086/151398},
	journal = {\apj},
	month = apr,
	pages = {25},
	title = {{Inferences from the Composition of Two Dwarf Blue Galaxies}},
	volume = {173},
	year = 1972}

@article{Talbot:1971aa,
	adsurl = {https://ui.adsabs.harvard.edu/abs/1971ApJ...170..409T},
	author = {{Talbot}, Jr., Raymond J. and {Arnett}, W. David},
	doi = {10.1086/151228},
	journal = {\apj},
	month = dec,
	pages = {409},
	title = {{The Evolution of Galaxies. I. Formulation and Mathematical Behavior of the One-Zone Model}},
	volume = {170},
	year = 1971}

@article{Nakajima:2023aa,
	adsurl = {https://ui.adsabs.harvard.edu/abs/2023ApJS..269...33N},
	archiveprefix = {arXiv},
	author = {{Nakajima}, Kimihiko and {Ouchi}, Masami and {Isobe}, Yuki and {Harikane}, Yuichi and {Zhang}, Yechi and {Ono}, Yoshiaki and {Umeda}, Hiroya and {Oguri}, Masamune},
	doi = {10.3847/1538-4365/acd556},
	eid = {33},
	eprint = {2301.12825},
	journal = {\apjs},
	month = dec,
	number = {2},
	pages = {33},
	primaryclass = {astro-ph.GA},
	title = {{JWST Census for the Mass-Metallicity Star Formation Relations at z = 4-10 with Self-consistent Flux Calibration and Proper Metallicity Calibrators}},
	volume = {269},
	year = 2023}

@article{Cameron:2023ac,
	adsurl = {https://ui.adsabs.harvard.edu/abs/2023MNRAS.522L..89C},
	archiveprefix = {arXiv},
	author = {{Cameron}, Alex J. and {Katz}, Harley and {Rey}, Martin P.},
	doi = {10.1093/mnrasl/slad046},
	eprint = {2210.14234},
	journal = {\mnras},
	month = jun,
	number = {1},
	pages = {L89-L94},
	primaryclass = {astro-ph.GA},
	title = {{A novel approach to correcting T$_{e}$-based mass-metallicity relations}},
	volume = {522},
	year = 2023}

@article{Yates:2020aa,
	adsurl = {https://ui.adsabs.harvard.edu/abs/2020A&A...634A.107Y},
	archiveprefix = {arXiv},
	author = {{Yates}, R.~M. and {Schady}, P. and {Chen}, T. -W. and {Schweyer}, T. and {Wiseman}, P.},
	doi = {10.1051/0004-6361/201936506},
	eid = {A107},
	eprint = {1901.02890},
	journal = {\aap},
	month = feb,
	pages = {A107},
	primaryclass = {astro-ph.GA},
	title = {{Present-day mass-metallicity relation for galaxies using a new electron temperature method}},
	volume = {634},
	year = 2020}

@article{Bastian:2018aa,
	adsurl = {https://ui.adsabs.harvard.edu/abs/2018ARA&A..56...83B},
	archiveprefix = {arXiv},
	author = {{Bastian}, Nate and {Lardo}, Carmela},
	doi = {10.1146/annurev-astro-081817-051839},
	eprint = {1712.01286},
	journal = {\araa},
	month = sep,
	pages = {83-136},
	primaryclass = {astro-ph.SR},
	title = {{Multiple Stellar Populations in Globular Clusters}},
	volume = {56},
	year = 2018}

@article{Vasiliev:2021aa,
	adsurl = {https://ui.adsabs.harvard.edu/abs/2021MNRAS.505.5978V},
	archiveprefix = {arXiv},
	author = {{Vasiliev}, Eugene and {Baumgardt}, Holger},
	doi = {10.1093/mnras/stab1475},
	eprint = {2102.09568},
	journal = {\mnras},
	month = aug,
	number = {4},
	pages = {5978-6002},
	primaryclass = {astro-ph.GA},
	title = {{Gaia EDR3 view on galactic globular clusters}},
	volume = {505},
	year = 2021}

@article{Abdurrouf:2022aa,
	adsurl = {https://ui.adsabs.harvard.edu/abs/2022ApJS..259...35A},
	archiveprefix = {arXiv},
	author = {{Abdurro'uf} and {Accetta}, Katherine and {Aerts}, Conny and {Silva Aguirre}, V{\'\i}ctor and {Ahumada}, Romina and {Ajgaonkar}, Nikhil and {Filiz Ak}, N. and {Alam}, Shadab and {Allende Prieto}, Carlos and {Almeida}, Andr{\'e}s and {Anders}, Friedrich and {Anderson}, Scott F. and {Andrews}, Brett H. and {Anguiano}, Borja and {Aquino-Ort{\'\i}z}, Erik and {Arag{\'o}n-Salamanca}, Alfonso and {Argudo-Fern{\'a}ndez}, Maria and {Ata}, Metin and {Aubert}, Marie and {Avila-Reese}, Vladimir and {Badenes}, Carles and {Barb{\'a}}, Rodolfo H. and {Barger}, Kat and {Barrera-Ballesteros}, Jorge K. and {Beaton}, Rachael L. and {Beers}, Timothy C. and {Belfiore}, Francesco and {Bender}, Chad F. and {Bernardi}, Mariangela and {Bershady}, Matthew A. and {Beutler}, Florian and {Bidin}, Christian Moni and {Bird}, Jonathan C. and {Bizyaev}, Dmitry and {Blanc}, Guillermo A. and {Blanton}, Michael R. and {Boardman}, Nicholas Fraser and {Bolton}, Adam S. and {Boquien}, M{\'e}d{\'e}ric and {Borissova}, Jura and {Bovy}, Jo and {Brandt}, W.~N. and {Brown}, Jordan and {Brownstein}, Joel R. and {Brusa}, Marcella and {Buchner}, Johannes and {Bundy}, Kevin and {Burchett}, Joseph N. and {Bureau}, Martin and {Burgasser}, Adam and {Cabang}, Tuesday K. and {Campbell}, Stephanie and {Cappellari}, Michele and {Carlberg}, Joleen K. and {Wanderley}, F{\'a}bio Carneiro and {Carrera}, Ricardo and {Cash}, Jennifer and {Chen}, Yan-Ping and {Chen}, Wei-Huai and {Cherinka}, Brian and {Chiappini}, Cristina and {Choi}, Peter Doohyun and {Chojnowski}, S. Drew and {Chung}, Haeun and {Clerc}, Nicolas and {Cohen}, Roger E. and {Comerford}, Julia M. and {Comparat}, Johan and {da Costa}, Luiz and {Covey}, Kevin and {Crane}, Jeffrey D. and {Cruz-Gonzalez}, Irene and {Culhane}, Connor and {Cunha}, Katia and {Dai}, Y. Sophia and {Damke}, Guillermo and {Darling}, Jeremy and {Davidson}, Jr., James W. and {Davies}, Roger and {Dawson}, Kyle and {De Lee}, Nathan and {Diamond-Stanic}, Aleksandar M. and {Cano-D{\'\i}az}, Mariana and {S{\'a}nchez}, Helena Dom{\'\i}nguez and {Donor}, John and {Duckworth}, Chris and {Dwelly}, Tom and {Eisenstein}, Daniel J. and {Elsworth}, Yvonne P. and {Emsellem}, Eric and {Eracleous}, Mike and {Escoffier}, Stephanie and {Fan}, Xiaohui and {Farr}, Emily and {Feng}, Shuai and {Fern{\'a}ndez-Trincado}, Jos{\'e} G. and {Feuillet}, Diane and {Filipp}, Andreas and {Fillingham}, Sean P. and {Frinchaboy}, Peter M. and {Fromenteau}, Sebastien and {Galbany}, Llu{\'\i}s and {Garc{\'\i}a}, Rafael A. and {Garc{\'\i}a-Hern{\'a}ndez}, D.~A. and {Ge}, Junqiang and {Geisler}, Doug and {Gelfand}, Joseph and {G{\'e}ron}, Tobias and {Gibson}, Benjamin J. and {Goddy}, Julian and {Godoy-Rivera}, Diego and {Grabowski}, Kathleen and {Green}, Paul J. and {Greener}, Michael and {Grier}, Catherine J. and {Griffith}, Emily and {Guo}, Hong and {Guy}, Julien and {Hadjara}, Massinissa and {Harding}, Paul and {Hasselquist}, Sten and {Hayes}, Christian R. and {Hearty}, Fred and {Hern{\'a}ndez}, Jes{\'u}s and {Hill}, Lewis and {Hogg}, David W. and {Holtzman}, Jon A. and {Horta}, Danny and {Hsieh}, Bau-Ching and {Hsu}, Chin-Hao and {Hsu}, Yun-Hsin and {Huber}, Daniel and {Huertas-Company}, Marc and {Hutchinson}, Brian and {Hwang}, Ho Seong and {Ibarra-Medel}, H{\'e}ctor J. and {Chitham}, Jacob Ider and {Ilha}, Gabriele S. and {Imig}, Julie and {Jaekle}, Will and {Jayasinghe}, Tharindu and {Ji}, Xihan and {Johnson}, Jennifer A. and {Jones}, Amy and {J{\"o}nsson}, Henrik and {Katkov}, Ivan and {Khalatyan}, Dr., Arman and {Kinemuchi}, Karen and {Kisku}, Shobhit and {Knapen}, Johan H. and {Kneib}, Jean-Paul and {Kollmeier}, Juna A. and {Kong}, Miranda and {Kounkel}, Marina and {Kreckel}, Kathryn and {Krishnarao}, Dhanesh and {Lacerna}, Ivan and {Lane}, Richard R. and {Langgin}, Rachel and {Lavender}, Ramon and {Law}, David R. and {Lazarz}, Daniel and {Leung}, Henry W. and {Leung}, Ho-Hin and {Lewis}, Hannah M. and {Li}, Cheng and {Li}, Ran and {Lian}, Jianhui and {Liang}, Fu-Heng and {Lin}, Lihwai and {Lin}, Yen-Ting and {Lin}, Sicheng and {Lintott}, Chris and {Long}, Dan and {Longa-Pe{\~n}a}, Pen{\'e}lope and {L{\'o}pez-Cob{\'a}}, Carlos and {Lu}, Shengdong and {Lundgren}, Britt F. and {Luo}, Yuanze and {Mackereth}, J. Ted and {de la Macorra}, Axel and {Mahadevan}, Suvrath and {Majewski}, Steven R. and {Manchado}, Arturo and {Mandeville}, Travis and {Maraston}, Claudia and {Margalef-Bentabol}, Berta and {Masseron}, Thomas and {Masters}, Karen L. and {Mathur}, Savita and {McDermid}, Richard M. and {Mckay}, Myles and {Merloni}, Andrea and {Merrifield}, Michael and {Meszaros}, Szabolcs and {Miglio}, Andrea and {Di Mille}, Francesco and {Minniti}, Dante and {Minsley}, Rebecca and {Monachesi}, Antonela},
	doi = {10.3847/1538-4365/ac4414},
	eid = {35},
	eprint = {2112.02026},
	journal = {\apjs},
	month = apr,
	number = {2},
	pages = {35},
	primaryclass = {astro-ph.GA},
	title = {{The Seventeenth Data Release of the Sloan Digital Sky Surveys: Complete Release of MaNGA, MaStar, and APOGEE-2 Data}},
	volume = {259},
	year = 2022}

@article{Calabro:2024aa,
	adsurl = {https://ui.adsabs.harvard.edu/abs/2024ApJ...975..245C},
	archiveprefix = {arXiv},
	author = {{Calabr{\`o}}, Antonello and {Castellano}, Marco and {Zavala}, Jorge A. and {Pentericci}, Laura and {Arrabal Haro}, Pablo and {Bakx}, Tom J.~L.~C. and {Burgarella}, Denis and {Casey}, Caitlin M. and {Dickinson}, Mark and {Finkelstein}, Steven L. and {Fontana}, Adriano and {Llerena}, Mario and {Mascia}, Sara and {Merlin}, Emiliano and {Mitsuhashi}, Ikki and {Napolitano}, Lorenzo and {Paris}, Diego and {P{\'e}rez-Gonz{\'a}lez}, Pablo G. and {Roberts-Borsani}, Guido and {Santini}, Paola and {Treu}, Tommaso and {Vanzella}, Eros},
	doi = {10.3847/1538-4357/ad7602},
	eid = {245},
	eprint = {2403.12683},
	journal = {\apj},
	month = nov,
	number = {2},
	pages = {245},
	primaryclass = {astro-ph.GA},
	title = {{Evidence of Extreme Ionization Conditions and Low Metallicity in GHZ2/GLASS-Z12 from a Combined Analysis of NIRSpec and MIRI Observations}},
	volume = {975},
	year = 2024}

@article{Castellano:2024aa,
	adsurl = {https://ui.adsabs.harvard.edu/abs/2024ApJ...972..143C},
	archiveprefix = {arXiv},
	author = {{Castellano}, Marco and {Napolitano}, Lorenzo and {Fontana}, Adriano and {Roberts-Borsani}, Guido and {Treu}, Tommaso and {Vanzella}, Eros and {Zavala}, Jorge A. and {Arrabal Haro}, Pablo and {Calabr{\`o}}, Antonello and {Llerena}, Mario and {Mascia}, Sara and {Merlin}, Emiliano and {Paris}, Diego and {Pentericci}, Laura and {Santini}, Paola and {Bakx}, Tom J.~L.~C. and {Bergamini}, Pietro and {Cupani}, Guido and {Dickinson}, Mark and {Filippenko}, Alexei V. and {Glazebrook}, Karl and {Grillo}, Claudio and {Kelly}, Patrick L. and {Malkan}, Matthew A. and {Mason}, Charlotte A. and {Morishita}, Takahiro and {Nanayakkara}, Themiya and {Rosati}, Piero and {Sani}, Eleonora and {Wang}, Xin and {Yoon}, Ilsang},
	doi = {10.3847/1538-4357/ad5f88},
	eid = {143},
	eprint = {2403.10238},
	journal = {\apj},
	month = sep,
	number = {2},
	pages = {143},
	primaryclass = {astro-ph.GA},
	title = {{JWST NIRSpec Spectroscopy of the Remarkable Bright Galaxy GHZ2/GLASS-z12 at Redshift 12.34}},
	volume = {972},
	year = 2024}

@article{Alvarez-Marquez:2025aa,
	adsurl = {https://ui.adsabs.harvard.edu/abs/2025A&A...695A.250A},
	archiveprefix = {arXiv},
	author = {{{\'A}lvarez-M{\'a}rquez}, J. and {Crespo G{\'o}mez}, A. and {Colina}, L. and {Langeroodi}, D. and {Marques-Chaves}, R. and {Prieto-Jim{\'e}nez}, C. and {Bik}, A. and {Alonso-Herrero}, A. and {Boogaard}, L. and {Costantin}, L. and {Garc{\'\i}a-Mar{\'\i}n}, M. and {Gillman}, S. and {Hjorth}, J. and {Iani}, E. and {Jermann}, I. and {Labiano}, A. and {Melinder}, J. and {Meyer}, R. and {{\"O}stlin}, G. and {P{\'e}rez-Gonz{\'a}lez}, P.~G. and {Rinaldi}, P. and {Walter}, F. and {van der Werf}, P. and {Wright}, G.},
	doi = {10.1051/0004-6361/202451731},
	eid = {A250},
	eprint = {2412.12826},
	journal = {\aap},
	month = mar,
	pages = {A250},
	primaryclass = {astro-ph.GA},
	title = {{Insight into the starburst nature of Galaxy GN-z11 with JWST MIRI spectroscopy}},
	volume = {695},
	year = 2025}

@article{Tacchella:2023ab,
	adsurl = {https://ui.adsabs.harvard.edu/abs/2023ApJ...952...74T},
	archiveprefix = {arXiv},
	author = {{Tacchella}, Sandro and {Eisenstein}, Daniel J. and {Hainline}, Kevin and {Johnson}, Benjamin D. and {Baker}, William M. and {Helton}, Jakob M. and {Robertson}, Brant and {Suess}, Katherine A. and {Chen}, Zuyi and {Nelson}, Erica and {Pusk{\'a}s}, D{\'a}vid and {Sun}, Fengwu and {Alberts}, Stacey and {Egami}, Eiichi and {Hausen}, Ryan and {Rieke}, George and {Rieke}, Marcia and {Shivaei}, Irene and {Williams}, Christina C. and {Willmer}, Christopher N.~A. and {Bunker}, Andrew and {Cameron}, Alex J. and {Carniani}, Stefano and {Charlot}, Stephane and {Curti}, Mirko and {Curtis-Lake}, Emma and {Looser}, Tobias J. and {Maiolino}, Roberto and {Maseda}, Michael V. and {Rawle}, Tim and {Rix}, Hans-Walter and {Smit}, Renske and {{\"U}bler}, Hannah and {Willott}, Chris and {Witstok}, Joris and {Baum}, Stefi and {Bhatawdekar}, Rachana and {Boyett}, Kristan and {Danhaive}, A. Lola and {de Graaff}, Anna and {Endsley}, Ryan and {Ji}, Zhiyuan and {Lyu}, Jianwei and {Sandles}, Lester and {Saxena}, Aayush and {Scholtz}, Jan and {Topping}, Michael W. and {Whitler}, Lily},
	doi = {10.3847/1538-4357/acdbc6},
	eid = {74},
	eprint = {2302.07234},
	journal = {\apj},
	month = jul,
	number = {1},
	pages = {74},
	primaryclass = {astro-ph.GA},
	title = {{JADES Imaging of GN-z11: Revealing the Morphology and Environment of a Luminous Galaxy 430 Myr after the Big Bang}},
	volume = {952},
	year = 2023}

@article{Cameron:2023ab,
	adsurl = {https://ui.adsabs.harvard.edu/abs/2023MNRAS.523.3516C},
	archiveprefix = {arXiv},
	author = {{Cameron}, Alex J. and {Katz}, Harley and {Rey}, Martin P. and {Saxena}, Aayush},
	doi = {10.1093/mnras/stad1579},
	eprint = {2302.10142},
	journal = {\mnras},
	month = aug,
	number = {3},
	pages = {3516-3525},
	primaryclass = {astro-ph.GA},
	title = {{Nitrogen enhancements 440 Myr after the big bang: supersolar N/O, a tidal disruption event, or a dense stellar cluster in GN-z11?}},
	volume = {523},
	year = 2023}

@article{Maiolino:2024aa,
	adsurl = {https://ui.adsabs.harvard.edu/abs/2024Natur.627...59M},
	archiveprefix = {arXiv},
	author = {{Maiolino}, Roberto and {Scholtz}, Jan and {Witstok}, Joris and {Carniani}, Stefano and {D'Eugenio}, Francesco and {de Graaff}, Anna and {{\"U}bler}, Hannah and {Tacchella}, Sandro and {Curtis-Lake}, Emma and {Arribas}, Santiago and {Bunker}, Andrew and {Charlot}, St{\'e}phane and {Chevallard}, Jacopo and {Curti}, Mirko and {Looser}, Tobias J. and {Maseda}, Michael V. and {Rawle}, Timothy D. and {Rodr{\'\i}guez del Pino}, Bruno and {Willott}, Chris J. and {Egami}, Eiichi and {Eisenstein}, Daniel J. and {Hainline}, Kevin N. and {Robertson}, Brant and {Williams}, Christina C. and {Willmer}, Christopher N.~A. and {Baker}, William M. and {Boyett}, Kristan and {DeCoursey}, Christa and {Fabian}, Andrew C. and {Helton}, Jakob M. and {Ji}, Zhiyuan and {Jones}, Gareth C. and {Kumari}, Nimisha and {Laporte}, Nicolas and {Nelson}, Erica J. and {Perna}, Michele and {Sandles}, Lester and {Shivaei}, Irene and {Sun}, Fengwu},
	doi = {10.1038/s41586-024-07052-5},
	eprint = {2305.12492},
	journal = {\nat},
	month = mar,
	number = {8002},
	pages = {59-63},
	primaryclass = {astro-ph.GA},
	title = {{A small and vigorous black hole in the early Universe}},
	volume = {627},
	year = 2024}

@article{Schaerer:2024aa,
	adsurl = {https://ui.adsabs.harvard.edu/abs/2024A&A...687L..11S},
	archiveprefix = {arXiv},
	author = {{Schaerer}, D. and {Marques-Chaves}, R. and {Xiao}, M. and {Korber}, D.},
	doi = {10.1051/0004-6361/202450721},
	eid = {L11},
	eprint = {2406.08408},
	journal = {\aap},
	month = jul,
	pages = {L11},
	primaryclass = {astro-ph.GA},
	title = {{Discovery of a new N-emitter in the epoch of reionization}},
	volume = {687},
	year = 2024}

@article{Curti:2025aa,
	adsurl = {https://ui.adsabs.harvard.edu/abs/2025A&A...697A..89C},
	archiveprefix = {arXiv},
	author = {{Curti}, Mirko and {Witstok}, Joris and {Jakobsen}, Peter and {Kobayashi}, Chiaki and {Curtis-Lake}, Emma and {Hainline}, Kevin and {Ji}, Xihan and {D'Eugenio}, Francesco and {Chevallard}, Jacopo and {Maiolino}, Roberto and {Scholtz}, Jan and {Carniani}, Stefano and {Arribas}, Santiago and {Baker}, William M. and {Bhatawdekar}, Rachana and {Boyett}, Kristan and {Bunker}, Andrew J. and {Cameron}, Alex and {Cargile}, Phillip A. and {Charlot}, St{\'e}phane and {Eisenstein}, Daniel J. and {Ji}, Zhiyuan and {Johnson}, Benjamin D. and {Kumari}, Nimisha and {Maseda}, Michael V. and {Robertson}, Brant and {Silcock}, Maddie S. and {Tacchella}, Sandro and {{\"U}bler}, Hannah and {Venturi}, Giacomo and {Williams}, Christina C. and {Willmer}, Christopher N.~A. and {Willott}, Chris},
	doi = {10.1051/0004-6361/202451410},
	eid = {A89},
	eprint = {2407.02575},
	journal = {\aap},
	month = may,
	pages = {A89},
	primaryclass = {astro-ph.GA},
	title = {{JADES: The star formation and chemical enrichment history of a luminous galaxy at z {\ensuremath{\sim}} 9.43 probed by ultra-deep JWST/NIRSpec spectroscopy}},
	volume = {697},
	year = 2025}

@article{Larson:2023aa,
	adsurl = {https://ui.adsabs.harvard.edu/abs/2023ApJ...953L..29L},
	archiveprefix = {arXiv},
	author = {{Larson}, Rebecca L. and {Finkelstein}, Steven L. and {Kocevski}, Dale D. and {Hutchison}, Taylor A. and {Trump}, Jonathan R. and {Arrabal Haro}, Pablo and {Bromm}, Volker and {Cleri}, Nikko J. and {Dickinson}, Mark and {Fujimoto}, Seiji and {Kartaltepe}, Jeyhan S. and {Koekemoer}, Anton M. and {Papovich}, Casey and {Pirzkal}, Nor and {Tacchella}, Sandro and {Zavala}, Jorge A. and {Bagley}, Micaela and {Behroozi}, Peter and {Champagne}, Jaclyn B. and {Cole}, Justin W. and {Jung}, Intae and {Morales}, Alexa M. and {Yang}, Guang and {Zhang}, Haowen and {Zitrin}, Adi and {Amor{\'\i}n}, Ricardo O. and {Burgarella}, Denis and {Casey}, Caitlin M. and {Ch{\'a}vez Ortiz}, {\'O}scar A. and {Cox}, Isabella G. and {Chworowsky}, Katherine and {Fontana}, Adriano and {Gawiser}, Eric and {Grazian}, Andrea and {Grogin}, Norman A. and {Harish}, Santosh and {Hathi}, Nimish P. and {Hirschmann}, Michaela and {Holwerda}, Benne W. and {Juneau}, St{\'e}phanie and {Leung}, Gene C.~K. and {Lucas}, Ray A. and {McGrath}, Elizabeth J. and {P{\'e}rez-Gonz{\'a}lez}, Pablo G. and {Rigby}, Jane R. and {Seill{\'e}}, Lise-Marie and {Simons}, Raymond C. and {de La Vega}, Alexander and {Weiner}, Benjamin J. and {Wilkins}, Stephen M. and {Yung}, L.~Y. Aaron and {Ceers Team}},
	doi = {10.3847/2041-8213/ace619},
	eid = {L29},
	eprint = {2303.08918},
	journal = {\apjl},
	month = aug,
	number = {2},
	pages = {L29},
	primaryclass = {astro-ph.GA},
	title = {{A CEERS Discovery of an Accreting Supermassive Black Hole 570 Myr after the Big Bang: Identifying a Progenitor of Massive z > 6 Quasars}},
	volume = {953},
	year = 2023}

@article{Cameron:2024aa,
	adsurl = {https://ui.adsabs.harvard.edu/abs/2024MNRAS.534..523C},
	archiveprefix = {arXiv},
	author = {{Cameron}, Alex J. and {Katz}, Harley and {Witten}, Callum and {Saxena}, Aayush and {Laporte}, Nicolas and {Bunker}, Andrew J.},
	doi = {10.1093/mnras/stae1547},
	eprint = {2311.02051},
	journal = {\mnras},
	month = oct,
	number = {1},
	pages = {523-543},
	primaryclass = {astro-ph.GA},
	title = {{Nebular dominated galaxies: insights into the stellar initial mass function at high redshift}},
	volume = {534},
	year = 2024}

@article{Ubler:2023aa,
	adsurl = {https://ui.adsabs.harvard.edu/abs/2023A&A...677A.145U},
	archiveprefix = {arXiv},
	author = {{{\"U}bler}, Hannah and {Maiolino}, Roberto and {Curtis-Lake}, Emma and {P{\'e}rez-Gonz{\'a}lez}, Pablo G. and {Curti}, Mirko and {Perna}, Michele and {Arribas}, Santiago and {Charlot}, St{\'e}phane and {Marshall}, Madeline A. and {D'Eugenio}, Francesco and {Scholtz}, Jan and {Bunker}, Andrew and {Carniani}, Stefano and {Ferruit}, Pierre and {Jakobsen}, Peter and {Rix}, Hans-Walter and {Rodr{\'\i}guez Del Pino}, Bruno and {Willott}, Chris J. and {Boeker}, Torsten and {Cresci}, Giovanni and {Jones}, Gareth C. and {Kumari}, Nimisha and {Rawle}, Tim},
	doi = {10.1051/0004-6361/202346137},
	eid = {A145},
	eprint = {2302.06647},
	journal = {\aap},
	month = sep,
	pages = {A145},
	primaryclass = {astro-ph.GA},
	title = {{GA-NIFS: A massive black hole in a low-metallicity AGN at z {\ensuremath{\sim}} 5.55 revealed by JWST/NIRSpec IFS}},
	volume = {677},
	year = 2023}

@article{Vanzella:2010aa,
	adsurl = {https://ui.adsabs.harvard.edu/abs/2010A&A...513A..20V},
	archiveprefix = {arXiv},
	author = {{Vanzella}, E. and {Grazian}, A. and {Hayes}, M. and {Pentericci}, L. and {Schaerer}, D. and {Dickinson}, M. and {Cristiani}, S. and {Giavalisco}, M. and {Verhamme}, A. and {Nonino}, M. and {Rosati}, P.},
	doi = {10.1051/0004-6361/200913042},
	eid = {A20},
	eprint = {0912.3007},
	journal = {\aap},
	month = apr,
	pages = {A20},
	primaryclass = {astro-ph.CO},
	title = {{The unusual N IV] -emitter galaxy GDS J033218.92-275302.7: star formation or AGN-driven winds from a massive galaxy at z = 5.56}},
	volume = {513},
	year = 2010}

@article{Zhang:2025aa,
	adsurl = {https://ui.adsabs.harvard.edu/abs/2025arXiv250204817Z},
	archiveprefix = {arXiv},
	author = {{Zhang}, Yechi and {Morishita}, Takahiro and {Stiavelli}, Massimo},
	doi = {10.48550/arXiv.2502.04817},
	eid = {arXiv:2502.04817},
	eprint = {2502.04817},
	journal = {arXiv e-prints},
	month = feb,
	pages = {arXiv:2502.04817},
	primaryclass = {astro-ph.GA},
	title = {{Potential Nitrogen Enrichment via Direct-Collapse Wolf-Rayet Stars in a $z=4.7$ Star-Forming Galaxy}},
	year = 2025}

@article{Labbe:2024aa,
	adsurl = {https://ui.adsabs.harvard.edu/abs/2024arXiv241204557L},
	archiveprefix = {arXiv},
	author = {{Labbe}, Ivo and {Greene}, Jenny E. and {Matthee}, Jorryt and {Treiber}, Helena and {Kokorev}, Vasily and {Miller}, Tim B. and {Kramarenko}, Ivan and {Setton}, David J. and {Ma}, Yilun and {Goulding}, Andy D. and {Bezanson}, Rachel and {Naidu}, Rohan P. and {Williams}, Christina C. and {Atek}, Hakim and {Brammer}, Gabriel and {Cutler}, Sam E. and {Chemerynska}, Iryna and {Cloonan}, Aidan P. and {Dayal}, Pratika and {de Graaff}, Anna and {Fudamoto}, Yoshinobu and {Fujimoto}, Seiji and {Furtak}, Lukas J. and {Glazebrook}, Karl and {Heintz}, Kasper E. and {Leja}, Joel and {Marchesini}, Danilo and {Nanayakkara}, Themiya and {Nelson}, Erica J. and {Oesch}, Pascal A. and {Pan}, Richard and {Price}, Sedona H. and {Shivaei}, Irene and {Sobral}, David and {Suess}, Katherine A. and {van Dokkum}, Pieter and {Wang}, Bingjie and {Weaver}, John R. and {Whitaker}, Katherine E. and {Zitrin}, Adi},
	doi = {10.48550/arXiv.2412.04557},
	eid = {arXiv:2412.04557},
	eprint = {2412.04557},
	journal = {arXiv e-prints},
	month = dec,
	pages = {arXiv:2412.04557},
	primaryclass = {astro-ph.GA},
	title = {{An unambiguous AGN and a Balmer break in an Ultraluminous Little Red Dot at z=4.47 from Ultradeep UNCOVER and All the Little Things Spectroscopy}},
	year = 2024}

@article{Vink:2023aa,
	adsurl = {https://ui.adsabs.harvard.edu/abs/2023A&A...679L...9V},
	archiveprefix = {arXiv},
	author = {{Vink}, Jorick S.},
	doi = {10.1051/0004-6361/202347827},
	eid = {L9},
	eprint = {2310.10725},
	journal = {\aap},
	month = nov,
	pages = {L9},
	primaryclass = {astro-ph.GA},
	title = {{Very massive stars and nitrogen-emitting galaxies}},
	volume = {679},
	year = 2023}

@article{Tsiatsiou:2024aa,
	adsurl = {https://ui.adsabs.harvard.edu/abs/2024A&A...687A.307T},
	archiveprefix = {arXiv},
	author = {{Tsiatsiou}, Sophie and {Sibony}, Yves and {Nandal}, Devesh and {Sciarini}, Luca and {Hirai}, Yutaka and {Ekstr{\"o}m}, Sylvia and {Farrell}, Eoin and {Murphy}, Laura and {Choplin}, Arthur and {Hirschi}, Raphael and {Chiappini}, Cristina and {Liu}, Boyuan and {Bromm}, Volker and {Groh}, Jose and {Meynet}, Georges},
	doi = {10.1051/0004-6361/202449156},
	eid = {A307},
	eprint = {2404.16512},
	journal = {\aap},
	month = jul,
	pages = {A307},
	primaryclass = {astro-ph.SR},
	title = {{Rapidly rotating Population III stellar models as a source of primary nitrogen}},
	volume = {687},
	year = 2024}

@article{Nagele:2023aa,
	adsurl = {https://ui.adsabs.harvard.edu/abs/2023ApJ...949L..16N},
	archiveprefix = {arXiv},
	author = {{Nagele}, Chris and {Umeda}, Hideyuki},
	doi = {10.3847/2041-8213/acd550},
	eid = {L16},
	eprint = {2304.05013},
	journal = {\apjl},
	month = may,
	number = {1},
	pages = {L16},
	primaryclass = {astro-ph.GA},
	title = {{Multiple Channels for Nitrogen Pollution by Metal-enriched Supermassive Stars and Implications for GN-z11}},
	volume = {949},
	year = 2023}

@article{Charbonnel:2023aa,
	adsurl = {https://ui.adsabs.harvard.edu/abs/2023A&A...673L...7C},
	archiveprefix = {arXiv},
	author = {{Charbonnel}, C. and {Schaerer}, D. and {Prantzos}, N. and {Ram{\'\i}rez-Galeano}, L. and {Fragos}, T. and {Kuruvanthodi}, A. and {Marques-Chaves}, R. and {Gieles}, M.},
	doi = {10.1051/0004-6361/202346410},
	eid = {L7},
	eprint = {2303.07955},
	journal = {\aap},
	month = may,
	pages = {L7},
	primaryclass = {astro-ph.GA},
	title = {{N-enhancement in GN-z11: First evidence for supermassive stars nucleosynthesis in proto-globular clusters-like conditions at high redshift?}},
	volume = {673},
	year = 2023}

@article{Nandal:2024ab,
	adsurl = {https://ui.adsabs.harvard.edu/abs/2024A&A...688A.142N},
	archiveprefix = {arXiv},
	author = {{Nandal}, Devesh and {Sibony}, Yves and {Tsiatsiou}, Sophie},
	doi = {10.1051/0004-6361/202348866},
	eid = {A142},
	eprint = {2405.11235},
	journal = {\aap},
	month = aug,
	pages = {A142},
	primaryclass = {astro-ph.GA},
	title = {{Fast-rotating massive Population III stars as possible sources of extreme N enrichment in high-redshift galaxies}},
	volume = {688},
	year = 2024}

@article{Nandal:2024aa,
	adsurl = {https://ui.adsabs.harvard.edu/abs/2024A&A...683A.156N},
	archiveprefix = {arXiv},
	author = {{Nandal}, Devesh and {Regan}, John A. and {Woods}, Tyrone E. and {Farrell}, Eoin and {Ekstr{\"o}m}, Sylvia and {Meynet}, Georges},
	doi = {10.1051/0004-6361/202348035},
	eid = {A156},
	eprint = {2402.03428},
	journal = {\aap},
	month = mar,
	pages = {A156},
	primaryclass = {astro-ph.GA},
	title = {{Explaining the high nitrogen abundances observed in high-z galaxies via population III stars of a few thousand solar masses}},
	volume = {683},
	year = 2024}

@article{Watanabe:2024aa,
	adsurl = {https://ui.adsabs.harvard.edu/abs/2024ApJ...962...50W},
	archiveprefix = {arXiv},
	author = {{Watanabe}, Kuria and {Ouchi}, Masami and {Nakajima}, Kimihiko and {Isobe}, Yuki and {Tominaga}, Nozomu and {Suzuki}, Akihiro and {Ishigaki}, Miho N. and {Nomoto}, Ken'ichi and {Takahashi}, Koh and {Harikane}, Yuichi and {Hatano}, Shun and {Kusakabe}, Haruka and {Moriya}, Takashi J. and {Nishigaki}, Moka and {Ono}, Yoshiaki and {Onodera}, Masato and {Sugahara}, Yuma},
	doi = {10.3847/1538-4357/ad13ff},
	eid = {50},
	eprint = {2305.02078},
	journal = {\apj},
	month = feb,
	number = {1},
	pages = {50},
	primaryclass = {astro-ph.GA},
	title = {{EMPRESS. XIII. Chemical Enrichment of Young Galaxies Near and Far at z {\ensuremath{\sim}} 0 and 4{\textendash}10: Fe/O, Ar/O, S/O, and N/O Measurements with a Comparison of Chemical Evolution Models}},
	volume = {962},
	year = 2024}

@article{Rizzuti:2025aa,
	adsurl = {https://ui.adsabs.harvard.edu/abs/2025A&A...697A..96R},
	archiveprefix = {arXiv},
	author = {{Rizzuti}, F. and {Matteucci}, F. and {Molaro}, P. and {Cescutti}, G. and {Maiolino}, R.},
	doi = {10.1051/0004-6361/202453275},
	eid = {A96},
	eprint = {2412.05363},
	journal = {\aap},
	month = may,
	pages = {A96},
	primaryclass = {astro-ph.GA},
	title = {{High N/O ratio at high redshift as a result of a strong burst of star formation and differential galactic winds}},
	volume = {697},
	year = 2025}

@article{DAntona:2023aa,
	adsurl = {https://ui.adsabs.harvard.edu/abs/2023A&A...680L..19D},
	archiveprefix = {arXiv},
	author = {{D'Antona}, F. and {Vesperini}, E. and {Calura}, F. and {Ventura}, P. and {D'Ercole}, A. and {Caloi}, V. and {Marino}, A.~F. and {Milone}, A.~P. and {Dell'Agli}, F. and {Tailo}, M.},
	doi = {10.1051/0004-6361/202348240},
	eid = {L19},
	eprint = {2311.14558},
	journal = {\aap},
	month = dec,
	pages = {L19},
	primaryclass = {astro-ph.SR},
	title = {{GN-z11: Witnessing the formation of second-generation stars and an accreting massive black hole in a massive star cluster}},
	volume = {680},
	year = 2023}

@article{Curti:2024ab,
	adsurl = {https://ui.adsabs.harvard.edu/abs/2024A&A...684A..75C},
	archiveprefix = {arXiv},
	author = {{Curti}, Mirko and {Maiolino}, Roberto and {Curtis-Lake}, Emma and {Chevallard}, Jacopo and {Carniani}, Stefano and {D'Eugenio}, Francesco and {Looser}, Tobias J. and {Scholtz}, Jan and {Charlot}, Stephane and {Cameron}, Alex and {{\"U}bler}, Hannah and {Witstok}, Joris and {Boyett}, Kristian and {Laseter}, Isaac and {Sandles}, Lester and {Arribas}, Santiago and {Bunker}, Andrew and {Giardino}, Giovanna and {Maseda}, Michael V. and {Rawle}, Tim and {Rodr{\'\i}guez Del Pino}, Bruno and {Smit}, Renske and {Willott}, Chris J. and {Eisenstein}, Daniel J. and {Hausen}, Ryan and {Johnson}, Benjamin and {Rieke}, Marcia and {Robertson}, Brant and {Tacchella}, Sandro and {Williams}, Christina C. and {Willmer}, Christopher and {Baker}, William M. and {Bhatawdekar}, Rachana and {Egami}, Eiichi and {Helton}, Jakob M. and {Ji}, Zhiyuan and {Kumari}, Nimisha and {Perna}, Michele and {Shivaei}, Irene and {Sun}, Fengwu},
	doi = {10.1051/0004-6361/202346698},
	eid = {A75},
	eprint = {2304.08516},
	journal = {\aap},
	month = apr,
	pages = {A75},
	primaryclass = {astro-ph.GA},
	title = {{JADES: Insights into the low-mass end of the mass-metallicity-SFR relation at 3 < z < 10 from deep JWST/NIRSpec spectroscopy}},
	volume = {684},
	year = 2024}

@article{Katz:2023ab,
	adsurl = {https://ui.adsabs.harvard.edu/abs/2023OJAp....6E..44K},
	archiveprefix = {arXiv},
	author = {{Katz}, Harley and {Rosdahl}, Joki and {Kimm}, Taysun and {Blaizot}, Jeremy and {Choustikov}, Nicholas and {Farcy}, Marion and {Garel}, Thibault and {Haehnelt}, Martin G. and {Michel-Dansac}, Leo and {Ocvirk}, Pierre},
	doi = {10.21105/astro.2309.03269},
	eid = {44},
	eprint = {2309.03269},
	journal = {The Open Journal of Astrophysics},
	month = dec,
	pages = {44},
	primaryclass = {astro-ph.GA},
	title = {{The SPHINX Public Data Release: Forward Modelling High-Redshift JWST Observations with Cosmological Radiation Hydrodynamics Simulations}},
	volume = {6},
	year = 2023}

@article{Yanagisawa:2024aa,
	adsurl = {https://ui.adsabs.harvard.edu/abs/2024ApJ...974..180Y},
	archiveprefix = {arXiv},
	author = {{Yanagisawa}, Hiroto and {Ouchi}, Masami and {Nakajima}, Kimihiko and {Yajima}, Hidenobu and {Umeda}, Hiroya and {Baba}, Shunsuke and {Nakagawa}, Takao and {Nakane}, Minami and {Matsumoto}, Akinori and {Ono}, Yoshiaki and {Harikane}, Yuichi and {Isobe}, Yuki and {Xu}, Yi and {Zhang}, Yechi},
	doi = {10.3847/1538-4357/ad7097},
	eid = {180},
	eprint = {2403.20118},
	journal = {\apj},
	month = oct,
	number = {2},
	pages = {180},
	primaryclass = {astro-ph.GA},
	title = {{Balmer Decrement Anomalies in Galaxies at z {\ensuremath{\sim}} 6 Found by JWST Observations: Density-bounded Nebulae or Excited H I Clouds?}},
	volume = {974},
	year = 2024}

@article{Ji:2024aa,
	adsurl = {https://ui.adsabs.harvard.edu/abs/2024MNRAS.535..881J},
	archiveprefix = {arXiv},
	author = {{Ji}, Xihan and {{\"U}bler}, Hannah and {Maiolino}, Roberto and {D'Eugenio}, Francesco and {Arribas}, Santiago and {Bunker}, Andrew J. and {Charlot}, St{\'e}phane and {Perna}, Michele and {Rodr{\'\i}guez Del Pino}, Bruno and {B{\"o}ker}, Torsten and {Cresci}, Giovanni and {Curti}, Mirko and {Kumari}, Nimisha and {Lamperti}, Isabella},
	doi = {10.1093/mnras/stae2375},
	eprint = {2404.04148},
	journal = {\mnras},
	month = nov,
	number = {1},
	pages = {881-908},
	primaryclass = {astro-ph.GA},
	title = {{GA-NIFS: an extremely nitrogen-loud and chemically stratified galaxy at z 5.55}},
	volume = {535},
	year = 2024}

@article{Shen:2024aa,
	adsurl = {https://ui.adsabs.harvard.edu/abs/2024MNRAS.534.1433S},
	archiveprefix = {arXiv},
	author = {{Shen}, Xuejian and {Vogelsberger}, Mark and {Borrow}, Josh and {Hu}, Yongao and {Erickson}, Evan and {Kannan}, Rahul and {Smith}, Aaron and {Garaldi}, Enrico and {Hernquist}, Lars and {Morishita}, Takahiro and {Tacchella}, Sandro and {Zier}, Oliver and {Sun}, Guochao and {Eilers}, Anna-Christina and {Wang}, Hui},
	doi = {10.1093/mnras/stae2156},
	eprint = {2402.08717},
	journal = {\mnras},
	month = oct,
	number = {2},
	pages = {1433-1458},
	primaryclass = {astro-ph.GA},
	title = {{The THESAN project: galaxy sizes during the epoch of reionization}},
	volume = {534},
	year = 2024}

@article{Seab:1983aa,
	adsurl = {https://ui.adsabs.harvard.edu/abs/1983ApJ...275..652S},
	author = {{Seab}, C.~G. and {Shull}, J.~M.},
	doi = {10.1086/161563},
	journal = {\apj},
	month = dec,
	pages = {652-660},
	title = {{Shock processing of interstellar grains.}},
	volume = {275},
	year = 1983}

@article{Draine:1979aa,
	adsurl = {https://ui.adsabs.harvard.edu/abs/1979ApJ...231..438D},
	author = {{Draine}, B.~T. and {Salpeter}, E.~E.},
	doi = {10.1086/157206},
	journal = {\apj},
	month = jul,
	pages = {438-455},
	title = {{Destruction mechanisms for interstellar dust.}},
	volume = {231},
	year = 1979}

@article{Dwek:1998aa,
	adsurl = {https://ui.adsabs.harvard.edu/abs/1998ApJ...501..643D},
	archiveprefix = {arXiv},
	author = {{Dwek}, Eli},
	doi = {10.1086/305829},
	eprint = {astro-ph/9707024},
	journal = {\apj},
	month = jul,
	pages = {643},
	primaryclass = {astro-ph},
	title = {{The Evolution of the Elemental Abundances in the Gas and Dust Phases of the Galaxy}},
	volume = {501},
	year = 1998}

@article{Travaglio:2004aa,
	adsurl = {https://ui.adsabs.harvard.edu/abs/2004A&A...425.1029T},
	archiveprefix = {arXiv},
	author = {{Travaglio}, C. and {Hillebrandt}, W. and {Reinecke}, M. and {Thielemann}, F. -K.},
	doi = {10.1051/0004-6361:20041108},
	eprint = {astro-ph/0406281},
	journal = {\aap},
	month = oct,
	pages = {1029-1040},
	primaryclass = {astro-ph},
	title = {{Nucleosynthesis in multi-dimensional SN Ia explosions}},
	volume = {425},
	year = 2004}

@article{Thielemann:2003aa,
	adsurl = {https://ui.adsabs.harvard.edu/abs/2003NuPhA.718..139T},
	author = {{Thielemann}, F. -K. and {Argast}, D. and {Brachwitz}, F. and {Hix}, W.~R. and {H{\"o}flich}, P. and {Liebend{\"o}rfer}, M. and {Martinez-Pinedo}, G. and {Mezzacappa}, A. and {Panov}, I. and {Rauscher}, T.},
	doi = {10.1016/S0375-9474(03)00704-8},
	journal = {\nphysa},
	month = may,
	pages = {139-146},
	title = {{Nuclear cross sections, nuclear structure and stellar nucleosynthesis}},
	volume = {718},
	year = 2003}

@article{Karakas:2010aa,
	adsurl = {https://ui.adsabs.harvard.edu/abs/2010MNRAS.403.1413K},
	archiveprefix = {arXiv},
	author = {{Karakas}, A.~I.},
	doi = {10.1111/j.1365-2966.2009.16198.x},
	eprint = {0912.2142},
	journal = {\mnras},
	month = apr,
	number = {3},
	pages = {1413-1425},
	primaryclass = {astro-ph.SR},
	title = {{Updated stellar yields from asymptotic giant branch models}},
	volume = {403},
	year = 2010}

@article{Portinari:1998aa,
	adsurl = {https://ui.adsabs.harvard.edu/abs/1998A&A...334..505P},
	archiveprefix = {arXiv},
	author = {{Portinari}, L. and {Chiosi}, C. and {Bressan}, A.},
	doi = {10.48550/arXiv.astro-ph/9711337},
	eprint = {astro-ph/9711337},
	journal = {\aap},
	month = jun,
	pages = {505-539},
	primaryclass = {astro-ph},
	title = {{Galactic chemical enrichment with new metallicity dependent stellar yields}},
	volume = {334},
	year = 1998}

@article{Witten:2025aa,
	adsurl = {https://ui.adsabs.harvard.edu/abs/2025MNRAS.537..112W},
	archiveprefix = {arXiv},
	author = {{Witten}, Callum and {McClymont}, William and {Laporte}, Nicolas and {Roberts-Borsani}, Guido and {Sijacki}, Debora and {Tacchella}, Sandro and {Simmonds}, Charlotte and {Katz}, Harley and {Ellis}, Richard S. and {Witstok}, Joris and {Maiolino}, Roberto and {Ji}, Xihan and {Hayes}, Billy R. and {Looser}, Tobias J. and {D'Eugenio}, Francesco},
	doi = {10.1093/mnras/staf001},
	eprint = {2407.07937},
	journal = {\mnras},
	month = feb,
	number = {1},
	pages = {112-126},
	primaryclass = {astro-ph.GA},
	title = {{Rising from the ashes: evidence of old stellar populations and rejuvenation events in the very early Universe}},
	volume = {537},
	year = 2025}

@article{Scarlata:2024aa,
	adsurl = {https://ui.adsabs.harvard.edu/abs/2024arXiv240409015S},
	archiveprefix = {arXiv},
	author = {{Scarlata}, C. and {Hayes}, M. and {Panagia}, N. and {Mehta}, V. and {Haardt}, F. and {Bagley}, M.},
	doi = {10.48550/arXiv.2404.09015},
	eid = {arXiv:2404.09015},
	eprint = {2404.09015},
	journal = {arXiv e-prints},
	month = apr,
	pages = {arXiv:2404.09015},
	primaryclass = {astro-ph.GA},
	title = {{On the universal validity of Case B recombination theory}},
	year = 2024}

@article{McClymont:2024aa,
	adsurl = {https://ui.adsabs.harvard.edu/abs/2024MNRAS.532.2016M},
	archiveprefix = {arXiv},
	author = {{McClymont}, William and {Tacchella}, Sandro and {Smith}, Aaron and {Kannan}, Rahul and {Maiolino}, Roberto and {Belfiore}, Francesco and {Hernquist}, Lars and {Li}, Hui and {Vogelsberger}, Mark},
	doi = {10.1093/mnras/stae1587},
	eprint = {2403.03243},
	journal = {\mnras},
	month = aug,
	number = {2},
	pages = {2016-2031},
	primaryclass = {astro-ph.GA},
	title = {{The nature of diffuse ionized gas in star-forming galaxies}},
	volume = {532},
	year = 2024}

@article{Sanders:2017aa,
	adsurl = {https://ui.adsabs.harvard.edu/abs/2017ApJ...850..136S},
	archiveprefix = {arXiv},
	author = {{Sanders}, Ryan L. and {Shapley}, Alice E. and {Zhang}, Kai and {Yan}, Renbin},
	doi = {10.3847/1538-4357/aa93e4},
	eid = {136},
	eprint = {1708.04625},
	journal = {\apj},
	month = dec,
	number = {2},
	pages = {136},
	primaryclass = {astro-ph.GA},
	title = {{Biases in Metallicity Measurements from Global Galaxy Spectra: The Effects of Flux Weighting and Diffuse Ionized Gas Contamination}},
	volume = {850},
	year = 2017}

@article{McCallum:2024ab,
	adsurl = {https://ui.adsabs.harvard.edu/abs/2024MNRAS.535.2889M},
	archiveprefix = {arXiv},
	author = {{McCallum}, Lewis and {Wood}, Kenneth and {Benjamin}, Robert and {Krishnarao}, Dhanesh and {Vandenbroucke}, Bert},
	doi = {10.1093/mnras/stae2549},
	eprint = {2411.07108},
	journal = {\mnras},
	month = dec,
	number = {3},
	pages = {2889-2902},
	primaryclass = {astro-ph.GA},
	title = {{Time-dependent metal ionization and the persistence of collisionally excited emission lines in the diffuse ionized gas of star-forming galaxies}},
	volume = {535},
	year = 2024}

@article{McCallum:2024aa,
	adsurl = {https://ui.adsabs.harvard.edu/abs/2024MNRAS.530.2548M},
	archiveprefix = {arXiv},
	author = {{McCallum}, Lewis and {Wood}, Kenneth and {Benjamin}, Robert and {Pe{\~n}aloza}, Camilo and {Krishnarao}, Dhanesh and {Smith}, Rowan and {Vandenbroucke}, Bert},
	doi = {10.1093/mnras/stae988},
	eprint = {2404.05651},
	journal = {\mnras},
	month = may,
	number = {3},
	pages = {2548-2564},
	primaryclass = {astro-ph.GA},
	title = {{The persistence of high altitude non-equilibrium diffuse ionized gas in simulations of star-forming galaxies}},
	volume = {530},
	year = 2024}

@article{Harikane:2025aa,
	adsurl = {https://ui.adsabs.harvard.edu/abs/2025ApJ...980..138H},
	archiveprefix = {arXiv},
	author = {{Harikane}, Yuichi and {Inoue}, Akio K. and {Ellis}, Richard S. and {Ouchi}, Masami and {Nakazato}, Yurina and {Yoshida}, Naoki and {Ono}, Yoshiaki and {Sun}, Fengwu and {Sato}, Riku A. and {Ferrami}, Giovanni and {Fujimoto}, Seiji and {Kashikawa}, Nobunari and {McLeod}, Derek J. and {P{\'e}rez-Gonz{\'a}lez}, Pablo G. and {Sawicki}, Marcin and {Sugahara}, Yuma and {Xu}, Yi and {Yamanaka}, Satoshi and {Carnall}, Adam C. and {Cullen}, Fergus and {Dunlop}, James S. and {Egami}, Eiichi and {Grogin}, Norman and {Isobe}, Yuki and {Koekemoer}, Anton M. and {Laporte}, Nicolas and {Lee}, Chien-Hsiu and {Magee}, Dan and {Matsuo}, Hiroshi and {Matsuoka}, Yoshiki and {Mawatari}, Ken and {Nakajima}, Kimihiko and {Nakane}, Minami and {Tamura}, Yoichi and {Umeda}, Hiroya and {Yanagisawa}, Hiroto},
	doi = {10.3847/1538-4357/ad9b2c},
	eid = {138},
	eprint = {2406.18352},
	journal = {\apj},
	month = feb,
	number = {1},
	pages = {138},
	primaryclass = {astro-ph.GA},
	title = {{JWST, ALMA, and Keck Spectroscopic Constraints on the UV Luminosity Functions at z {\ensuremath{\sim}} 7{\textendash}14: Clumpiness and Compactness of the Brightest Galaxies in the Early Universe}},
	volume = {980},
	year = 2025}

@article{Yanagisawa:2024ab,
	adsurl = {https://ui.adsabs.harvard.edu/abs/2024ApJ...974..266Y},
	archiveprefix = {arXiv},
	author = {{Yanagisawa}, Hiroto and {Ouchi}, Masami and {Watanabe}, Kuria and {Matsumoto}, Akinori and {Nakajima}, Kimihiko and {Yajima}, Hidenobu and {Nagamine}, Kentaro and {Takahashi}, Koh and {Nakane}, Minami and {Tominaga}, Nozomu and {Umeda}, Hiroya and {Fukushima}, Hajime and {Harikane}, Yuichi and {Isobe}, Yuki and {Ono}, Yoshiaki and {Xu}, Yi and {Zhang}, Yechi},
	doi = {10.3847/1538-4357/ad72ec},
	eid = {266},
	eprint = {2405.01823},
	journal = {\apj},
	month = oct,
	number = {2},
	pages = {266},
	primaryclass = {astro-ph.GA},
	title = {{Strong He I Emission Lines in High N/O Galaxies at z {\ensuremath{\sim}} 6 Identified in JWST Spectra: High He/H Abundance Ratios or High Electron Densities?}},
	volume = {974},
	year = 2024}

@article{Isobe:2023aa,
	adsurl = {https://ui.adsabs.harvard.edu/abs/2023ApJ...959..100I},
	archiveprefix = {arXiv},
	author = {{Isobe}, Yuki and {Ouchi}, Masami and {Tominaga}, Nozomu and {Watanabe}, Kuria and {Nakajima}, Kimihiko and {Umeda}, Hiroya and {Yajima}, Hidenobu and {Harikane}, Yuichi and {Fukushima}, Hajime and {Xu}, Yi and {Ono}, Yoshiaki and {Zhang}, Yechi},
	doi = {10.3847/1538-4357/ad09be},
	eid = {100},
	eprint = {2307.00710},
	journal = {\apj},
	month = dec,
	number = {2},
	pages = {100},
	primaryclass = {astro-ph.GA},
	title = {{JWST Identification of Extremely Low C/N Galaxies with [N/O] {\ensuremath{\gtrsim}} 0.5 at z 6-10 Evidencing the Early CNO-cycle Enrichment and a Connection with Globular Cluster Formation}},
	volume = {959},
	year = 2023}

@article{Bunker:2023aa,
	adsurl = {https://ui.adsabs.harvard.edu/abs/2023A&A...677A..88B},
	archiveprefix = {arXiv},
	author = {{Bunker}, Andrew J. and {Saxena}, Aayush and {Cameron}, Alex J. and {Willott}, Chris J. and {Curtis-Lake}, Emma and {Jakobsen}, Peter and {Carniani}, Stefano and {Smit}, Renske and {Maiolino}, Roberto and {Witstok}, Joris and {Curti}, Mirko and {D'Eugenio}, Francesco and {Jones}, Gareth C. and {Ferruit}, Pierre and {Arribas}, Santiago and {Charlot}, Stephane and {Chevallard}, Jacopo and {Giardino}, Giovanna and {de Graaff}, Anna and {Looser}, Tobias J. and {L{\"u}tzgendorf}, Nora and {Maseda}, Michael V. and {Rawle}, Tim and {Rix}, Hans-Walter and {Del Pino}, Bruno Rodr{\'\i}guez and {Alberts}, Stacey and {Egami}, Eiichi and {Eisenstein}, Daniel J. and {Endsley}, Ryan and {Hainline}, Kevin and {Hausen}, Ryan and {Johnson}, Benjamin D. and {Rieke}, George and {Rieke}, Marcia and {Robertson}, Brant E. and {Shivaei}, Irene and {Stark}, Daniel P. and {Sun}, Fengwu and {Tacchella}, Sandro and {Tang}, Mengtao and {Williams}, Christina C. and {Willmer}, Christopher N.~A. and {Baker}, William M. and {Baum}, Stefi and {Bhatawdekar}, Rachana and {Bowler}, Rebecca and {Boyett}, Kristan and {Chen}, Zuyi and {Circosta}, Chiara and {Helton}, Jakob M. and {Ji}, Zhiyuan and {Kumari}, Nimisha and {Lyu}, Jianwei and {Nelson}, Erica and {Parlanti}, Eleonora and {Perna}, Michele and {Sandles}, Lester and {Scholtz}, Jan and {Suess}, Katherine A. and {Topping}, Michael W. and {{\"U}bler}, Hannah and {Wallace}, Imaan E.~B. and {Whitler}, Lily},
	doi = {10.1051/0004-6361/202346159},
	eid = {A88},
	eprint = {2302.07256},
	journal = {\aap},
	month = sep,
	pages = {A88},
	primaryclass = {astro-ph.GA},
	title = {{JADES NIRSpec Spectroscopy of GN-z11: Lyman-{\ensuremath{\alpha}} emission and possible enhanced nitrogen abundance in a z = 10.60 luminous galaxy}},
	volume = {677},
	year = 2023}

@article{Kobayashi:2024aa,
	adsurl = {https://ui.adsabs.harvard.edu/abs/2024ApJ...962L...6K},
	archiveprefix = {arXiv},
	author = {{Kobayashi}, Chiaki and {Ferrara}, Andrea},
	doi = {10.3847/2041-8213/ad1de1},
	eid = {L6},
	eprint = {2308.15583},
	journal = {\apjl},
	month = feb,
	number = {1},
	pages = {L6},
	primaryclass = {astro-ph.GA},
	title = {{Rapid Chemical Enrichment by Intermittent Star Formation in GN-z11}},
	volume = {962},
	year = 2024}

@article{Ellison:2008aa,
	adsurl = {https://ui.adsabs.harvard.edu/abs/2008ApJ...672L.107E},
	archiveprefix = {arXiv},
	author = {{Ellison}, Sara L. and {Patton}, David R. and {Simard}, Luc and {McConnachie}, Alan W.},
	doi = {10.1086/527296},
	eprint = {0711.4833},
	journal = {\apjl},
	month = jan,
	number = {2},
	pages = {L107},
	primaryclass = {astro-ph},
	title = {{Clues to the Origin of the Mass-Metallicity Relation: Dependence on Star Formation Rate and Galaxy Size}},
	volume = {672},
	year = 2008}

@article{Maoz:2010aa,
	adsurl = {https://ui.adsabs.harvard.edu/abs/2010ApJ...722.1879M},
	archiveprefix = {arXiv},
	author = {{Maoz}, Dan and {Sharon}, Keren and {Gal-Yam}, Avishay},
	doi = {10.1088/0004-637X/722/2/1879},
	eprint = {1006.3576},
	journal = {\apj},
	month = oct,
	number = {2},
	pages = {1879-1894},
	primaryclass = {astro-ph.CO},
	title = {{The Supernova Delay Time Distribution in Galaxy Clusters and Implications for Type-Ia Progenitors and Metal Enrichment}},
	volume = {722},
	year = 2010}

@article{Ma:2016aa,
	adsurl = {https://ui.adsabs.harvard.edu/abs/2016MNRAS.456.2140M},
	archiveprefix = {arXiv},
	author = {{Ma}, Xiangcheng and {Hopkins}, Philip F. and {Faucher-Gigu{\`e}re}, Claude-Andr{\'e} and {Zolman}, Nick and {Muratov}, Alexander L. and {Kere{\v{s}}}, Du{\v{s}}an and {Quataert}, Eliot},
	doi = {10.1093/mnras/stv2659},
	eprint = {1504.02097},
	journal = {\mnras},
	month = feb,
	number = {2},
	pages = {2140-2156},
	primaryclass = {astro-ph.GA},
	title = {{The origin and evolution of the galaxy mass-metallicity relation}},
	volume = {456},
	year = 2016}

@article{Marszewski:2024aa,
	adsurl = {https://ui.adsabs.harvard.edu/abs/2024ApJ...967L..41M},
	archiveprefix = {arXiv},
	author = {{Marszewski}, Andrew and {Sun}, Guochao and {Faucher-Gigu{\`e}re}, Claude-Andr{\'e} and {Hayward}, Christopher C. and {Feldmann}, Robert},
	doi = {10.3847/2041-8213/ad4cee},
	eid = {L41},
	eprint = {2403.08853},
	journal = {\apjl},
	month = jun,
	number = {2},
	pages = {L41},
	primaryclass = {astro-ph.GA},
	title = {{The High-Redshift Gas-Phase Mass{\textendash}Metallicity Relation in FIRE-2}},
	volume = {967},
	year = 2024}

@article{Dave:2012aa,
	adsurl = {https://ui.adsabs.harvard.edu/abs/2012MNRAS.421...98D},
	archiveprefix = {arXiv},
	author = {{Dav{\'e}}, Romeel and {Finlator}, Kristian and {Oppenheimer}, Benjamin D.},
	doi = {10.1111/j.1365-2966.2011.20148.x},
	eprint = {1108.0426},
	journal = {\mnras},
	month = mar,
	number = {1},
	pages = {98-107},
	primaryclass = {astro-ph.CO},
	title = {{An analytic model for the evolution of the stellar, gas and metal content of galaxies}},
	volume = {421},
	year = 2012}

@article{Peng:2014aa,
	adsurl = {https://ui.adsabs.harvard.edu/abs/2014MNRAS.443.3643P},
	archiveprefix = {arXiv},
	author = {{Peng}, Ying-jie and {Maiolino}, Roberto},
	doi = {10.1093/mnras/stu1288},
	eprint = {1402.5964},
	journal = {\mnras},
	month = oct,
	number = {4},
	pages = {3643-3664},
	primaryclass = {astro-ph.CO},
	title = {{From haloes to Galaxies - I. The dynamics of the gas regulator model and the implied cosmic sSFR history}},
	volume = {443},
	year = 2014}

@article{Sanders:2021aa,
	adsurl = {https://ui.adsabs.harvard.edu/abs/2021ApJ...914...19S},
	archiveprefix = {arXiv},
	author = {{Sanders}, Ryan L. and {Shapley}, Alice E. and {Jones}, Tucker and {Reddy}, Naveen A. and {Kriek}, Mariska and {Siana}, Brian and {Coil}, Alison L. and {Mobasher}, Bahram and {Shivaei}, Irene and {Dav{\'e}}, Romeel and {Azadi}, Mojegan and {Price}, Sedona H. and {Leung}, Gene and {Freeman}, William R. and {Fetherolf}, Tara and {de Groot}, Laura and {Zick}, Tom and {Barro}, Guillermo},
	doi = {10.3847/1538-4357/abf4c1},
	eid = {19},
	eprint = {2009.07292},
	journal = {\apj},
	month = jun,
	number = {1},
	pages = {19},
	primaryclass = {astro-ph.GA},
	title = {{The MOSDEF Survey: The Evolution of the Mass-Metallicity Relation from z = 0 to z 3.3}},
	volume = {914},
	year = 2021}

@article{Kobayashi:2020aa,
	adsurl = {https://ui.adsabs.harvard.edu/abs/2020ApJ...900..179K},
	archiveprefix = {arXiv},
	author = {{Kobayashi}, Chiaki and {Karakas}, Amanda I. and {Lugaro}, Maria},
	doi = {10.3847/1538-4357/abae65},
	eid = {179},
	eprint = {2008.04660},
	journal = {\apj},
	month = sep,
	number = {2},
	pages = {179},
	primaryclass = {astro-ph.GA},
	title = {{The Origin of Elements from Carbon to Uranium}},
	volume = {900},
	year = 2020}

@article{Chiappini:2003aa,
	adsurl = {https://ui.adsabs.harvard.edu/abs/2003MNRAS.339...63C},
	archiveprefix = {arXiv},
	author = {{Chiappini}, Cristina and {Romano}, Donatella and {Matteucci}, Francesca},
	doi = {10.1046/j.1365-8711.2003.06154.x},
	eprint = {astro-ph/0209627},
	journal = {\mnras},
	month = feb,
	number = {1},
	pages = {63-81},
	primaryclass = {astro-ph},
	title = {{Oxygen, carbon and nitrogen evolution in galaxies}},
	volume = {339},
	year = 2003}

@article{Woosley:1995aa,
	adsurl = {https://ui.adsabs.harvard.edu/abs/1995ApJS..101..181W},
	author = {{Woosley}, S.~E. and {Weaver}, Thomas A.},
	doi = {10.1086/192237},
	journal = {\apjs},
	month = nov,
	pages = {181},
	title = {{The Evolution and Explosion of Massive Stars. II. Explosive Hydrodynamics and Nucleosynthesis}},
	volume = {101},
	year = 1995}

@article{Andrews:2013aa,
	adsurl = {https://ui.adsabs.harvard.edu/abs/2013ApJ...765..140A},
	archiveprefix = {arXiv},
	author = {{Andrews}, Brett H. and {Martini}, Paul},
	doi = {10.1088/0004-637X/765/2/140},
	eid = {140},
	eprint = {1211.3418},
	journal = {\apj},
	month = mar,
	number = {2},
	pages = {140},
	primaryclass = {astro-ph.CO},
	title = {{The Mass-Metallicity Relation with the Direct Method on Stacked Spectra of SDSS Galaxies}},
	volume = {765},
	year = 2013}

@book{Osterbrock:2006aa,
	author = {{Osterbrock}, Donald E. and {Ferland}, Gary J.},
	title = {Astrophysics of gaseous nebulae and active galactic nuclei},
	year = {2006}}




\appendix

\section{Fundamental metallicity relation parametrization}
\label{sec:Fundamental metallicity relation parametrization}

\begin{figure} 
\centering
	\includegraphics[width=\columnwidth]{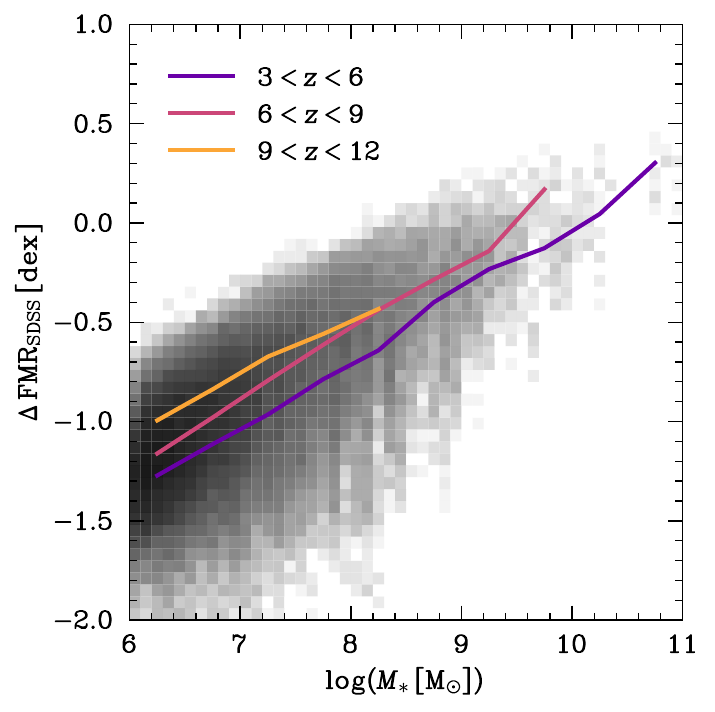}
    \caption{The same as Fig.~\ref{fig:fmr_offset_metallicity_curti}, except we use the \citet{Andrews:2013aa} FMR parametrization. This causes a weak, inverted redshift dependence and normalization offset, but the stellar mass-dependence is still dominant.}
    \label{fig:fmr_offset_metallicity_andrews}
\end{figure}

In Fig.~\ref{fig:fmr_offset_metallicity_andrews} we show the FMR offset using the \citet{Andrews:2013aa} parameterization, which causes a weak inverse redshift trend and change in normalization. The stellar mass is still the dominant factor determining offset from the FMR, possibly due to extrapolation to an SFR and mass range outside of the original data.

\section{Giant molecular cloud chemical abundance definitions}
\label{sec:Giant molecular cloud chemical abundance definitions}

\begin{figure} 
\centering
	\includegraphics[width=\columnwidth]{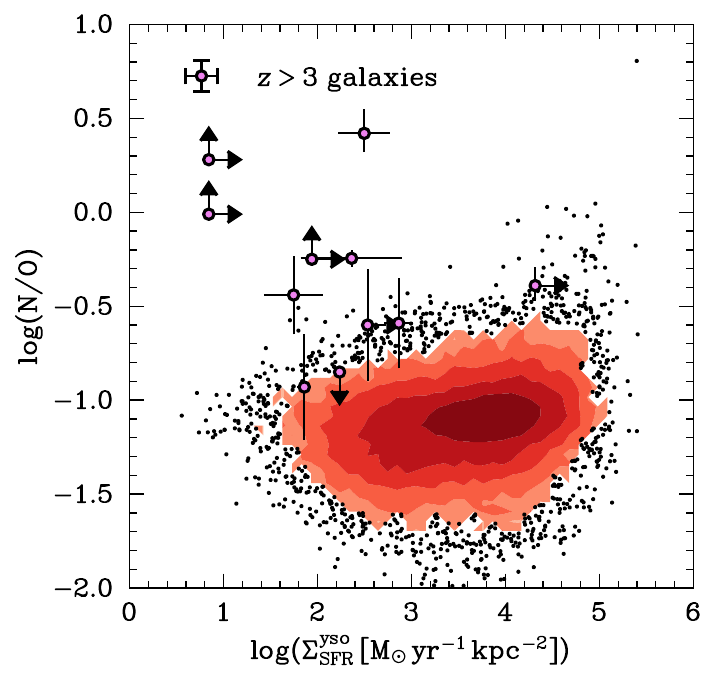}
    \caption{The gas-phase nitrogen to oxygen ratio as a function of SFR surface density in GMCs, as estimated from associated stellar particles younger than 3\,Myr.}
    \label{fig:gmc_sfr_density_no2}
\end{figure}

In Fig.~\ref{fig:gmc_sfr_density_no} we chose to use the instantaneous SFR to calculate $\Sigma_{\mathrm{SFR}}$ and to use SFR-weighted chemical abundances. This allows us to show that there are dense, nitrogen-rich GMCs in \thzoom which are actively forming nitrogen-rich stars.

For comparison, we show in Fig.~\ref{fig:gmc_sfr_density_no2} the same plot but choosing instead to use the SFR measured from associated stars averaged over 3\,Myr and with mass-weighted chemical abundances. $\Sigma_{\mathrm{SFR}}^\mathrm{yso}$ tends to be higher than $\Sigma_{\mathrm{SFR}}^\mathrm{inst}$. The mass-weighted N/O ratios tend to be lower than the SFR-weighted ratios in nitrogen-rich GMCs, albeit not drastically.


\bsp	
\label{lastpage}
\end{document}